\UseRawInputEncoding
\documentclass{./emulateapj}
\usepackage{color,epsfig,graphics,graphicx} 

\usepackage{natbib,ulem}
\definecolor{brown}{rgb}{0.6,0.4,0.2} 
\definecolor{purple}{rgb}{0.5,0,0.5} 
\shorttitle{Infrared Ejecta and Cold Dust in N132D } 
\shortauthors{Rho et al.}

\def\one{{\,\sc i}}
\def\two{{\,\sc ii}}

\newcommand{\kms}{km\,s$^{-1}$}

\newcommand{\spitzer}{\textit{Spitzer}} 
\newcommand{\herschel}{\textit{Herschel}}

\newcommand{\oiiif}{\ion{[O}{3}]}
\newcommand{\oivf}{\ion{[O}{4}]}

\newcommand{\siif}{\ion{[S}{2}]}
\newcommand{\siiif}{\ion{[S}{3}]}
\newcommand{\siliif}{\ion{[Si}{2}]}

\newcommand{\neiif}{\ion{[Ne}{2}]}
\newcommand{\neiiif}{\ion{[Ne}{3}]}

\newcommand{\nevf}{\ion{[Ne}{5}]}
\newcommand{\niciif}{\ion{[Ni}{2}]}
\newcommand{\ariif}{\ion{[Ar}{2}]}

\newcommand{\feiif}{\ion{[Fe}{2}]}

\newcommand{\mic}{$\mu$$m$}

\newcommand\new[1]{{\textbf{#1}}}
\shorttitle{Ejecta and Supernova-Dust in N132D}
 
\begin{document} 

\title{Infrared Ejecta and Cold Dust in the young supernova remnant N132D}

\author{
Jeonghee Rho\altaffilmark{1},
Aravind P. Ravi\altaffilmark{2},
Jonathan D. Slavin\altaffilmark{3}
Heechan Cha\altaffilmark{4, 5}
}
\altaffiltext{1}{SETI Institute, 189 N. Bernardo Ave., Mountain View, CA 94043; jrho@seti.org}
\altaffiltext{2}{Department of Physics, University of Texas at Arlington, Box 19059, 
Arlington, TX 76019, USA}
\altaffiltext{3}{Center for Astrophysics $|$ Harvard \& Smithsonian, 60 Garden Street, Cambridge, MA 02138, USA}
\altaffiltext{4}{Department of Physics and Astronomy, University of Texas at San Antonio, San Antonio, TX 78249, USA}
\altaffiltext{5}{Department of Astronomy and Space Science, Chungbuk National University, Cheongju, 28644, Republic of Korea}

\begin{abstract} 
We present {\it Spitzer}, {\it WISE}, and \herschel\ observations of the young supernova remnant (SNR) N132D in the LMC, including 3-40 \mic\ \spitzer\ IRS mapping, 12 \mic\ WISE and 70, 100, 160, 250, 350, and 500 \mic\ \herschel\ images. The high-velocity lines of [\ion{Ne}{2}] at 12.8 \mic, [\ion{Ne}{3}] at 15.5 \mic, and [\ion{O}{4}] 26 \mic\ reveal infrared ejecta concentrated in a central ring and coincide the optical and X-ray ejecta. \herschel\ images reveal far-IR emission coinciding with the central ejecta, which suggests that the IR emission is freshly formed, cold dust in the SN-ejecta. The infrared spectra are remarkably similar to those of another young SNR of 1E0102 with Ne and O lines. Shock modeling of the Ne ejecta emission suggests a gas temperature of 300 - 600 K and densities in the range $1000-2\times10^4$ cm$^{-3}$ in the post-shock photoionized region. The IR continuum from the ejecta shows an 18 \mic-peak dust feature. We performed spectral fitting to the IRS dust continuum and \herschel\ photometry. The dust mass associated with the central ejecta is 1.25$\pm$0.65 M$_\odot$, while the 18 \mic\ dust feature requires forsterite grains. The dust mass of the central ejecta region in N132D is higher than those of other young SNRs, which is likely associated with its higher progenitor mass. We discuss the dust productivity in the ejecta of N132D and infer its plausible implications for the dust in the early Universe.

\end{abstract} 

\section{Introduction} 

Core-collapse supernovae (ccSN) are the catastrophic deaths of massive stars. The most direct method of probing the nuclear physics that takes place in ccSNe is to examine their nucleosynthetic products and to constrain the explosion mechanism by observing their supernova remnants (SNRs). Dust formation in ccSNe, e.g. Types II, Ib and Ic, may account for the presence of large amounts of dust observed in high-z  galaxies, since their progenitors, stars much more massive than the sun, evolve on much shorter timescales (millions to a few tens of millions of years) than the other main source of dust in the current Universe, asymptotic giant branch stars, and can eject large amounts of heavy elements into the ISM. If a typical Type II SN were able to condense only 10\% of its heavy elements into dust grains, it could eject around half a solar mass of dust into the ISM. The predicted dust mass formed in a ccSN depends on  its progenitor mass; for a progenitor mass of 15 to 30 M$_{\odot}$, the predicted dust mass ranges from 0.1 to 1.0 M$_{\odot}$ \citep{nozawa03, todini01}. For a star formation rate (SFR) 100 times that of the Milky Way this would lead to a dust mass of order $10^8~\rm M_\odot$ at $z$ = 6.3-7.5 \citep{nozawa03, michalowski15}.

%Figure1
\begin{figure*}
\includegraphics[scale=0.75,angle=0,width=14.5truecm]{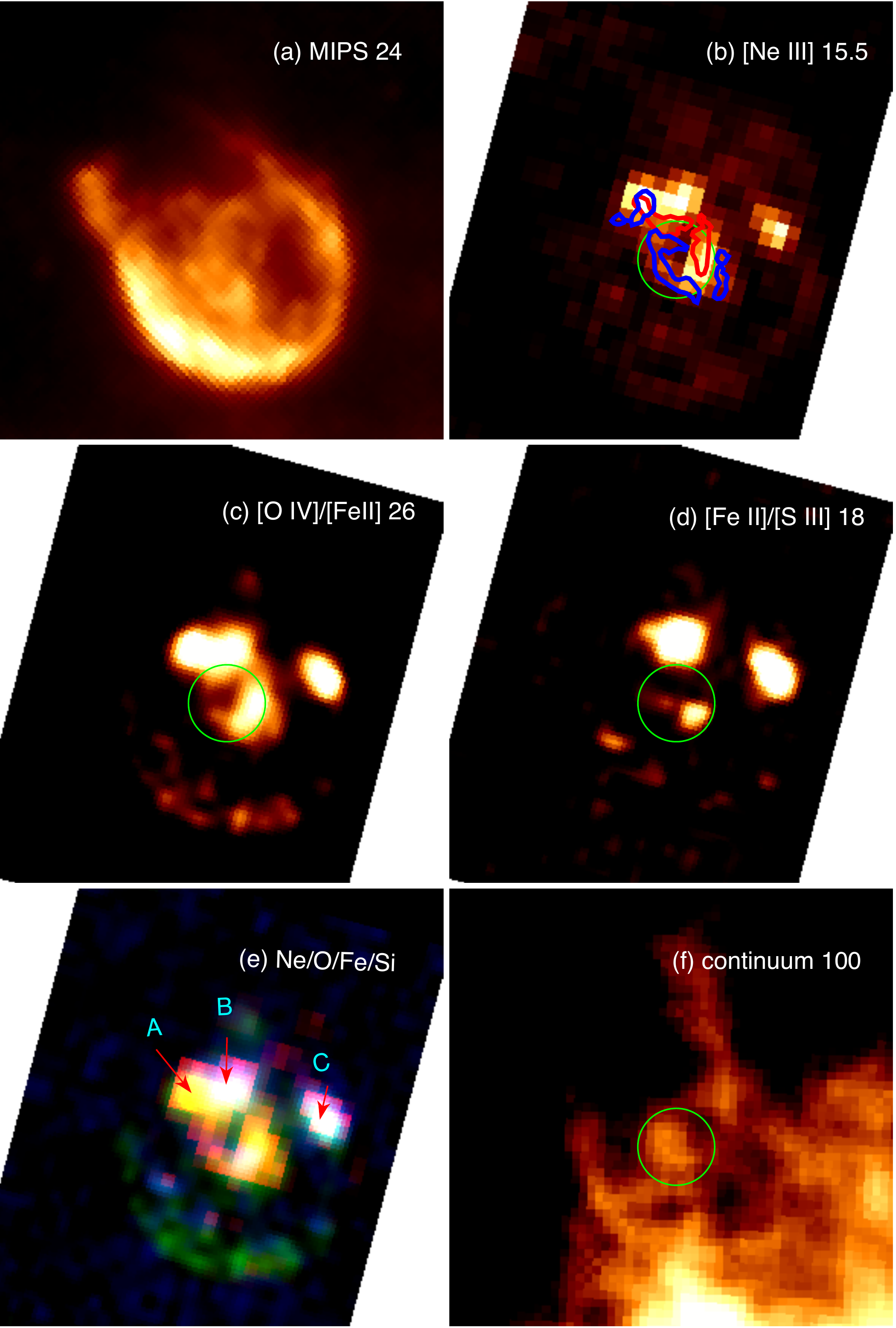}
\caption{\spitzer\ and \herschel\ images of N132D: $(a)$ MIPS 24 \mic; $(b)$ [\ion{Ne}{3}]15.5 \mic;
$(c)$ [\ion{O}{4}]/[\ion{Fe}{2}] 26 \mic\ (indicating [\ion{O}{4}] at 25.89 \mic\ and/or [\ion{Fe}{2}] at 25.98 \mic\ and the IRS spectrum implies that the emission is mainly from [\ion{O}{4}] as shown in Fig.~\ref{n132dirsspec1}), which coincides optical ejecta (contours are marked in green); $(d)$ [\ion{Fe}{2}]/[\ion{S}{3}] ([\ion{Fe}{2}] 17.9 \mic\ and/or
[\ion{S}{3}] at 18.7 \mic; $(e)$ three color images of ejecta (red, green, and blue representing [\ion{Ne}{3}], [\ion{O}{4}]/[\ion{Fe}{2}], [\ion{Fe}{2}]/[\ion{Si}{2}], respectively); $(f)$ \herschel\ 100 \mic. In b, c, d, and f the central emission is denoted by green circles. The optical \oiiif\ (5007 \AA) ejecta \citep{morse95} of blue-shifted (in cyan) and red-shifted (in red) ejecta are marked in the [\ion{Ne}{3}] image $(b)$. In panel $(e)$, regions of northeastern ejecta, Lasker's Bowl, and West Complex are marked as ``A", ``B" and ``C", respectively.
The images are centered on R.A.\ $5^{\rm h} 25^{\rm m} 03.23^{\rm s}$ and Dec.\ -69$+^\circ$38$^{\prime}$24.43$^{\prime \prime}$ (J2000) with each field of view (FOV) of 2.84$'$x2.84$'$. 
The north is up, and the east is to the left; all other images have the same direction.}
\label{n132dspitzerimages}
\end{figure*}

\herschel\ and \spitzer\ observations of four SNRs: Cas A \citep{rho08, delooze17}, SN 1987A \citep{matsuura11, matsuura15}, the Crab Nebula \citep{gomez12}, and G54.1+0.3 \citep{rho18, temim17} yielded significant amounts of dust (0.1-1 M$_\odot$ per SN). The longer wavelength far-infrared and submillimeter (FIR-submm) observations for G54.1+0.3 revealed a cooler component of dust at $\sim 35\,$K, and the Photodetector Array Camera (PACS) and the Spectral and Photometric
Imaging Receiver (SPIRE) images show ejecta emission. Both Cas A and G54.1+0.3 show 21 and 11 $\mu$m dust features which are attributed to silica (SiO$_2$) and SiC, respectively \citep{rho18}. \cite{delooze17} performed a spatially resolved analysis of ISM and SN dust emission using 17 - 500~$\mu$m maps that allowed them to exclude the ISM contribution. The derived dust masses for both SN were between 0.4 and 0.6 M$_\odot$. {\it Herschel} observations reveal significant quantities of cool dust (at 20-50 K) with masses of 0.1-0.65 M$_\odot$ in the pulsar-wind nebulae of G11.2-0.3, G21.5-0.9, and G29.7-0.3 \citep{chawner19} and MSH 15-52 \citep[G320.4-1.2,][]{millard21}. The dust masses of the eight young SNRs (YSNRs) observed with \herschel\ and \spitzer\ exceed by an order of magnitude the previously detected dust masses and are two orders of magnitude higher than those observed in SNe \citep{kotak06, gall11}. They also are in agreement with theoretical models of dust condensation \citep{nozawa03, todini01, sluder18}. These results support the theory that the huge quantities of SN-dust in high-redshift galaxies are produced in ccSNe \citep{isaak02, laporte17, spilker18, dayal22, sandro22, sommovigo22}.

Despite this evidence that large amounts of dust are created in ccSNe, uncertainties remain about the contribution of dust by such SNe to the ISM. The cold dust observed in young remnants is generally in the interior of the remnants \citep[see][]{delooze17} and has not yet encountered the reverse shock, which will eventually propagate to the center of the remnant. The amount of dust destruction, once the shock hits the dense clumps of ejecta, depends on a variety of parameters such as the grain size distribution, the grain material, and the density contrast between the clumps and lower density ejecta as well as the shock speed \citep{silvia12,kirchschlager19}. Because the dust formed in ccSNe includes sufficiently large grains (0.1 - 0.5 \mic),  a significant fraction of the grains can survive \citep[10 - 20\% for silicate dust and 30 - 50\% for carbon dust;][]{slavin20}.

Theoretical estimates of the destruction range between almost total to high surviving fractions (\citealt{kirchschlager19} and references therein). Observational estimates from Cas A for areas of the remnant where the newly-formed dust has already encountered the shock suggest a significant mass of dust can survive into the interstellar medium \citep{delooze17,priestley22}.

N132D (SNR B0525-69.6 or SNR J052501-693842) is one of the young SNRs in the Large Magellanic Cloud (LMC) with a diameter of 100$''$ (physical size of 25 pc) and age of  $\sim$2500 (1300 - 3500) yr \citep{law20, sutherland95, lasker80, morse95}. 
Despite its age, N132D shows its youth via ejecta emission in its central region. Optical observations identified O-rich ejecta emission (as in Cas A), an inner ring morphology which shows elevated metal abundances, and red and blue-shifted oxygen-rich ejecta with velocities as high as 4400 \kms\ \citep[][see blue- and red-contours in Fig.~\ref{n132dspitzerimages}b]{morse95}. The SNR's 3-D structures of optical emission observed by the 2.3m telescope at Siding Spring Observatory show that the majority of the ejecta emission comes from a ring of diameter $\sim$12 pc \citep{vogt11}. The 3-D kinematic reconstruction of the optically-emitting, oxygen-rich ejecta observed by the 6.5$m$ Magellan telescope shows a torus-like geometry \citep{law20}. Analysis of both the optical and X-ray emission \citep{borkowski07, hwang93} has shown that the ejecta are oxygen-rich.  N132D has been classified as a type Ib SNR and its progenitor has been suggested to be a Wolf-Rayet (W/O) star with a mass between 30 and 35 M$_\odot$ \citep{blair00}. The X-ray observations identified highly ionized oxygen ejecta \citep{borkowski07}. K-shell Fe emission in X-rays is detected at the center, and X-ray oxygen emission is detected from the NE ejecta knots (see Figure \ref{n132dspitzerimages}) and at the center \citep{behar01}.

The shell emission of N132D is bright in the southeast, where the dominant contributor is the shocked ISM \citep{blair00}. Our previous \spitzer\ observations toward the southeastern shell revealed a plateau of polycyclic aromatic hydrocarbon (PAH) emission at 15-20 $\mu$m, which is due to PAHs with a relatively large number of carbon atoms \citep{tappe06, tappe12}. 
The results suggest that PAH molecules in the surrounding medium are swept up and processed in the hot gas of the blast wave shock, where they survive the harsh conditions long enough to be detected. 

In this paper, we use archival \herschel, {\it WISE}, and \spitzer\ images and \spitzer\ Infrared Spectrograph (IRS) observations of N132D and present maps of infrared emission from the IR ejecta in the YSNR N132D and the cold dust likely associated with the IR ejecta. The \herschel\ and {\it WISE} observations of N132D are presented here for the first time. We present dynamic and physical properties of IR ejecta, including shock modeling (Sections \ref{sec:3.1} -- \ref{sec:3.2} and \ref{SDNeejectalineratio} -- \ref{Sshockmodelsejecta}). The \herschel\ images reveal emission from a cold blob that appears at the remnant's center and coincides with the IR/optical ejecta (Sections \ref{Scolddust} -- \ref{SSED}). We estimate the dust mass of the cold blob using \herschel\ and \spitzer\ spectral energy distributions (SEDs; Sections \ref{SSED} and \ref{Sdustfit1} -- \ref{Sdustmass}). We find that the dust mass associated with the ejecta of N132D is higher than those of other young SNRs and its implications for the dust in the early Universe (Section \ref{Sdustmass} for details).

\section{Observations}
\label{Sobs}

\begin{figure}
\includegraphics[scale=1,angle=0,width=8.2truecm]{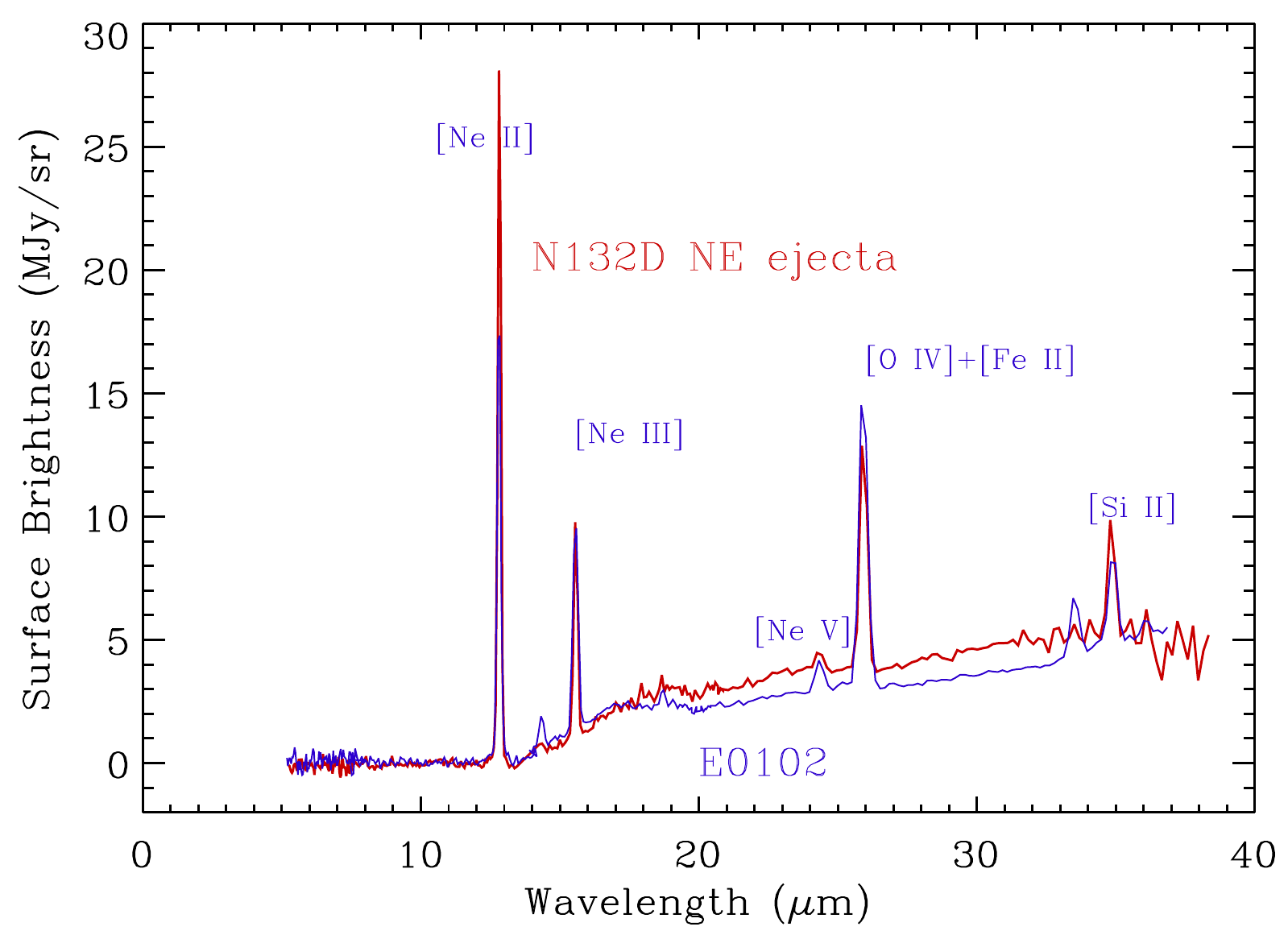}
\caption{\spitzer\ IRS spectrum of N132D (in red) toward NE ejecta (marked `A' in Figure \ref{n132dspitzerimages}e) is compared with that of 1E0102 ejecta emission 
\citep[in blue; Figure 3 of][]{rho09}, showing remarkably similar line strengths
and dust continuum shapes. 
One major difference is the strength of the \neiif\ line, resulting in the different ratio of \neiif\ to \neiiif.}
\label{n132dirsspec1}
\end{figure}

%Figure3
\begin{figure*}
\includegraphics[scale=1,angle=0,width=17.0truecm]{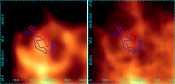}
\caption{{\it WISE} 12 \mic\ (band 3: $w3$) (left) and \herschel\ 100\mic\ (right) images of N132D. Both show central emission as well as circular shell-like emission. The optical \oiiif\ (5007 \AA) ejecta of blue-shifted (in blue) and red-shifted (in red) ejecta are superposed. The eastern red-shifted ejecta coincides with the {\it WISE} 12\mic\ and {\it Herschel} 100\mic\  emission.}
\label{n132dwise}
\end{figure*}

%Figure4
\begin{figure*}
\begin{center}
\includegraphics[scale=1,angle=0,width=15truecm]{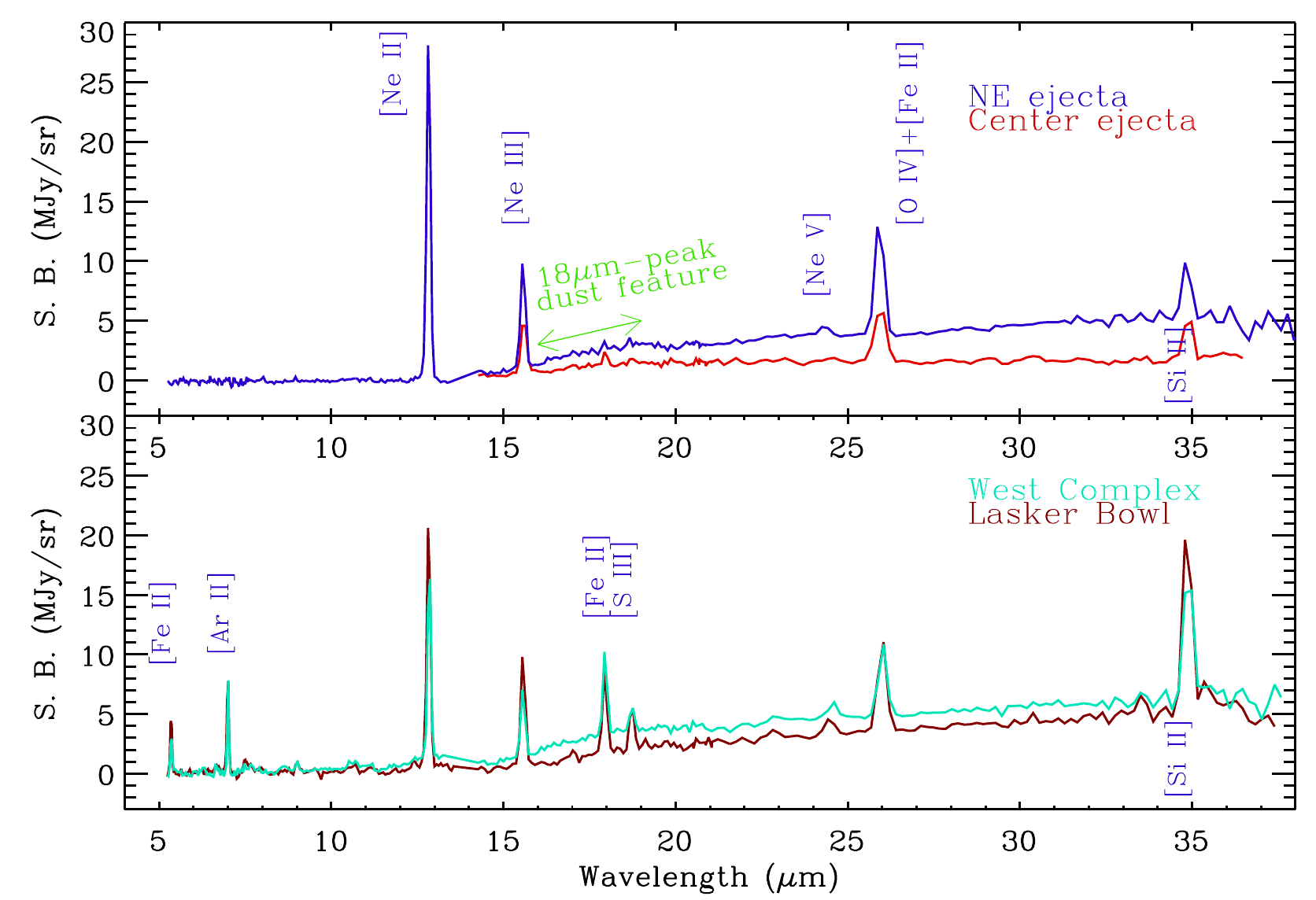}
\caption{Comparison of \spitzer\ IRS spectra (y-axis unit is Surface Brightness) of N132D toward NE Ejecta (the 18\,\mic-peak dust feature is marked), Lasker's Bowl (marked as `B' in Figure \ref{n132dspitzerimages}e), West Complex (`C'), and 
the central ejecta (marked as a circle in Figure \ref{n132dspitzerimages}f; see Section \ref{Scolddust} for details). The spectrum of the central ejecta in N132D has only long-wavelength ($>$15\mic) coverage.
The ejecta spectra are dominated by O and Ne lines, Lasker's Bowl and West Complex from the CSM/ISM dense knots include Fe, Ar, and S lines.
The position of NE ejecta (marked as `A' in Figure \ref{n132dspitzerimages}e) is at R. A.  $05^{\rm h} 25^{\rm m} 05.62^{\rm s}$ and Dec.\,$-69^\circ$38$^{\prime}12.79^{\prime \prime}$ (N132D-P3 in the HST spectroscopy observations \citet{blair00}), and the center of ejecta is at R. A. $05^{\rm h} 25^{\rm m} 02.03^{\rm s}$ and Dec.\,$-69^\circ$38$^{\prime}44.8^{\prime \prime}$ (approximate position is N132D-P2 in \citet{blair00}).  Lasker's Bowl is R. A. $05^{\rm h} 25^{\rm m} 02.40^{\rm s}$ and Dec.\ $-69^\circ$38$^{\prime}$ 12.7$^{\prime \prime}$, J2000) and West Complex is at R. A. 05$^{\rm h} 24^{\rm m} 55.15^{\rm s}$ and Dec.\ $-69^\circ$38$^{\prime}$ 26.5$^{\prime \prime}$.}
\label{n132dirsspecall}
\end{center}
\end{figure*}

\subsection{{\it Herschel} observations}
{\it Herschel} observations of N132D were taken as a part of \herschel\ Large Magellanic Cloud (LMC) GTO and Key project (Proposal\_ID of KPOT\_mmeixner\_1 and SDP\_mmeixner\_1) to cover the entire LMC with the Herschel Photodetector Array Camera \citep[PACS;][]{pogslitsch10} and Spectral and Photometric Imaging Receiver (SPIRE) images using the parallel mode. The observational IDs are 1342187188, 1342195707, 1342195708, 1342202224, and 1342202225. We have used both UPDP (User Provided Data Products) and HPDP (highly-processed data products) and cross-checked the photometry between them. The flux calibration uncertainty for PACS is less than 10\% \citep{pogslitsch10} and the expected color corrections are small compared to the calibration errors. We, therefore, adopt a 10\% calibration error for the 100 and 160 \mic\ data. The SPIRE calibration methods and accuracy are outlined by \cite{swinyard10} and are estimated to be 7\%. The spatial resolutions of \herschel\ images (Fig.~\ref{n132dherschelimages}) are 8$\arcsec$, 12$\arcsec$, 18.1$\arcsec$, 24.9$\arcsec$ and 36.4$\arcsec$ at 100, 160, 250, 350 and 500 \mic, respectively. Herschel observations of N132D have not been previously published. They enable us to access the coldest dust component, which typically makes up most of the dust mass in SNRs \citep{gomez12,delooze17}.

\subsection{{\it WISE} Observations}
We used archival Wide-field Infrared Survey Explorer ({\it WISE}) images of N132D. {\it WISE} is a whole sky mid-IR survey that uniquely provides data beyond the Galactic plane \citep{wright10}. {\it WISE} $w1$, $w2$, $w3$, and $w4$ images are centered at 3.4, 4.6, 12, and 22 $\mu$m, respectively. Unique to the {\it WISE} (i.e., not covered by \spitzer\ (3.4, 4.6, and 24 $\mu$m),  is its  broad-band filter ($w3$) centered at 12 $\mu$m. The observational parameters are listed in Table \ref{Tobs}. {\it WISE} $w1$ and $w2$ images did not detect IR emission from N132D. The {\it WISE} $w3$ image of N132D is shown in Figure \ref{n132dwise}, showing the central ejecta emission as well as the circular shell emission. The {\it WISE} $w4$ image is similar to \spitzer\ 24\mic\ image but 3 times lower spatial resolution.

\subsection{\spitzer\ Observations}

We used archival observations  (PID of 3483 and 30372) by \spitzer\ for IRS Long-Low (IRS-LL) mapping covering the entire SNR, and IRS staring of Short-Low (SL) observations toward 4 positions \citep[see][]{tappe06}. The staring mode was 12 cycles of 60 s exposures for SL. IRS-LL mapping is 2 cycles of 30 s exposure for 10$\arcsec$ stepsize, resulting in 4 minutes of integration per position. 
The observations are summarized in Table \ref{Tobs}. The images and spectra are shown in Figures \ref{n132dspitzerimages} - \ref{n132dirsspecall}. We used \spitzer\ IRS spectra of ejecta emission to construct the spectral energy distribution (SED) by combining them with the \herschel\ photometry. Spectral model fitting of dust emission have not been performed and analyzed previously. Note, however, that \cite{tappe06, tappe12} presented the shocked ISM emission from the southeastern shell and studied its PAH emission.

\section{Results}
\subsection{Infrared Images and Spectra}
\label{sec:3.1}

Figure \ref{n132dspitzerimages} shows \spitzer\ and \herschel\ images of N132D. Figure \ref{n132dirsspec1} is the \spitzer\ IRS spectrum of N132D  at the location of the NE ejecta (marked as ``A" in Fig.~\ref{n132dspitzerimages}e) and shows bright lines of \neiif\ (12.8 $\mu$m), \neiiif\ (15.5 $\mu$m), \oivf\ and/or \feiif\ (26 $\mu$m), weak lines of \nevf\ (14.3 $\mu$m), \nevf\ (24.3 $\mu$m), and \neiiif\ (36 $\mu$m), and possibly \siif, \siiif\ (33.5 $\mu$m), and \siliif\ (34.89 $\mu$m). The O and Ne lines dominate the IR spectrum of N132D, as they do in the optical ejecta \citep[P3 position in][]{blair00} and the X-ray ejecta \citep[green emission in Fig.~2 of][]{borkowski07}. The spectrum of N132D and  that of the 1E0102 ejecta are compared in Figure \ref{n132dirsspec1}. The spectra of the two SNRs are remarkably similar in their lines of Ne and O and their continuum shapes, indicating that the N132D emission in the NE position is also likely from the SN ejecta as it is in the case of 1E0102 \citep{rho09}.  The line intensity ratio between \neiif\ and \neiiif\ is much higher in N132D than in 1E0102, indicating that Ne is less ionized in N132D (see Sections \ref{SDNeejectalineratio} and \ref{Sshockmodelsejecta} for discussion).

From the \spitzer\ IRS spectral cube, we generate line maps of \neiiif\ at 15.5 \mic,  \oivf\ at 26\mic, and [Si\,II] at 34.8 \mic\ and present them in Figure \ref{n132dspitzerimages}(b-e). The three line maps each show strong emission at the center with the morphology of a central ring. The \neiiif\  and \oivf\ emission coincides with the known optical ejecta (see Fig.~\ref{n132dspitzerimages}b), including NE ejecta and the central, bright far-IR emission (marked as a circle in Fig.~\ref{n132dspitzerimages}f). Figure \ref{n132dwise} shows {\it WISE} 12 \mic\ image, which is essentially a [Ne~II] map since [Ne~II] (12.8 \mic) dominates the emission in the $w3$ band. The central emission in the {\it WISE} image coincides with that of the \herschel\ 100 \mic\ image. The CSM/ISM knots of Lasker's Bowl and the West Complex (marked as `B' and `C' in Figure~\ref{n132dspitzerimages}e) produce Fe/Si line emission as well as the Ne and O lines. The Fe/Si emission (from the CSM/ISM knots) at the center (the southwestern part of the circled region in Figure~\ref{n132dspitzerimages}d) is anti-correlated with the continuum emission at 100 \mic\ (eastern part of the circled region of Fig.~\ref{n132dspitzerimages}f).

%Figure5
\begin{figure}
\includegraphics[scale=0.75,angle=0,width=8.5truecm]{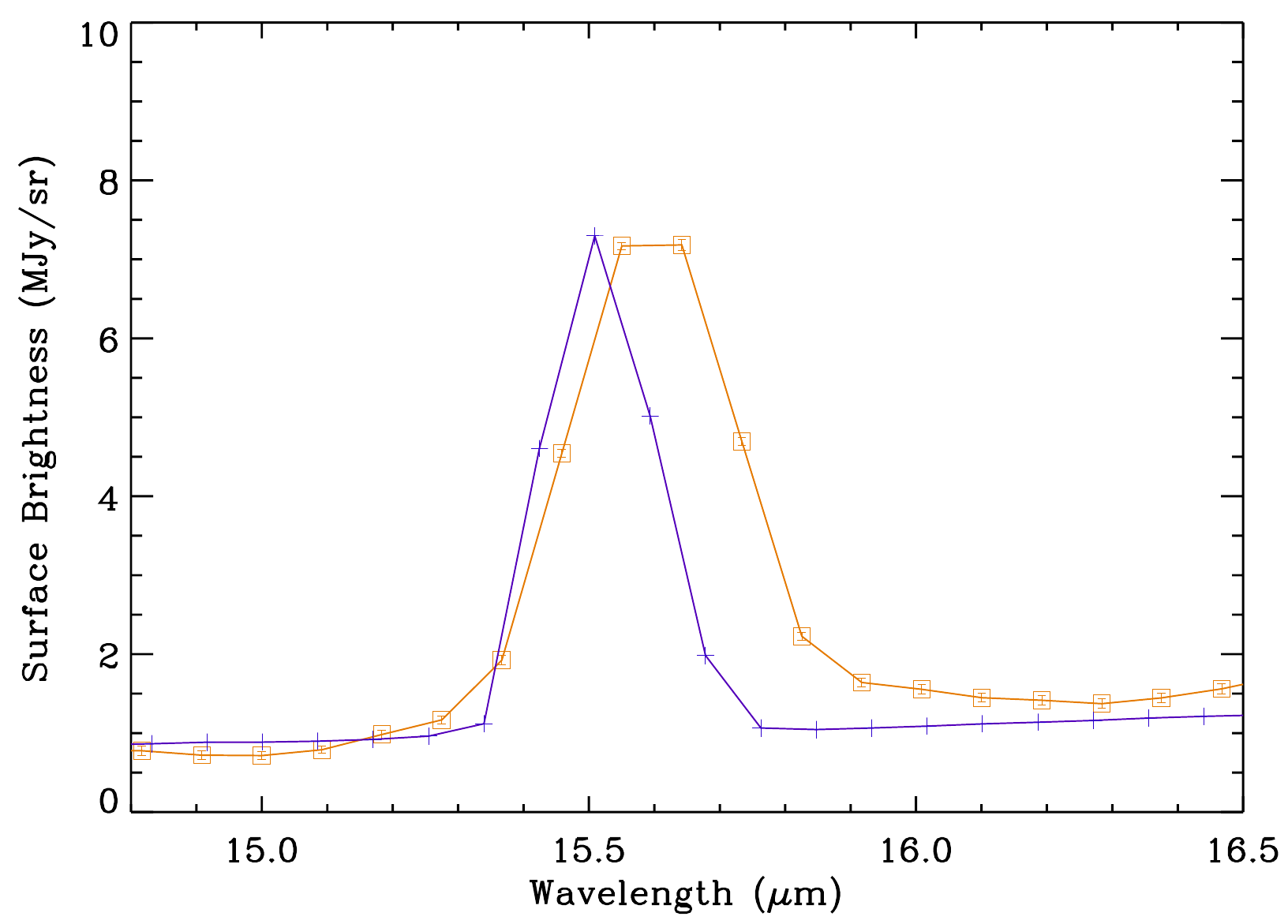}
\caption{Spectral line profile of \neiiif\ toward the central ejecta (in orange) compared with the unresolved line profile of Planetary Nebula (PN) NGC6543 (in blue). 
\label{n132dneiiivel}}
\end{figure}

%Figure6
\begin{figure*}
\includegraphics[scale=1,angle=0,width=16truecm]{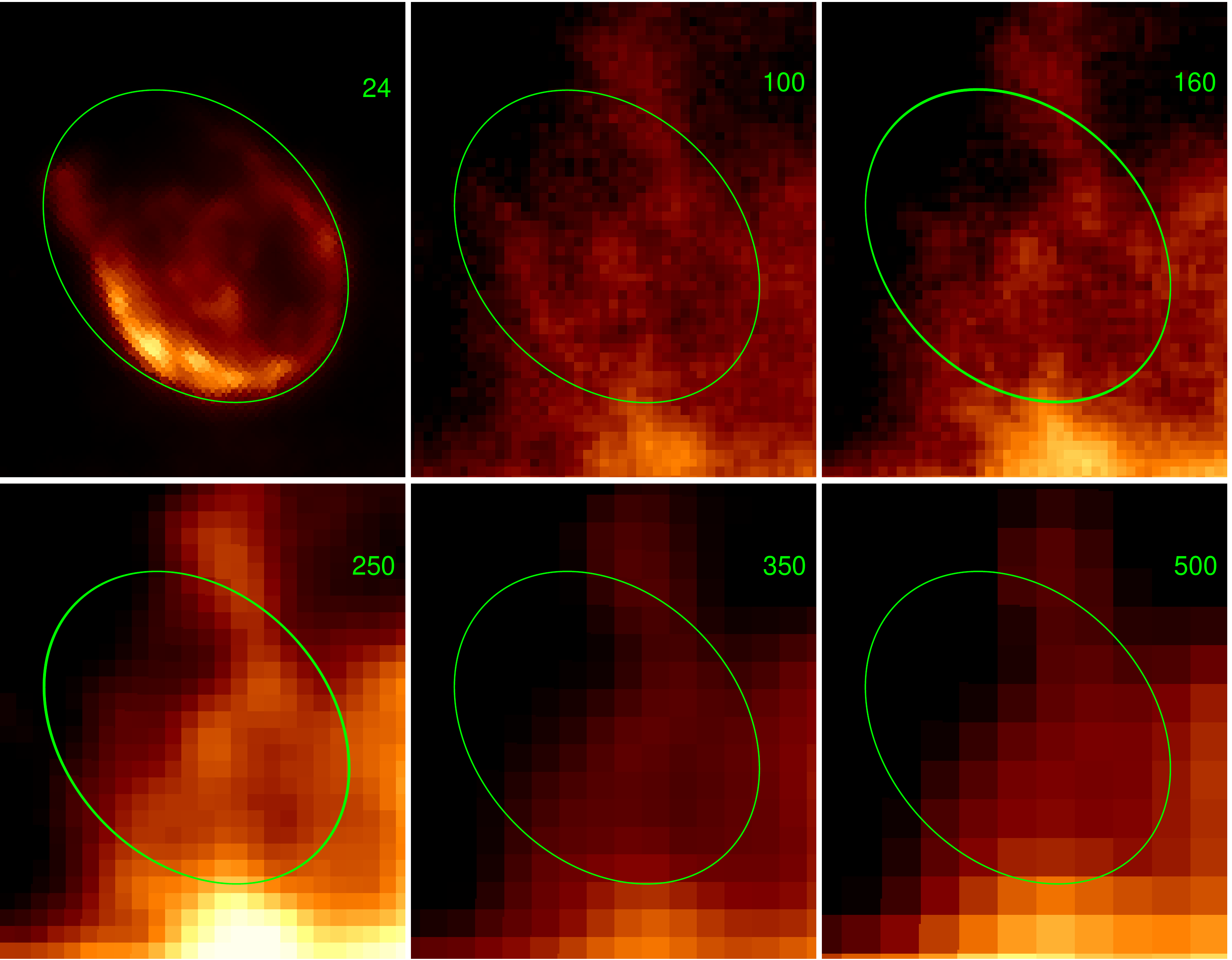}
\caption{\spitzer\ and \herschel\ images towards the SNR N132D with the wavelength in \mic\ unit: \spitzer\ 24 $\mu$m $(a)$, \herschel\ PACS 100 $\mu$m $(b)$,  \herschel\ PACS 160 $\mu$m $(c)$, \herschel\ SPIRE 
250 $\mu$m $(d)$, \herschel\ SPIRE 350 \mic\ $(e)$, \herschel\ SPIRE 500 \mic\ $(f)$ where the wavelength of each image is labeled. The central ejecta blobs (marked as a circle or ellipse in Figures \ref{n132dspitzerimages} and \ref{n132dphotometryregionmark}) are detected from 24 to 350 \mic, demonstrating the presence of both warm and cold dust; interpretation of the 500 \mic\ image is unclear due to low angular resolution. 
The image is centered on R.A.\ $5^{\rm h} 25^{\rm m} 03.16^{\rm s}$ and Dec.\ -69$+^\circ$38$^{\prime}$24.13$^{\prime \prime}$ (J2000) with a FOV of 2.35$'$x2.89$'$.}
\label{n132dherschelimages}
\end{figure*}

%Figure7
\begin{figure*}
\includegraphics[scale=1,angle=0,height=6.3truecm]{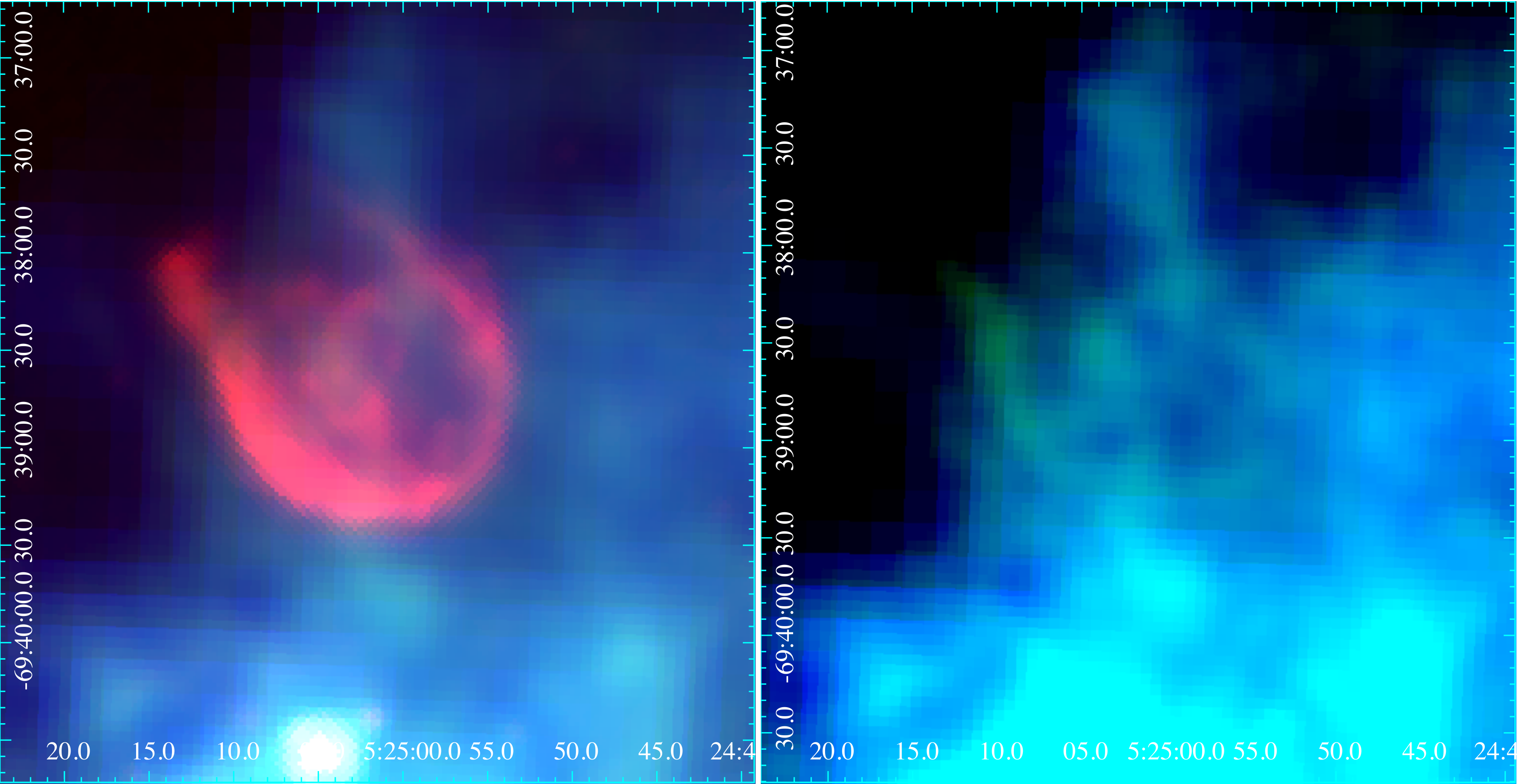}
\includegraphics[scale=1,angle=0,height=6.3truecm]{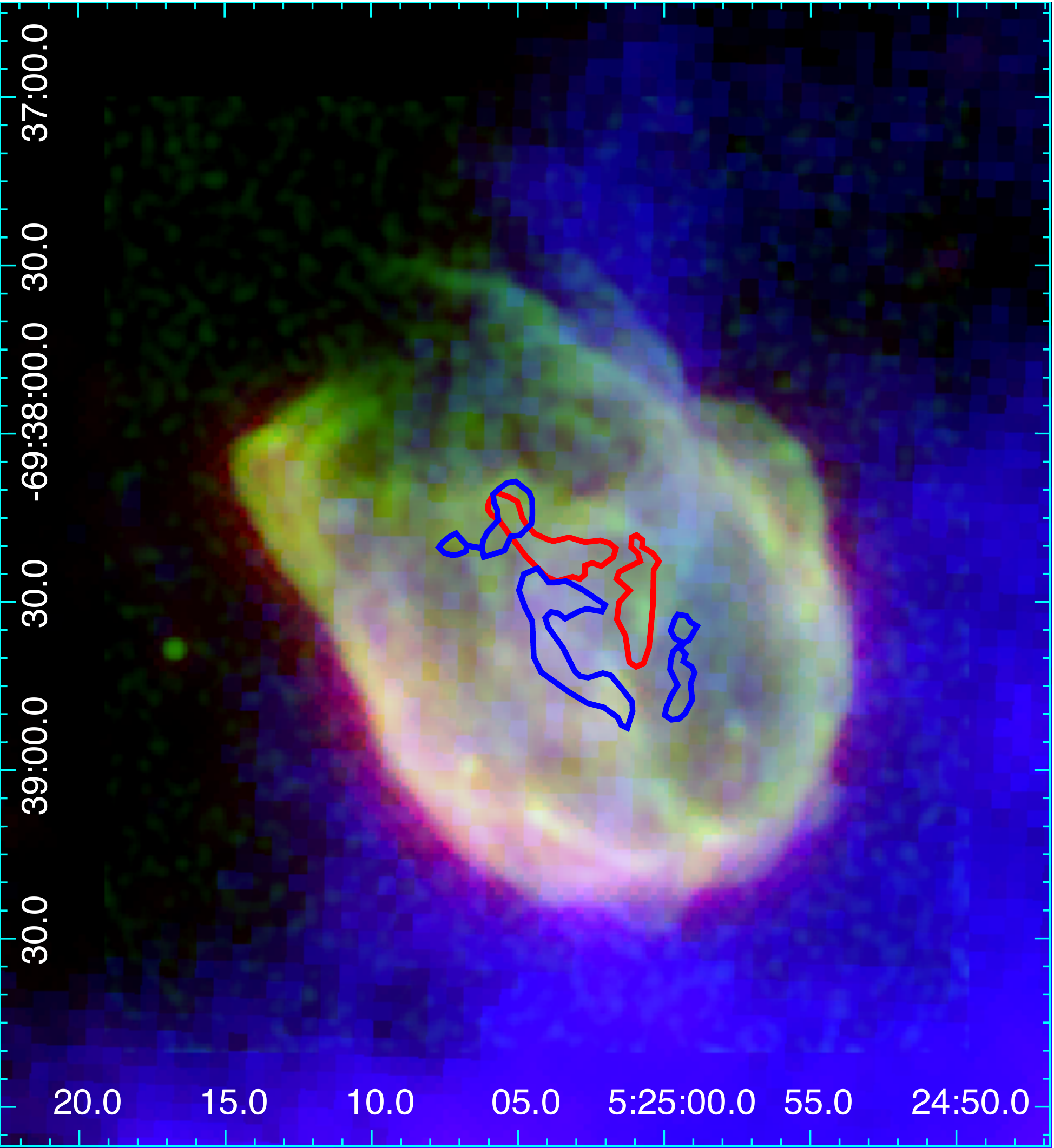}
\caption{Superimposed \spitzer\ and \herschel\ images of SNR N132D: {\it (Left: a)} red, green, and blue represent \spitzer\ 24 $\mu$m, \herschel\ PACS 100 $\mu$m, and  \herschel\ SPIRE 350 $\mu$m, respectively. {\it (Middle: b)} same images as (a) but without the 24 \mic\ image. The SNR and the central blob appear brighter at 100 \mic\ emission (green) than at 350 \mic\ (blue). {\it (Right: c)} Multi-wavelength three-color image of N132D at \herschel\ 350 \mic\ (blue), \spitzer\ 24 $\mu$m (red), and {\it chandra} X-rays (green).
The high-velocity blue- and red-shifted optical ejecta are shown as contours (blue and red, respectively).}
\label{n132dherschelimagescolor}
\end{figure*}

IRS spectra at four locations in N132D are compared in Figure \ref{n132dirsspecall}. The lines of \neiif\ at 12 \mic\ and [O~IV]+[Fe~II] (likely dominated by [O~IV]) in the NE ejecta spectrum are stronger than the CSM/ISM lines at the ``Lasker Bowl" and the ``West Complex". The detected lines and their brightnesses are summarized in Table \ref{Tlinefluxes}. We measured the line brightnesses using two methods: Gaussian fitting and numerical integration of line fluxes over their wavelength ranges. The latter is required in order to account for red- and blue-shifted velocity dispersions. We made extinction corrections using the LMC extinction curve (Fitzpatrick 1985) and assuming E(B-V) = 0.12 \citep{blair00}, which is equivalent to an interstellar atomic column density of  8$\times$10$^{20}$ cm$^{-2}$. We then fitted the continuum excluding the wavelengths with ionic lines. Distinct infrared emission is observed from the ejecta and the shock-heated interstellar material.

\subsection{Ejecta Lines}
\label{sec:3.2}

We compare the line widths of \neiif\ at 12.8 \mic\ and \neiiif\ at 15.57 \mic\ in Table \ref{Tlinevel}. The widths of [Ne~II] and [Ne~III] are higher than the instrumental spectral width \citep{houck04}. The instrument line width of \neiif\  is 0.11 - 0.12 \mic\ (2340 - 2800 \kms)\footnote[1]{https://irsa.ipac.caltech.edu/data/SPITZER/docs/irs/} from the measurements of the \neiif\ calibration sources of G333.9, NGC6720, and NGC7293 \citep[see also][]{smith07, rho09, lai20}. The \neiif\ line width from the NE ejecta is $\sim$10 (7 - 27) \% broader than the instrument line profile.
After taking into account the instrument spectral resolution,
we can estimate the true velocity dispersion of [Ne~II] line for each position
(Table \ref{Tlinevel}). The NE ejecta shows the true velocity dispersion  of
1513$\pm$91 \kms\ higher than those (720 - 1120 \kms) of Lasker's Bowl and West Complex positions.
From the [Ne~III] lines, the ejecta (NE and center) show the true velocity dispersion  of 
1610 - 1660 \kms\ higher than those (1260 - 1390 \kms) of Lasker's Bowl and West Complex positions. 
 The \neiiif\ line width  in the central ejecta is 3620$\pm$47 \kms. Figure \ref{n132dneiiivel} shows the line profile of the central ejecta is broader than that of the planetary nebula NGC 6543. The IRS calibration shows the resolving power at 15.55 \mic\ is 110, equivalent to a width of 0.14 \mic, equivalent to $\sim$2730 \kms\ \citep{smith07}. The width of \neiiif\ line of the central ejecta is 3620$\pm$47 \kms (see Table \ref{Tlinevel}), and it is $\sim$30 \% broader than the instrument line profile. The broad line is evidence that the Ne line emission is from SN-ejecta.

%\end{document}

%Table1
\begin{table*}
\caption{Summary of Infrared and Sub-millimeter Observations.}\label{Tobs}
\begin{center}
\begin{tabular}{llll}
\hline \hline
Date  & Telescope &  prog\_id (PI) & Wavelengths (mode or resolution) \\ 
\hline
2004 November 7 & \spitzer\ MIPS &    3483 (Rho) & 24 \mic\        \\
2004 December 14 & \spitzer\ IRS & 3483 (Rho) & 3 - 40 \mic\ (staring)    \\
2007 June 23 & \spitzer\ IRS     & 30372 (Tappe) & 5- 40 \mic\ (staring), 14-40 \mic\ (mapping)   \\
2010 April 30   & PACS+SPIRE& 1342202224 (Meixner)&  70(6$''$), 100(8$''$), 160(12$''$), 250(18$''$), 350(24.9$''$) \& 500 \mic(36.4$''$)\\
2010 May 4  & {\it WISE} &survey & 3.4(6$''$), 4.6(6$''$), 12(6$''$), 22$\mu$m(6$''$) \\
\hline \hline
\end{tabular}
\end{center}
\end{table*}

%Table2
\begin{deluxetable*}{llccc}
\scriptsize
\tablewidth{0pt}
\tablecaption{Observed Spectral Line Brightness}
\startdata
\hline
Wavelength & Line &  Line Width & Surface Brightness& de-reddened S.B. \\
($\mu$m)         && (\mic)      & (erg s$^{-1}$ cm$^{-2}$ sr$^{-1}$)&
(erg s$^{-1}$ cm$^{-2}$ sr$^{-1}$)\\
\hline
NE ejecta &&&&\\
   12.83$\pm$    0.01& [Ne~II]   &   0.127$\pm$    0.003& 6.90E-05$\pm$ 1.08E-06&6.95E-05$\pm$ 1.09E-06\\
   15.57$\pm$    0.01& [Ne~III] &    0.164$\pm$    0.003& 2.07E-05$\pm$ 3.89E-07&2.08E-05$\pm$ 3.91E-07\\
   14.36$\pm$    0.03&  [Ne~V]         &    0.414$\pm$    0.067& 3.74E-06$\pm$ 8.23E-07&3.76E-06$\pm$ 8.27E-07\\
   24.32$\pm$    0.02& [Ne~V]  &    0.411$\pm$    0.035& 1.94E-06$\pm$ 1.67E-07&1.95E-06$\pm$ 1.67E-07\\
   25.92$\pm$    0.01&[O~IV]   &    0.300$\pm$    0.005& 1.32E-05$\pm$ 1.96E-07&1.33E-05$\pm$ 1.97E-07\\
   34.79$\pm$    0.01&  [Si ~II]    &    0.206$\pm$    0.013& 3.60E-06$\pm$ 2.28E-07&3.62E-06$\pm$ 2.29E-07\\
   36.07$\pm$    0.02& [Ne III] &    0.171$\pm$    0.023& 6.97E-07$\pm$ 1.89E-07&6.99E-07$\pm$ 1.90E-07\\
\hline
{\bf Central Ejecta} &&&&\\   
  15.59$\pm$    0.01& [Ne~III]  &    0.188$\pm$    0.003& 1.15E-05$\pm$ 2.01E-07&1.16E-05$\pm$   2.02E-07\\
   25.95$\pm$    0.01&  [O~IV] &  0.384$\pm$    0.005& 8.40E-06$\pm$ 1.51E-07& 8.40E-06$\pm$ 1.52E-07\\
   34.89$\pm$    0.01& [Si~II]          &   0.253$\pm$    0.014& 2.82E-06$\pm$ 5.38E-07& 2.85E-06$\pm$ 5.41E-07\\
\hline
Lasker's Bowl &&&&\\
 5.35$\pm$    0.01& [Fe~II]  &    0.059$\pm$    0.002&  2.90E-05$\pm$  6.70E-07&2.92E-05$\pm$  6.74E-07\\
    6.99$\pm$    0.01& [Ar~II]  &    0.059$\pm$    0.001&  2.65E-05$\pm$  3.75E-07&2.66E-05$\pm$  3.76E-07\\
   12.83$\pm$    0.01&[Ne~II]   &    0.121$\pm$    0.002&  4.23E-05$\pm$  4.34E-07&4.26E-05$\pm$  4.37E-07\\
   14.40$\pm$    0.01&  [Cl~II]        &    0.324$\pm$    0.010&  4.15E-06$\pm$  1.37E-07&4.17E-06$\pm$  1.38E-07\\
   15.57$\pm$    0.01&[Ne~III]  &    0.157$\pm$    0.001&  1.89E-05$\pm$  8.89E-08&1.90E-05$\pm$  8.95E-08\\
   17.94$\pm$    0.01&[Fe~II]   &    0.159$\pm$    0.001 &  1.01E-05$\pm$  5.90E-08&1.01E-05$\pm$  5.95E-08\\
   18.72$\pm$    0.01&[S~III]   &    0.147$\pm$    0.004&  3.68E-06$\pm$  8.09E-08&3.71E-06$\pm$  8.16E-08\\
   24.50$\pm$    0.07& [Fe II]         &    0.391$\pm$    0.018&  3.11E-06$\pm$  1.48E-07&3.13E-06$\pm$  1.49E-07\\
   25.98$\pm$    0.01&[Fe~II]   &    0.297$\pm$    0.003&  1.24E-05$\pm$  1.20E-07&1.24E-05$\pm$  1.20E-07\\
   33.45$\pm$    0.01&[S~III]   &    0.305$\pm$    0.007&  2.22E-06$\pm$  5.50E-08&2.23E-06$\pm$  5.52E-08\\
   34.80$\pm$    0.01&[Si~II]          &    0.264$\pm$    0.002&  1.36E-05$\pm$  9.77E-08&1.37E-05$\pm$  9.80E-08\\
\hline
W complex &&&&\\
    5.36$\pm$    0.01& [Fe~II]  &    0.060$\pm$    0.002&  1.55E-05$\pm$  5.13E-07& 1.56E-05$\pm$  5.16E-07\\
    6.66$\pm$    0.03& [Ni ~II] &    0.067$\pm$    0.007&  3.20E-06$\pm$  3.34E-07& 3.22E-06$\pm$  3.35E-07\\
    7.02$\pm$    0.01& [Ar~II]  &    0.053$\pm$    0.001&  1.95E-05$\pm$  2.64E-07& 1.96E-05$\pm$  2.65E-07\\
   10.52$\pm$    0.01& [S IV]   &    0.101$\pm$    0.017&  1.20E-06$\pm$  2.60E-07& 1.22E-06$\pm$  2.65E-07\\
   10.69$\pm$    0.01& [Ni~II]  &    0.168$\pm$    0.027&  2.24E-06$\pm$  4.21E-07& 2.28E-06$\pm$  4.28E-07\\
   12.85$\pm$    0.01&[Ne~II]   &    0.114$\pm$    0.001&  3.25E-05$\pm$  2.19E-07& 3.27E-05$\pm$  2.21E-07\\
   15.57$\pm$    0.02&[Ne~III]  &    0.155$\pm$    0.003&  1.52E-05$\pm$  3.08E-07& 1.53E-05$\pm$  3.10E-07\\
   17.95$\pm$    0.01&[Fe~II]   &    0.141$\pm$    0.002&  1.14E-05$\pm$  1.04E-07& 1.14E-05$\pm$  1.05E-07\\
   18.74$\pm$    0.03&  [S~III] &    0.162$\pm$    0.006&  3.37E-06$\pm$  1.04E-07& 3.40E-06$\pm$  1.05E-07\\
   24.57$\pm$    0.05& [Fe~II]     &    0.249$\pm$    0.011&  1.75E-06$\pm$  7.63E-08& 1.76E-06$\pm$  7.67E-08\\
   26.01$\pm$    0.01&[Fe~II]   &    0.288$\pm$    0.003&  8.13E-06$\pm$  5.93E-08& 8.17E-06$\pm$  5.96E-08\\
   33.56$\pm$    0.02&  [S~III] &    0.586$\pm$    0.068&  2.32E-06$\pm$  2.69E-07& 2.33E-06$\pm$  2.70E-07\\
   34.84$\pm$    0.01&  [Si~II]        &    0.289$\pm$    0.007&  9.52E-06$\pm$  2.25E-07& 9.55E-06$\pm$  2.26E-07\\
   \hline \hline
\label{Tlinefluxes}
\end{deluxetable*}

The infrared ejecta emission is mainly clustered at the central, inner shell where the optical ejecta are present (see Figures \ref{n132dspitzerimages}b, \ref{n132dspitzerimages}c and \ref{n132dspitzerimages}e). The IRS spectrum of NE ejecta shows strong O and Ne line emission with 18\,$\mu$m-peak dust feature, which is a broad bump between 16 and 19 \mic\ as shown in Figure \ref{n132dirsspecall} (see also Figure \ref{N132DSED} in y-log scale). The infrared emission (central and NE ejecta) coincides with the optical ejecta positions at the center \citep[around the blue-shifted region of B2;][]{morse95, law20} and NE ejecta (B4 in \cite{morse95} and N132D-P3 in \cite{blair00}). The O and Ne lines are dominant in the optical ejecta  \citep[see Table 9 of][]{blair00} as well as in the infrared.

%Figure8
\begin{figure*}
\includegraphics[scale=1,angle=0,width=16.5truecm]{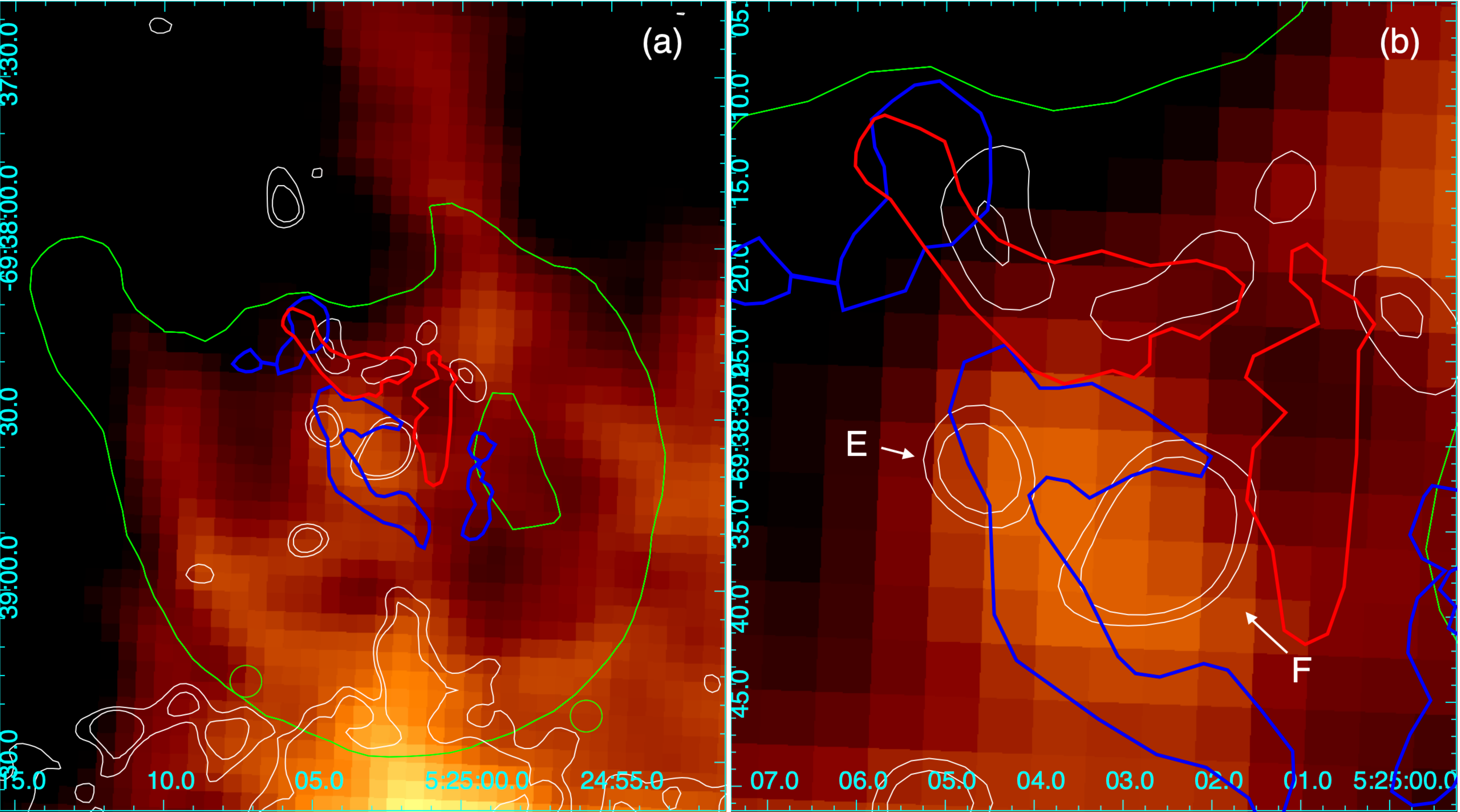}
\caption{(a) The central blob of the \herschel\ 100 \mic\ image of N132D coincides with the blue-shifted (blue contours) optical \oiiif\ and is offset from the ALMA CO emission (white contours). The red-shifted optical \oiiif\ emission is marked in red contours. The boundary of N132D 24 \mic\ emission is marked as green contours. (b) Zoomed view of image (a) shows an offset between the \herschel\ dust blob and ALMA CO clumps labelled E and F.}
\label{n132dSpitzerHerschelALMA2}
\end{figure*}

%Figure9
\begin{figure*}
\includegraphics[scale=1,angle=0,width=10.8truecm]{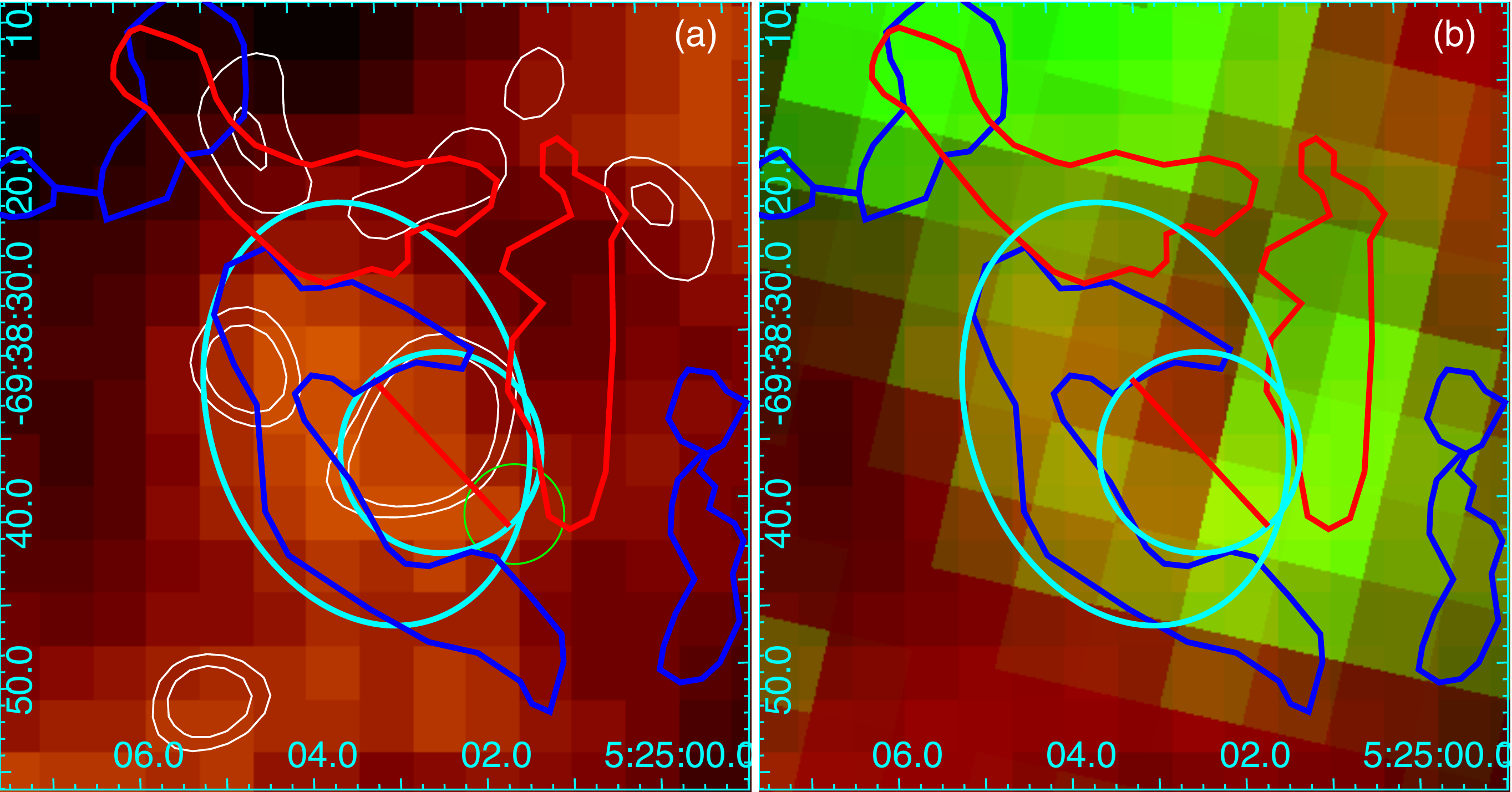}
\includegraphics[scale=1,angle=0,width=6.8truecm]{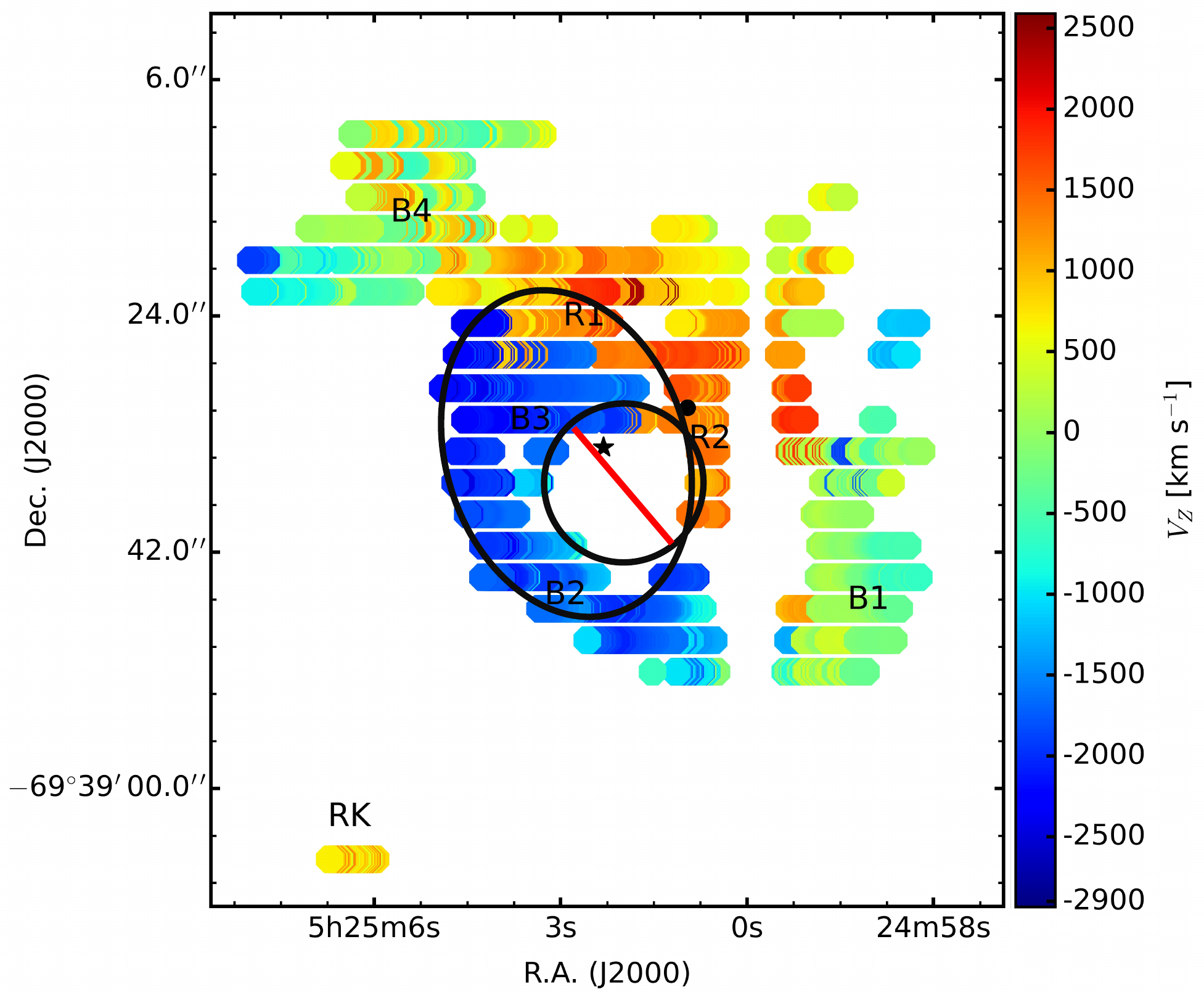}
\caption{{\it (a)} The region used for \herschel\ photometry toward the central emission in constructing the SED (Fig.~\ref{N132DSED}) is marked as an ellipse (in cyan), but excludes a circular region (in cyan with a red diagonal solid line) toward the central blob on the \herschel\ 100 \mic\ image. The exact region is an ellipse of (R.A.,\,Dec.,\,$a$,\,$b$,\,PA)=(81.2635$^{\circ}$, -69.6429$^{\circ}$, 9$\arcsec$, 13$\arcsec$, 17$^{\circ}$) where $a$ is a major axis, $b$ is a minor axis, and P.A. is a position angle from the north; the excluded circular region is centered on (R.A., Dec.) = (81.260$^{\circ}$ -69.643$^{\circ}$) and has a radius of 6$\arcsec$. Excluded regions include the CO $`$cloud F' on the ALMA image (see Fig. \ref{n132dSpitzerHerschelALMA2}) and shocked cloud, P12 (marked as a green circle from \cite{dopita18}).
{\it (b)} Infrared [Ne~III] image in green is added to the image $(a)$ without the CO contours. 
{\it (c)} The same  regions (in black) are marked on a high-velocity optical ejecta image \cite[adapted from Fig.~5 of][]{law20} with color-coded velocity information.}
\label{n132dphotometryregionmark}
\end{figure*}

%Figure10
\begin{figure}
\includegraphics[scale=1,angle=0,width=8truecm]{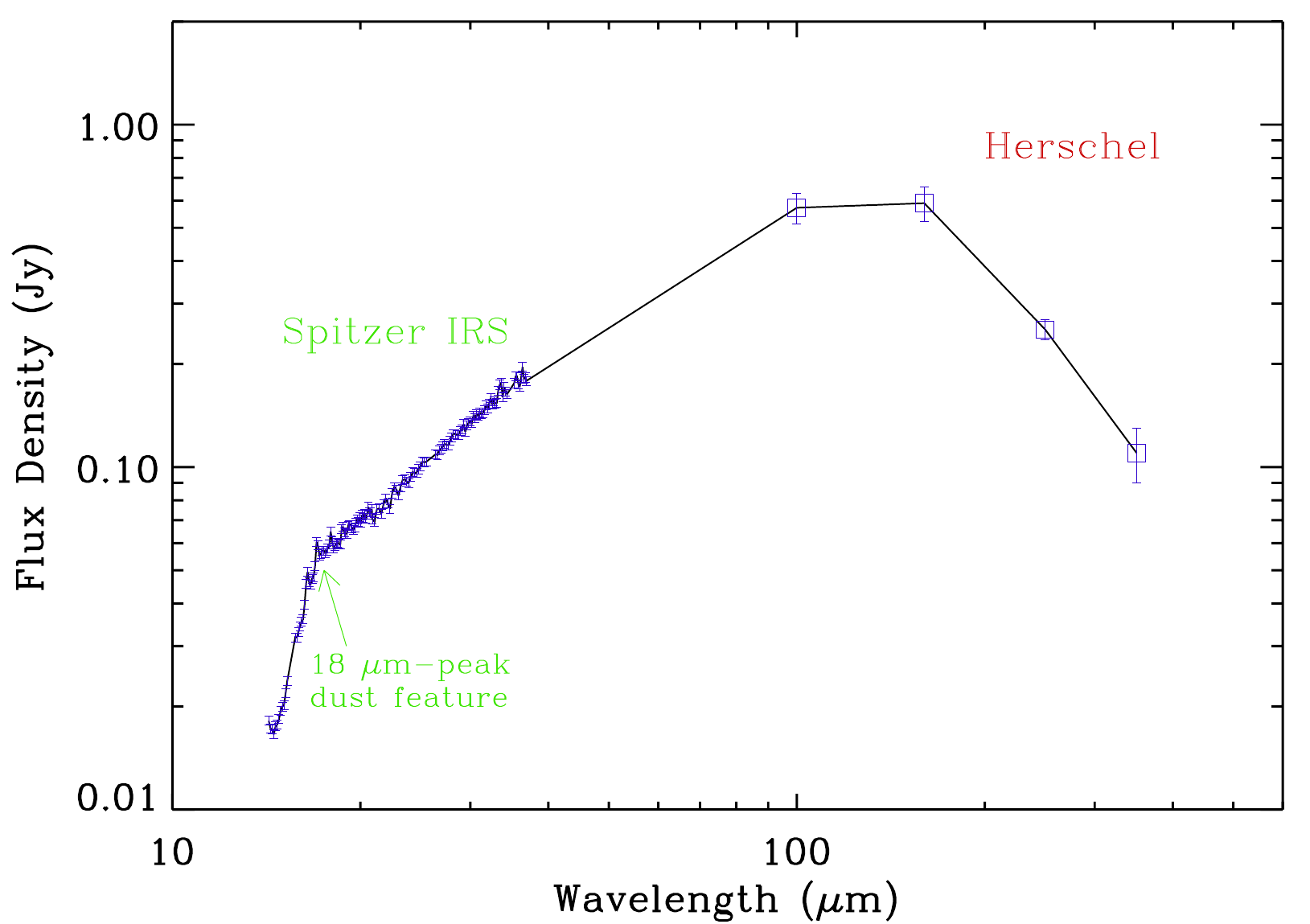}
\caption{Spectral energy distribution of N132D towards the central 
ejecta emission. The data (in blue) include \spitzer\ IRS and \herschel\ photometry. 
The 18 \mic-peak dust feature is marked.
The detection of the emission at 500 $\mu$m is uncertain; see text for details.}
\label{N132DSED}
\end{figure}

%Figure11
\begin{figure}
\includegraphics[scale=1,angle=0,width=8.2truecm]{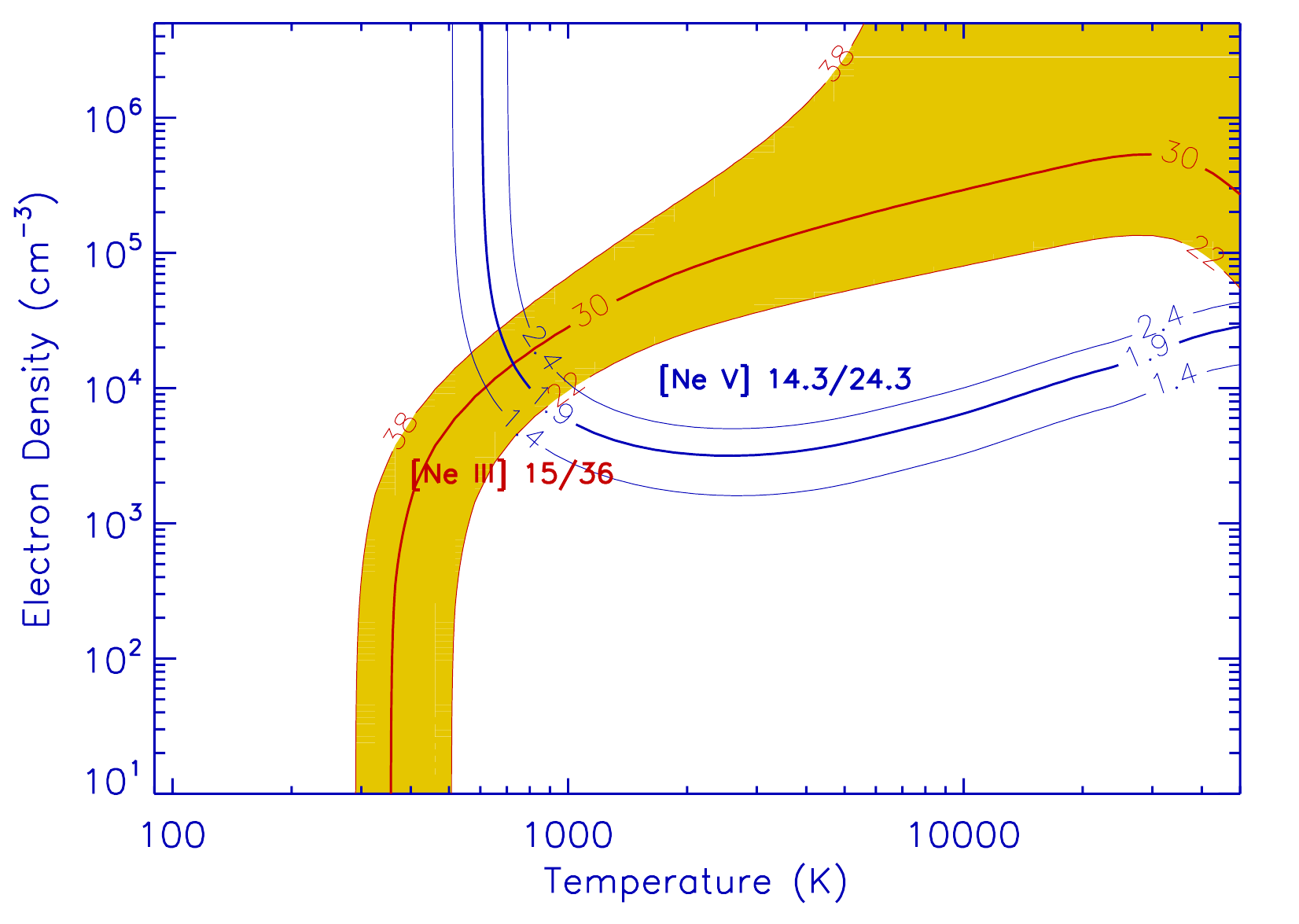}
\caption{Line diagnostic contours of the ratios of \neiiif\ 15/35 \mic\ (red) and \nevf\ 14.3/24.3 \mic\ (blue) at location of the NE ejecta. The observed ratios are denoted by thick solid lines imply most likely values of \neiiif\ = 30 and \nevf\ = 1.92. Note that the density is an electron density. The shaded region (in yellow) shows the range of temperatures and densities allowed by errors for the \neiiif\ ratio, whereas the allowed physical conditions of \nevf\ lines are marked with the thick blue contour lines.}
\label{n132dNediag}
\end{figure}

%Figure12
\begin{figure}
\includegraphics[scale=1,angle=90,width=8.7truecm]{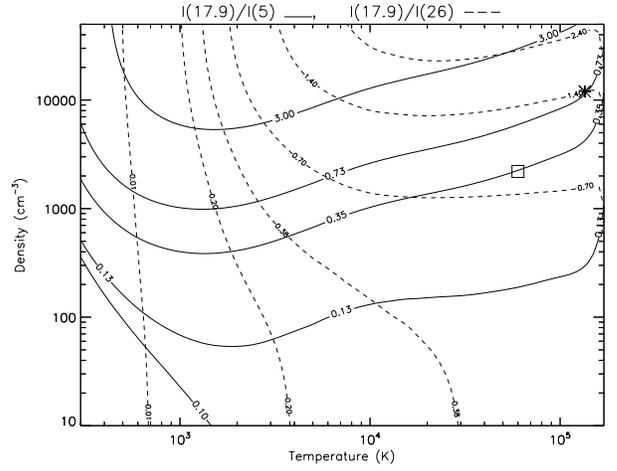}
\caption{Contour plots of \feiif\ diagnostic line ratios using 5.3, 17.9, and 26 \mic\ lines.  Lasker's Bowl has a temperature of $6\times10^4$ K and a density of 2200 cm$^{-3}$ (marked as a square) and the
West Complex has a temperature of $1.3\times10^5$ K and a density of 12,000 cm$^{-3}$ (marked as an asterisk).}
\label{n132dfeiidiag}
\end{figure}

%Figure13
\begin{figure}
\includegraphics[scale=1,angle=0,width=8truecm]{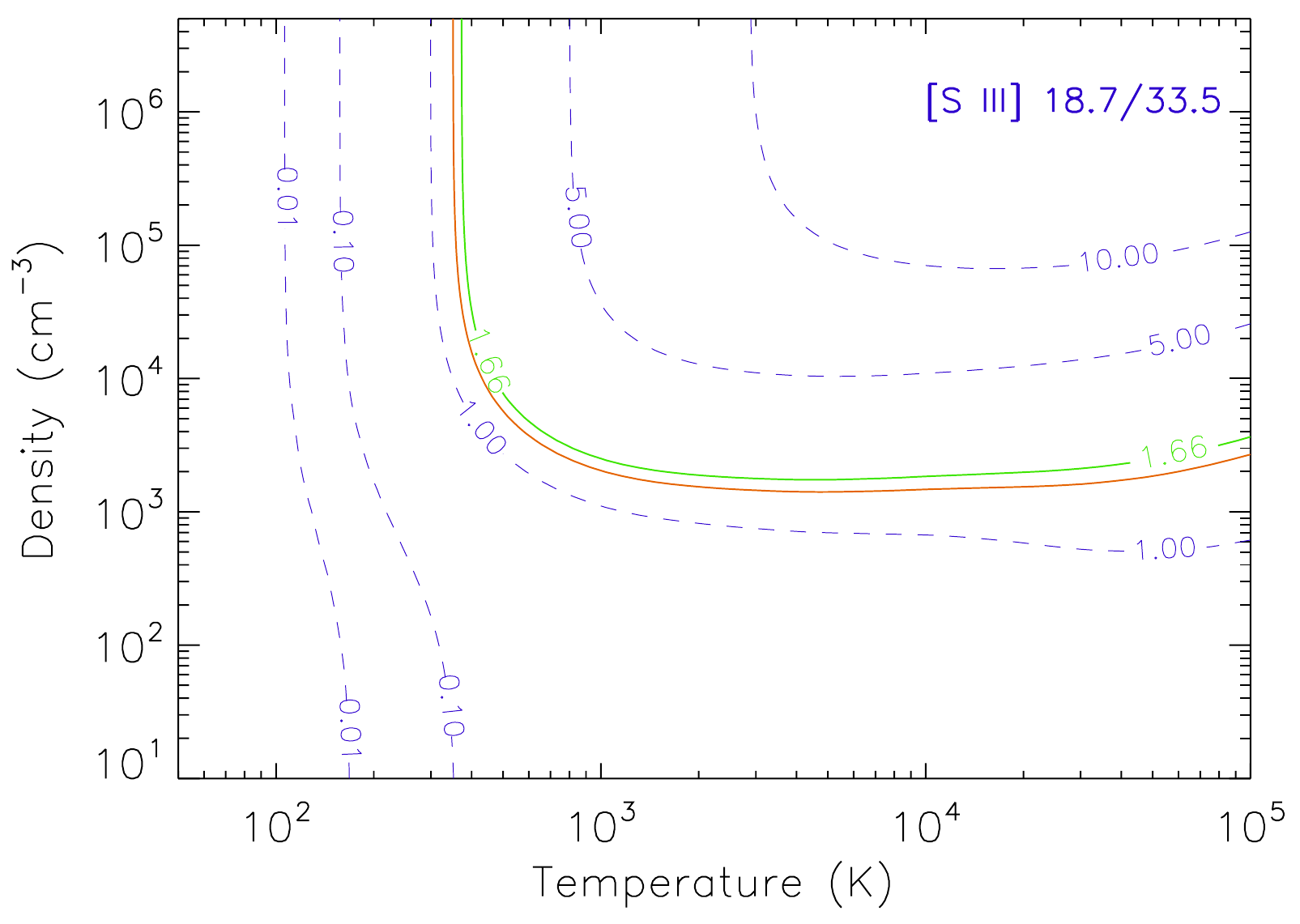}
\caption{ 
Line diagnostic contours of the ratios of \siiif\ 18.7/33.5 \mic. The contours for the Lasker's Bowl (with the ratio of 1.66) and West Complex (1.46) are marked in green and brown, respectively.} 
\label{n132dsiiidiag}
\end{figure}

%Figure14
\begin{figure}
\includegraphics[scale=1,angle=0,width=8truecm]{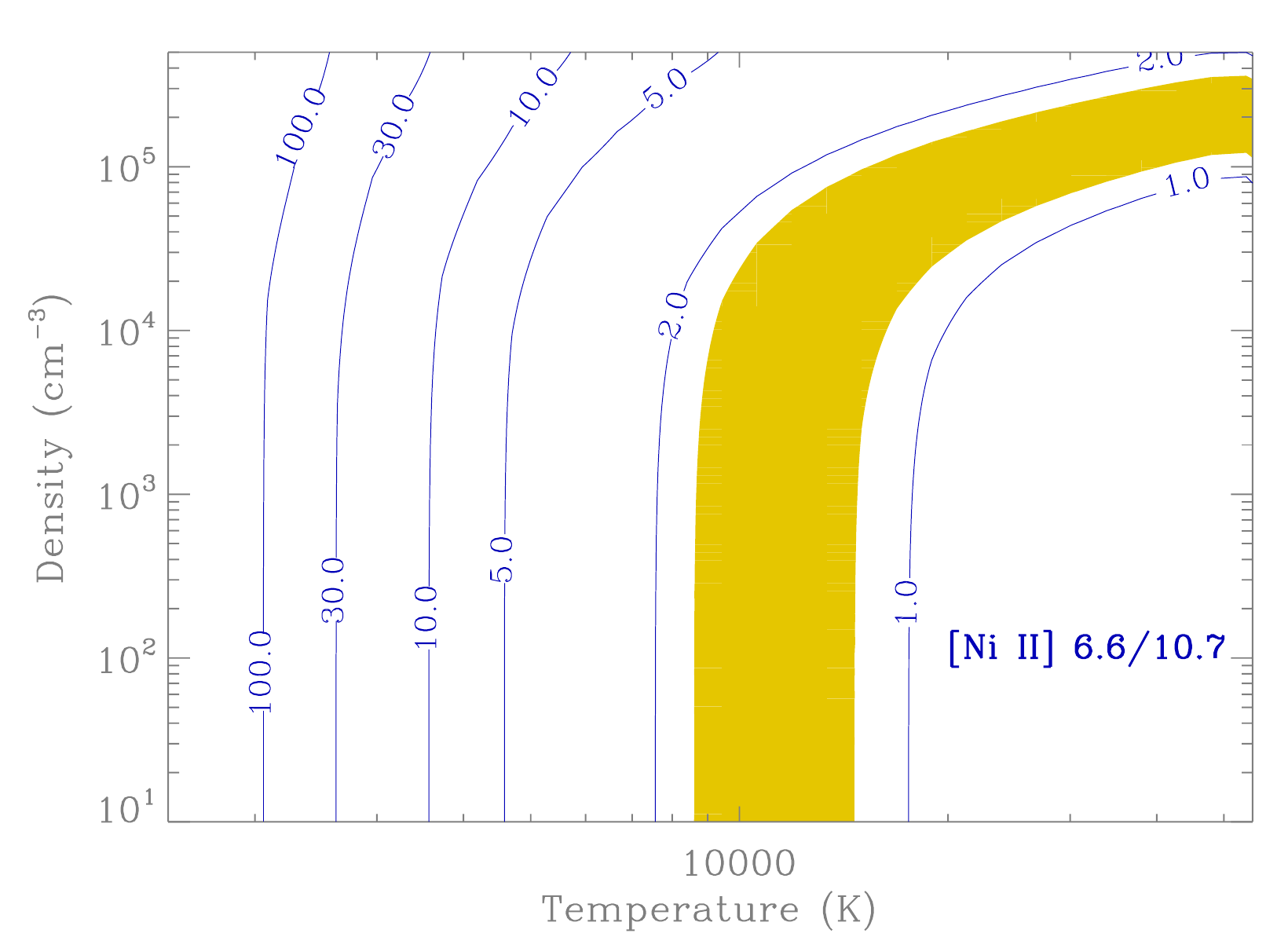}
\caption{Line diagnostic contours of the ratios of \niciif\ 6.6/10.7 \mic. The shaded region (in yellow) shows the range of temperatures and densities allowed by errors for the \niciif\ ratio. The \niciif\ ratio is a temperature indicator and implies $T$ $>$8000 K.}
\label{n132dniciidiag}
\end{figure}

%Table3
\begin{table*}
\caption{Comparison of the Line Widths from Different Regions}\label{Tlinevel}
\begin{center}
\begin{tabular}{lllllll}
\hline \hline
Region & Wavelength & FWHM & Velocity & Vel. shift. & Velocity$_{\rm {true}}$\\
       & (\mic)     & (\mic) & (\kms) & (\kms) & (\kms) \\
\neiif\ &&\\
NE ejecta & 12.827$\pm$0.001 & 0.127$\pm$0.003  & 2987$\pm$47 & 319$\pm$16 & 1513$\pm$91 \\
Lasker's Bowl & 12.828$\pm$0.001 & 0.121$\pm$0.002 & 2840$\pm$28 & 340$\pm$12 & 1197$\pm$65  \\ 
West Complex  & 12.850$\pm$0.001 & 0.114$\pm$0.001 & 2676$\pm$16 & 806$\pm$7 & 726$\pm$57 \\
\hline
\neiiif\ &&\\
NE ejecta & 15.572$\pm$    0.002 & 0.164$\pm$    0.004   & 3167$\pm$60 & 410$\pm$22 & 1655$\pm$112 \\
Center ejecta & 15.595$\pm$    0.002&  0.187$\pm$0.003 & 3620$\pm$47 & 781$\pm$20 & 2411$\pm$70 \\ 
Lasker's Bowl & 15.572$\pm$    0.001  &    0.157$\pm$    0.001 &3037$\pm$14&410$\pm$6 & 1390$\pm$31 \\
West Complex  & 15.568$\pm$    0.002  &    0.154$\pm$    0.004 & 2982$\pm$60 & 307$\pm$25 & 1265$\pm$136\\
\hline \hline
\end{tabular}
\end{center}
\end{table*}

\subsection{Cold Dust in the SN ejecta} 
\label{Scolddust}

Figure \ref{n132dherschelimages} shows \herschel\ images at 100, 160, 250, 350, and 500 \mic\ compared with the \spitzer\ MIPS 24 \mic\ image \citep{tappe06}. N132D has both ejecta emission and a shell of ISM emission. The far-IR maps show cold clouds in the south of the SNR. The PACS and SPIRE images contain infrared (e.g., at 100 \mic\ as shown in Fig.~\ref{n132dspitzerimages}f) and submm emission at the center of the SNR, which coincides with the optical/infrared ejecta as shown in Figure \ref{n132dspitzerimages}. Figure \ref{n132dherschelimagescolor} shows that N132D is bright at {\it Spitzer} MIPS 24 \mic\ (in red in Figure \ref{n132dherschelimagescolor}a) and that the SNR including the central blob is brighter at 100 \mic\ than the 350 \mic\ (greener emission from nearby infrared cirrus in Figure \ref{n132dherschelimagescolor}b). It indicates that N132D, including the central blob, has a hotter temperature than the surrounding cirrus clouds. The SED of the central blob peaks between 100 and 160 \mic\ (see Fig.~\ref{N132DSEDfitA4}).  Fits to the SED yield a dust temperature ($T_d$) of the central blob of 23 - 26 K (see Table \ref{Tdustmass}), which is hotter than the temperatures of infrared cirrus and ISM clouds \citep[15 - 20 K;][]{reach95, boulanger96, lagache98, millard21}. Details are given in Sections \ref{SSED} and \ref{Sdustfit1}. In contrast, the shock-heated ISM dust in the southeastern shell of N132D has $T_d$  = 110 K and 58 K \citep{tappe06}. The lower temperature component ($T_d$ = 58 K) is much hotter than the dust temperature ($T_d$ = 23 - 26 K) in the central blob.

Figure \ref{n132dSpitzerHerschelALMA2} compares the \herschel\ 100 \mic\ image with ALMA CO (J=1-0 at $\lambda$ = 2.6 $mm$) contours from \cite{sano20} and the blue- and red-shifted optical ejecta. The eight new molecular clouds resolved by ALMA are likely interacting with shock waves and lie inside a wind-blown bubble. The X-ray emission from Cloud F shows enhanced abundances (e.g., Ne and O), indicating some clumps may be associated with the ejecta. Moreover, the CO $`$cloud F' \citep{sano20} in the ALMA image is somewhat offset from the optical and infrared neon ejecta (see Figures~\ref{n132dSpitzerHerschelALMA2}b and \ref{n132dphotometryregionmark}), and lies between the blue- and red-shifted IR/optical ejecta. The angular resolutions of the ALMA image, 5$\arcsec$, and the \neiif\ map, 4-6$\arcsec$, are comparable. The peak of the dust emission in the 100 \mic\ \herschel\ image (resolution 8$\arcsec$) is located between CO $`$clouds E and F'. The central dust emission coincides with the neon infrared line emission from the ejecta and also with the optical ejecta (e.g., blue-shifted in blue contours). The peak of the dust emission lies between the ALMA CO $`$clouds E and F'. The angular separation between the central dust peak at 100 \mic\ and the larger CO $`$cloud F' is 23$\arcsec$, which is 3 times larger than the \herschel\ beam size.

%Figure15
\begin{figure}
\includegraphics[scale=0.8,angle=0,width=8.5truecm]{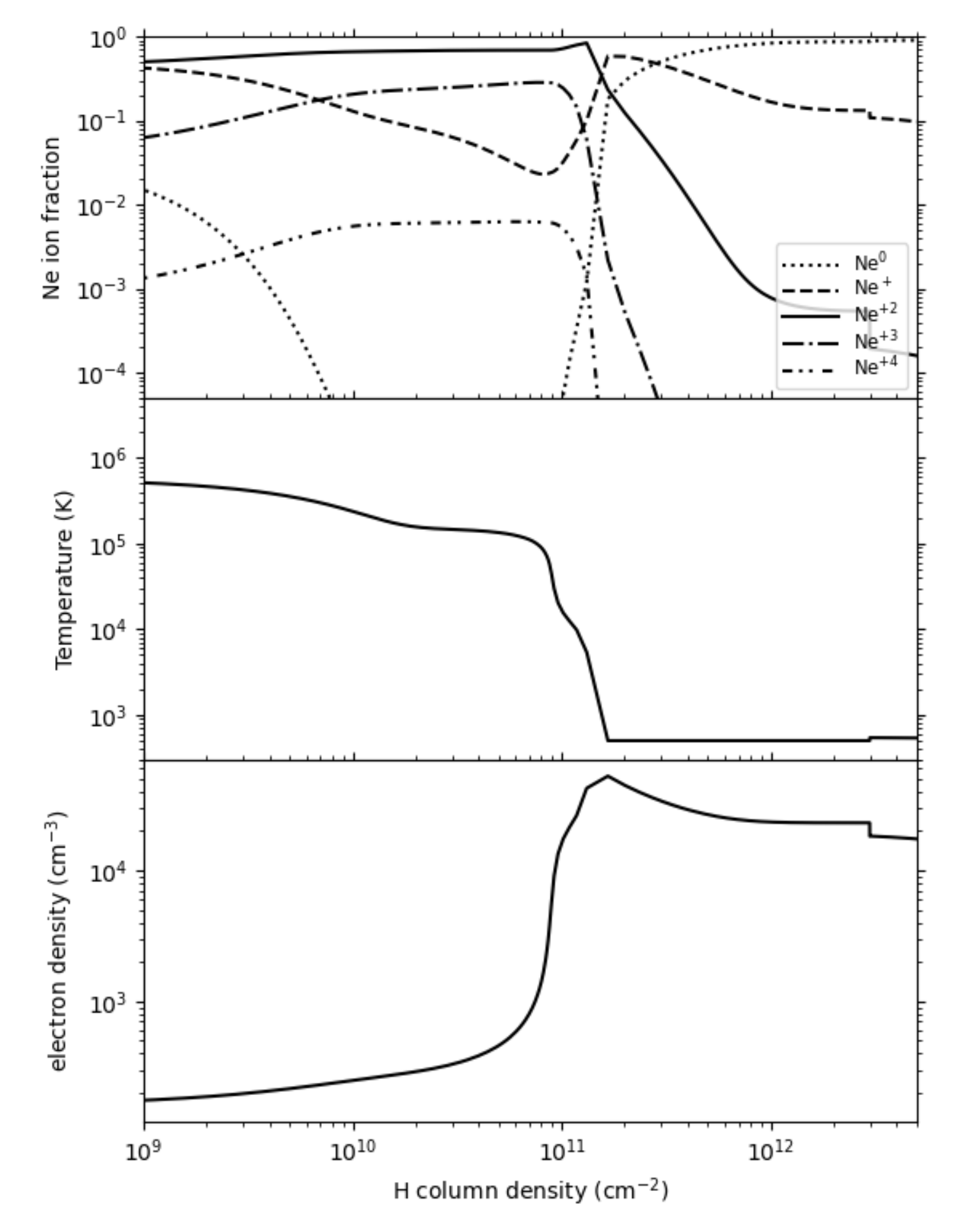}
\caption{Neon ionization (above) and gas temperature (below) in the shock for our model with shock speed 90 km s$^{-1}$ and ram pressure 20$\times$ the \citet{blair00} assumed value.  The ionization of Ne clearly shows the lag in ionization and recombination in the ionization ($T \geq 10^5$ K) and cooling ($10^3 <\leq T \leq 10^5$ K) zones. The region on the right, where $T \sim 540$ K, is photoionized and assumed to be in thermal and ionization equilibrium.  The Ne$^{+4}$ ion fraction is very low in the photoionized zone and so the observed [\ion{Ne}{5}] 14.3 $\mu$m and 24.3 $\mu$m emission come from the cooling zone (with electron density $10^3-5\times10^4$cm$^{-3}$) whereas most of the [\ion{Ne}{2}] and [\ion{Ne}{3}] emission comes from the photoionized zone.}
\label{fig:Ne_ioniz}
\end{figure}

\subsection{The Broad-band Spectral Energy Distribution of the Central Cold Dust}
\label{SSED}

To derive dust masses associated with the central ejecta of N132D, we generated a spectral energy distribution (SED) from the MIR to submm for N132D in Figure~\ref{N132DSED}. We chose the central ejecta only, since the \herschel\ images reveal clear detection from the central ejecta, whereas far-IR emission from NE ejecta are not distinct (see Figure \ref{n132dherschelimages}). We combined \spitzer\ IRS spectra and \herschel\ photometry. The \spitzer\ spectra towards the center of the SNR  only covered the 14-40 \mic\ range and highlighted the broad dust features in the MIR. 

We performed \herschel\ 100, 160, 250, and 350 \mic\ photometry toward the central region because the emission there is clearly present at all wavelengths as shown in Figures \ref{n132dspitzerimages}f (marked as a circle) and \ref{n132dherschelimages}. The exact area measured is an ellipse excluding a circular region covering CO $`$cloud F' \citep{sano20} and the shocked cloud \citep[P12 in][]{dopita18} as shown in Figure \ref{n132dphotometryregionmark}. Excluding this region minimizes the contribution from the ISM clouds, although the cloud's contribution to far-IR is expected to be small. The shocked cloud ``P7" \citep{dopita18} region has strong [O~III] optical emission and highly blue-shifted O-rich knot emission, as shown in Figure \ref{n132dphotometryregionmark}c as well as infrared neon emission (green emission in Figure~\ref{n132dphotometryregionmark}b). Note that the ``P7" region is small, and its flux is less than 5\% of the total \herschel\ flux in the ellipse. Since N132D is known to be interacting with the surrounding ISM (the SNR shell emission is largely from shocked gas), we carefully selected the source and the background regions. For the latter, we averaged three elliptical regions, to the northwest, to the northeast and to the east. Their central coordinates are (R.A.,\,Dec.,\,$a$,\,$b$,\,PA) = (81.2111$^{\circ}$, -69.6216$^{\circ}$, 32$\arcsec$, 27$\arcsec$, 357$^{\circ}$), (81.3460$^{\circ}$, -69.6441$^{\circ}$, 25$\arcsec$, 25$\arcsec$, 357$^{\circ}$), and (81.3627$^{\circ}$, -69.6169$^{\circ}$, 42$\arcsec$, 33$\arcsec$, 357$^{\circ}$). We avoided the region south of the SNR, since it is bright in far-IR from  giant molecular clouds.

The 500 \mic\ flux density is not included because the image does not resolve the central emission or the emission from N132D. It is not clear if the far IR emission at the position of the SNR is associated with the SNR or a part of a large ISM structure, probably due to the limited \herschel\ resolution (36.4$\arcsec$ at 500 \mic). The largest contribution to the flux uncertainty can originate from the background variation at the long IR wavelengths. We varied the choice of background region when measuring the flux density of N132D at various wavelengths, and found that it causes an increase of uncertainties from a few percent to a few tens of percent. The estimated flux densities of N132D in far-IR and submm are summarized in Table \ref{THerschelphotometry}.

\section{Discussion}  

\subsection {Neon Ejecta and Line-Flux Ratio} 
\label{SDNeejectalineratio}

In the line list in Table \ref{Tlinefluxes}, the ions \neiiif\ and \nevf\ are suitable diagnostics for constraining densities and temperatures of the ionized gas.

The measured line flux ratios for [\ion{Ne}{5}] 14.3/24.3 $\mu$m are 1.92$\pm$0.45, and 29.8$\pm$12 for [\ion{Ne}{3}] 15.6/36 $\mu$m. The [\ion{Ne}{5}] ratio in N132D is similar to, but slightly higher than that in 1E0102 \citep[1.76$\pm$0.11][]{rho09} while the [\ion{Ne}{3}] ratio in N132D is almost a factor of 2 smaller than that in E0102 (54$\pm$28). The ratio $I_{15.6}/I_{36}$ $\mu$m is assigned an additional 50\% systematic error for the 36 $\mu$m line intensity, because this line falls on the degrading part of the array. To constrain temperatures and densities, we calculated the line intensities and ratios of \nevf\ and \neiiif.  We solved the excitation-rate equations, including collisional and radiative processes, as a matrix using 5 energy levels for \nevf\ and 3 levels for \neiiif.  The atomic data are described in \citet{rho09}. 
The contours of their ratios are marked in Figure \ref{n132dNediag}
and Table \ref{Tratioproperties} lists the inferred density and temperature
ranges.

Radiative shock models for a range of shock velocities, including the effects of photoionization, were produced by \citet{rho09} to understand the \neiiif\ and \nevf\ lines of E0102. The shock models indicate that the \nevf\ lines come mainly from the cooling zone, in which the temperature is falling sharply from hot to cold, whereas \neiiif\ comes mainly from the photoionization zone, which has a low temperature. 
We present and discuss the shock models and derive the physical conditions of Ne ejecta of N132D in Section \ref{Sshockmodelsejecta}.

%Table4

\begin{table*}
\caption{Physical properties from same-ion line flux ratios in dense knots}\label{Tratioproperties}
\begin{center}
\begin{tabular}{lllllllll}
\hline \hline
Region & Line &  Wavelengths  & Ratio & T (K)$^a$ & n$_e$ ($cm^{-3}$)$^a$ & \\
\hline
Ejecta & \nevf\  & 24.3/14.3 $\mu$m  & 1.92$\pm$0.45  &500 - 10$^5$ K & $>$10$^3$ (e.g., { 10$^3$ - 10$^{4}$}) &  & \\ 
       &         &                   & shock model$^b$    & {\bf 10$^3$ - 10$^{4}$ K} & {\bf 10$^3$ - 5$\times$10$^{4}$} \\  
       & \neiiif\ & 15.6/36 $\mu$m  & 29.8$\pm$12 & 300 - 5$\times$10$^4$ K  & 100 - 5$\times$10$^6$&   \\ 
       &          &                 &  shock model$^b$           & {\bf 300 - 600 K}  & {\bf 10$^3$ - 10$^4$} &\\  
\hline
Lasker's Bowl &\feiif\  & 17.9/5 $\mu$m   &   0.35$\pm$0.01 & \\
(CSM/ISM knots)          &        & 17.9/26$\mu$m    &   0.82$\pm$0.01 & {\bf 6$\times$10$^4$ K} &  {\bf 2200}\\

&\siiif\ & 18.71/33.48 $\mu$m &  1.66$\pm$0.06 & {\bf  $>$600 K} & {\bf 10$^3$ - 3$\times$10$^3$} & \\
      &                         &              &               & or 300-400 K & $>$10$^4$ \\
\hline
West Complex & \feiif\ & 17.9/5 $\mu$m   &  0.73$\pm$0.03 & \\
(ISM/CSM knots)             &        & 17.9/26 $\mu$m   & 1.40$\pm$0.02 &  {\bf 1.3$\times$10$^5$ K} &  {\bf 1.2$\times$10$^4$}\\
& \siiif\ & 18.71/33.48 $\mu$m & 1.46$\pm$0.18 &  {\bf $>$600 }& {\bf 10$^3$ - 3$\times$10$^3$} & \\
&                             &              &               & or 300-400 K & $>$10$^4$ \\
&\niciif\ & 6.63/10.7 $\mu$m & 1.4$\pm$0.3 & {\bf $>$8000 K} & any \\
\hline
\end{tabular}
\end{center}
\footnotesize{$^a$The best values are marked in bold characters.}
\footnotesize{$^b$See Section \ref{Sshockmodelsejecta} for details.}
\end{table*}

\subsection{Line diagnostics of \feiif, \siiif, and \niciif\ lines in dense knots}

\subsubsection{CSM and ISM dense knots}
Lasker's Bowl and the West Complex (see Figure \ref{n132dspitzerimages}) are known to be luminous and to contain dense shocked clouds (dense knots, hereafter) with interstellar composition and a $\sim$200 \kms\ velocity dispersion in optical lines \citep{blair00}. The dense knots are suggested to be either ISM or mass loss material (CSM) from the progenitor \citep{blair00, sutherland95}. The optical spectroscopy of the West Complex indicates that collisional de-excitation of 6716\AA\ is largely responsible for the relatively low [\ion{S}{2}] 6724\AA/H$\alpha$ ratio observed as compared with many SNRs. The optical [\ion{S}{2}] emission arises in the compressed postshock medium behind the shock and implies a preshock cloud density of greater than 100 cm$^{-3}$ \citep{blair00}. Shock velocities of order $\sim$200 \kms\ are needed to generate the emission from high-ionization species such as the UV lines of \ion{O}{4}], \ion{N}{5}, and [\ion{Ne}{4}]. Such shocks produce sufficient high-energy photons in the postshock region to ionize the upstream gas before it enters the shock \citep{shull79, hartigan87, sutherland93}. \citet{blair00} assumed complete preionization of H and 85\% preionization of He to He$^+$. Most H $\alpha$ emission (with $\sim$200 \kms) arises in the recombination zone of the cloud shocks, which have velocities similar to nearby \ion{H}{2} regions \citep{dopita18}.

The IRS spectra of Lasker's Bowl and the West Complex reveal that these regions emit strongly in \feiif,  \ariif, \siiif, and \niciif\ lines as shown in Figure \ref{n132dirsspecall}. In Lasker's Bowl the \feiif\ ratio between 17.9 \mic\ and 5 \mic\ lines (\feiif\ $I_{17.9}/I_{5}$) is 0.35$\pm$0.01 and the \feiif\ ratio between 17.9 \mic\ and 26 \mic\ (\feiif$I_{17.9}/I_{26}$) is 0.81$\pm$0.01.

\subsubsection{Line diagnostics of \feiif, \siiif, and \niciif\ lines}

Diagnostics of the \feiif\ lines use excitation rate equations presented in \citep{hewitt09, rho01}. Figure \ref{n132dfeiidiag} shows contours of line ratios 17.9/5.35 \mic\ and 17.9/26 \mic\ as functions of density and temperature; the former ratio is sensitive mainly to density, and the latter ratio is dependent on both density and temperature. We used temperature-dependent atomic data from \citet{ramsbottom07} and 46 levels of \feiif\ lines. The derived densities and temperatures for each of the Lasker's Bowl and West Complex positions are summarized in Table \ref{Tratioproperties}. The densities (2000 -- 12000 cm$^{-3}$) of both of these regions of CSM/ISM knots are higher than those ($<$ 1000 cm$^{-3}$) in SNRs interacting with molecular clouds \citep{hewitt09} and the temperatures (6 -- 13$\times$10$^4$ K) are also higher.

In the same manner, we construct line diagnostic plots for \siiif\ $I_{18.7}$/$I_{33.48}$ and \niciif\ $I_{18.7}$/$I_{33.48}$. The estimated ratios are listed in Table \ref{Tratioproperties}. For \siiif\ lines, we solve 4-level system excitation rate equations and use the temperature-dependent atomic data from \citet{galavis95} and \citet{galavis98}. Figure \ref{n132dsiiidiag} shows \siiif\ line diagnostics the measured ratio implies a density $>$ 10$^{3}$ $cm^{-3}$. For a gas temperature $>$600 K, the density is 10$^3$ - 3$\times$10$^3$ cm$^{-3}$, and for gas with a temperature of 300-400 K, the density is $>$10$^4$ cm$^{-3}$. For the \niciif\ lines, we solve 8-level system excitation rate equations and use the temperature dependent effective collision strengths from \citet{bautista04}. The \niciif\ lines are potentially good temperature indicators. Figure \ref{n132dniciidiag} shows \niciif\ line diagnostics; the observed ratio $I_{18.7}$/$I_{33.48}$ indicates a gas temperature $>$ 8000 K. The inferred densities and temperatures from various line ratios are summarized in Table \ref{Tratioproperties}.

%Table5
\begin{table}
\caption{FIR-submm Flux Densities.}
\label{THerschelphotometry}
\begin{center}
\begin{tabular}{lllll}
\\
\hline \hline
Data          &  {Wavelength} & {Flux Density} \\
               & ($\mu$m)    & (Jy) \\
\hline
{\it WISE} ($w3$)         & 12         & 0.015$\pm$0.010\\
\spitzer\ MIPS   & 24         & 0.090$\pm$0.006 \\
\herschel\ PACS  & 100        &0.572$\pm$0.058 \\
\herschel\ PACS  & 160        &0.590$\pm$0.068 \\
\herschel\ SPIRE & 250        &0.252$\pm$0.017 \\
\herschel\ SPIRE & 350        &0.110$\pm$0.020 \\
\hline \hline
\end{tabular}
\end{center}
\end{table}

\subsection{Application of shock models and physical conditions in ejecta}
\label{Sshockmodelsejecta}

We carried out calculations of the shock profiles similarly to the way we did for 1E0102 \citep{rho09} but using the lower ram pressure and shock speeds from \citet{blair00} for N132D. We have now also developed the capability to simulate shocks with only partial electron-ion temperature equilibration at the front, i.e.\ following the electron and ion temperatures separately including their evolution toward equilibration.  We made the calculations assuming an initial 50\% equilibration.  As in \citet{rho09} we combine the steady shock calculation of the hot post-shock region and cooling region with a calculation of the cooled photoionized zone that uses Cloudy \citep{ferland17}. The shocks produce hot gas just behind the front which ionizes the pre-shock gas. Further behind the shock, cooling leads to a sharp drop in temperature in the recombination zone, where the gas is overionized. Once the gas has cooled to low temperatures, it is maintained by the ionizing radiation from the upstream regions that heats the gas and balances cooling in the photoionized region. 
The steady-state shock calculation uses ionizing radiation only from the forward shock and calculates the non-equilibrium ionization.
The photoionized region calculation includes emission generated in the forward shock as well as that is generated in the reverse shock, but assumes equilibrium ionization.

We find that matching the emission ratio [\ion{Ne}{3}] 15.6~$\mu$m/[\ion{Ne}{2}] 12.8~$\mu$m as well as matching the flux of [\ion{Ne}{2}] 12.8 $\mu$m emission requires shocks with speeds of $\sim 90 - 150$ km s$^{-1}$. In order to match the optical HST data for the N132D ejecta, \citet{blair00} explored models with a range of shock velocities with a constant ram pressure of $n v_{sh}^2 = 100\,$cm$^{-3}$\,km$^{2}$\,s$^{-2}$, and magnetic field given by $B/n^{1/2} = 0.1$ (where the magnetic field, $B$, is in $\mu$G and density, $n$, is in cm$^{-3}$). We found that higher pressures and magnetic fields were needed than used by \citet{blair00}, and explored combinations of factors of 10 to 20 times the values for the ram pressure and magnetic field. We found acceptable solutions for several different shock velocities, although solutions that also matched the [\ion{Ne}{5}] 14.3~$\mu$m/24.3~$\mu$m ratio required high ram pressures, $\sim20 \times$ the \citet{blair00} value. The model that best matches our data has a pre-shock density $n$ of $\sim25$ cm$^{-3}$ and shock velocity v$_{sh}$ of $\sim100$ km~s$^{-1}$. We assumed the same abundances as \citet{blair00} (logarithmic relative to H with H = 12.00) O = 16.00, Ne = 15.25, C = 14.50, and Mg = 13.50.  Figure \ref{fig:Ne_ioniz} illustrates the Ne ionization and temperature for the two parts of the calculation for one model. In the post-shock photoionized region, these models lead to values of $\sim 1000 - 2\times10^4$ cm$^{-3}$ for the electron density and temperatures of $\sim 300 - 600$ K. Essentially all of the [\ion{Ne}{2}] emission comes from the photoionized zone in these models, while [\ion{Ne}{3}] emission comes from both the recombination zone and the photoionized zone. In the post-shock cooling region, the [\ion{Ne}{5}] emission peaks near the location where the electron density peaks and the temperature and ion fraction of Ne$^{+4}$ have begun to drop. None of the [\ion{Ne}{5}] emission comes from the photoionized zone.

\subsection{Synchrotron contribution to the dust emission}
\label{Ssynchrotron}

We estimated the contribution of  synchrotron emission to the infrared spectrum using the radio fluxes and spectral index. The radio flux is 1.5 Jy at 6 cm for the entire SNR \citep{dickel95}. We estimated the contribution of the flux to the emission from the central ejecta region using the 5 GHz radio image, and found that it is 12\% of the flux from the entire SNR. For a radio spectral index $\alpha=-0.70$ \citep{dickel95}, where $S_\nu\propto\nu^\alpha$, this is a very small contribution to the infrared continuum ($<$ 1\%, see the dotted line in Figures \ref{N132DSEDfitA4} and \ref{N132DSEDfitA5}).

\subsection{Dust model fitting}
\label{Sdustfit1}

To estimate the dust mass, spectral fitting was performed using a chi-squared-minimizing routine \citep[{\it mpfit} routine by][]{markwardt09} which was applied to the SED of N132D. The detailed method is described in \citet[][]{rho18}. The flux F$_{\nu}^i$ = $\Sigma_i \, C_i \, B_\nu \, 3\, Q_{abs,i} / (4\,a$\,$\rho_i$), where the Q($\lambda$, a) are calculated using Mie theory \citep{bohren83} and the continuous distributions of ellipsoidal (CDE) model for silicate dust, $a$ is the dust particle size, $\rho_i$ is the bulk density, and $B_\nu$ is the Planck function for each of grain species ($i$).  We derived the scale factor C$_i$ and estimated dust masses which are summarized in Table \ref{Tdustmass}.

We calculated Q$_{abs}$ from $n$ and $k$ values of Al$_2$O$_3$ \citep{begemann97}, but the values only cover up to 500 $\mu$m. Therefore, we extended the power law of the Q$_{abs}$ between 450 and 500 \mic\  to 1000 \mic, although it doesn't affect the quality of the fit since the observed data only cover up to 350 \mic. The same absorption coefficients from various grains were applied to G54.1+0.3 and Cas A \citep{rho18}.

\begin{figure}
\includegraphics[scale=1,angle=0,width=8truecm]{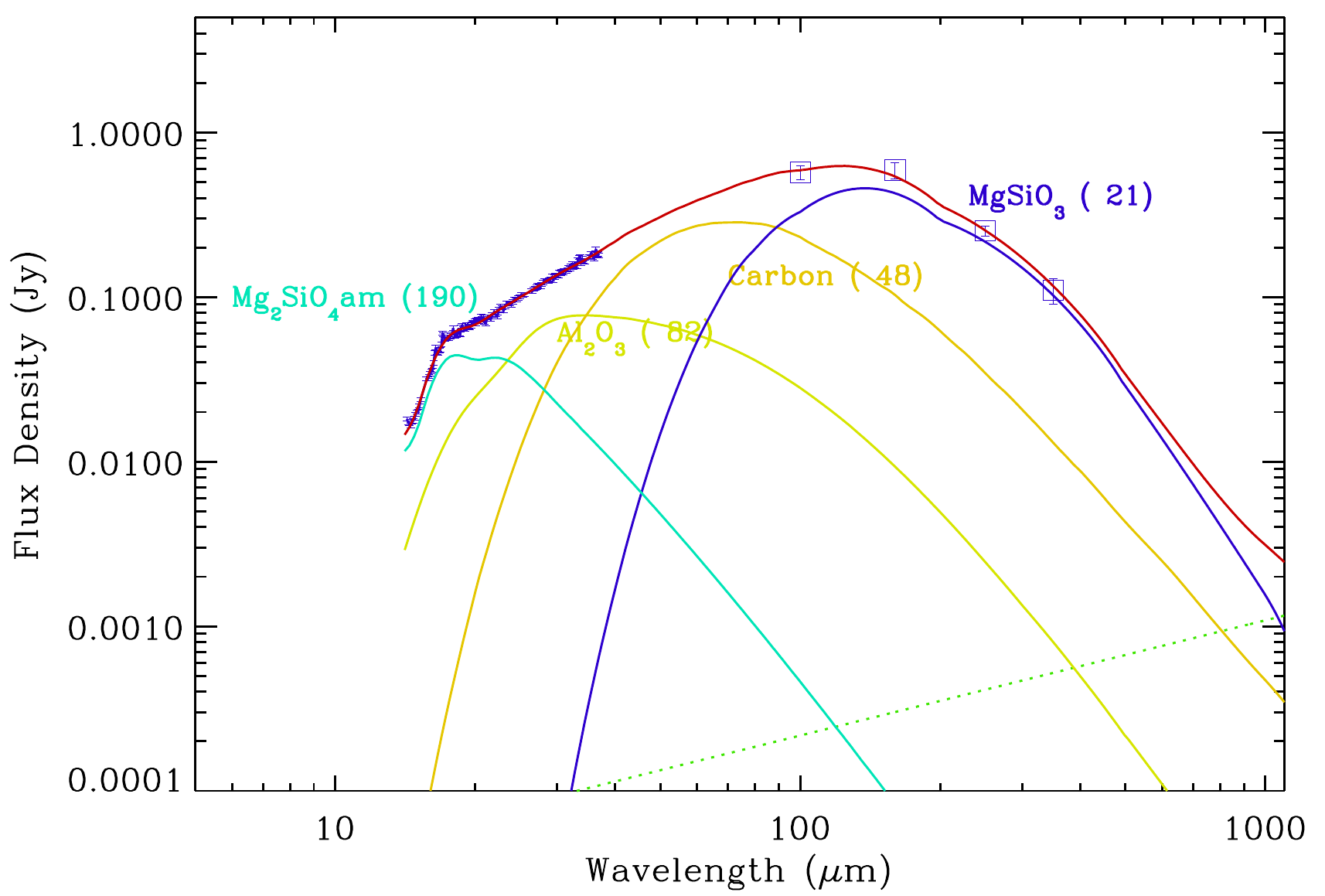}
\caption{Model A4 (Table \ref{Tdustmass}) is superposed on \spitzer\ IRS  spectra and \herschel\ photometry of N132D. The grains of Mg$_2$SiO$_4$ and MgSiO$_3$ reproduce the 18 \mic\ feature and cold dust, respectively. The number in parenthesis is dust temperature for each type of grain. Synchrotron radiation (green dotted line) is included in the fitting, but makes only a minor contribution.}
\label{N132DSEDfitA4}
\end{figure}

\begin{figure}
\includegraphics[scale=1,angle=0,width=8truecm]{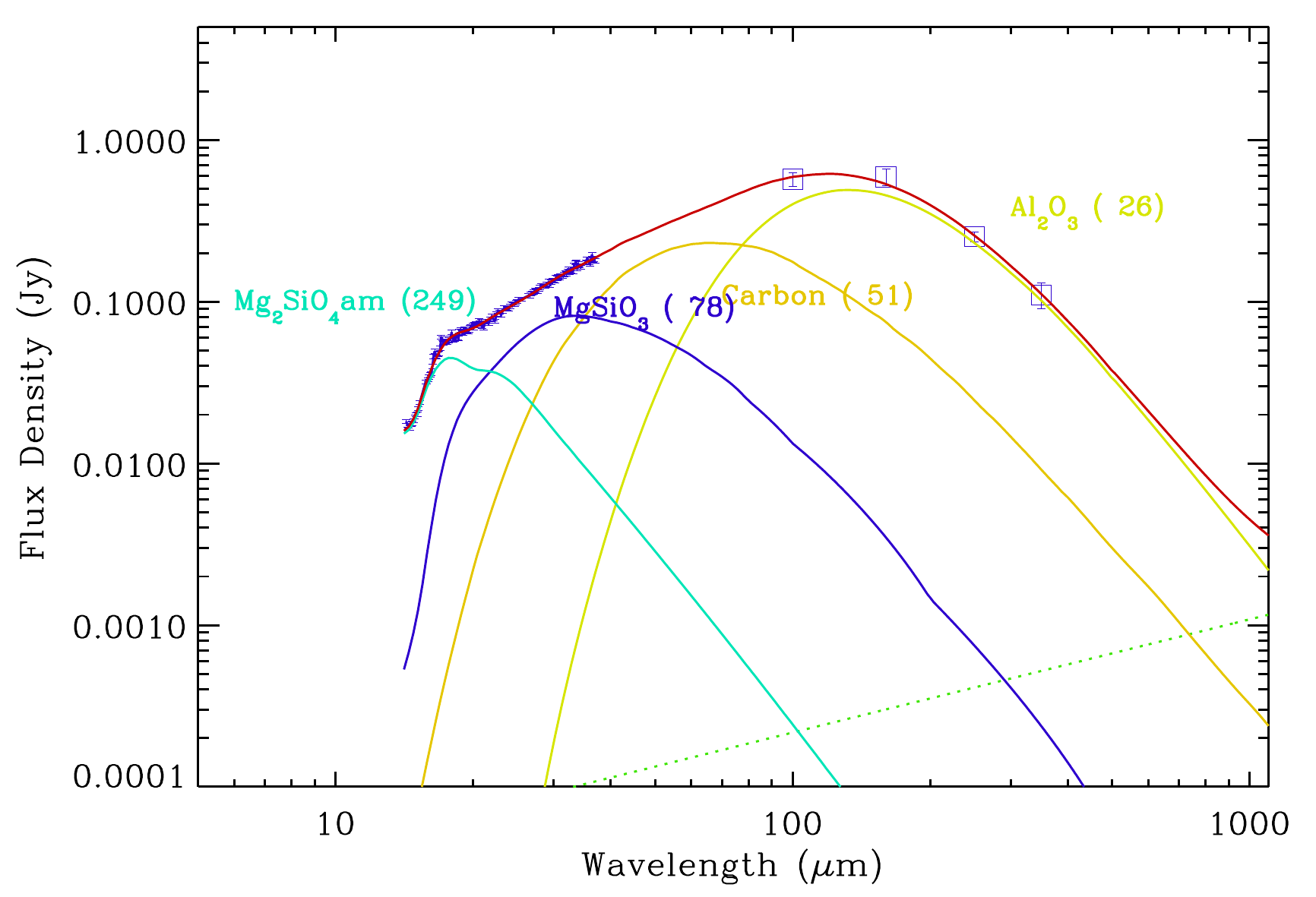}
\caption{Model A5: \spitzer\ IRS  spectra and \herschel\ photometry of N132D with one (Model A5 in Table \ref{Tdustmass}) of the best-fitting dust models superposed. The dust compositions include  Carbon/Mg$_2$SiO$_4$ and carbon dust to fit the 18 \mic\ feature. The dust compositions include MgSiO$_3$,  Mg$_2$SiO$_4$, Carbon  dust and Al$_2$O$_3$. Synchrotron radiation (green dotted line) is included in the fitting, but makes only a minor contribution.}
\label{N132DSEDfitA5}   
\end{figure}

We first performed spectral fitting of the 18 \mic-feature using carbon, enstatite (MgSiO$_3$) and forsterite (Mg$_2$SiO$_4$). Note that the synchrotron component is fixed as described in Section \ref{Ssynchrotron} (see the dotted green lines in Figures \ref{N132DSEDfitA4}  and \ref{N132DSEDfitA5}).  The feature is reasonably reproduced by Mg$_2$SiO$_4$ alone (Model A4 in Table \ref{Tdustmass} and Figure~\ref{N132DSEDfitA4}), a combination of MgSiO$_3$ and Mg$_2$SiO$_4$ (Models A2 and A4), or MgSiO$_3$ and Al$_2$O$_3$ (Model A5). Pure carbon dust fails to reproduce the 18 \mic\ feature. The result of the fitting is a smaller $\chi^2$  (Model A1 of Table \ref{Tdustmass}) than those of other models. Table \ref{Tdustmass} summarizes the dust spectral fitting results and dust masses (Models A1 to A6). The values of the reduced $\chi^2$ are small ($\sim$0.5) and they are similar to each other (Models A2-A6) except for Model A1, which is a much poorer fit. Model A1 is excluded because significant residuals appear around 100 \mic\ and the 18 \mic\ dust feature, which lead to a high reduced $\chi^2$. 

The cold component dominates the dust mass; the yielded dust mass is 1.60 M$_\odot$ for silicate dust of MgSiO$_3$ and Mg$_2$SiO$_4$ (Model A5), 0.79 M$_\odot$ for Carbon (Models A2 and A3), and 0.17 M$_\odot$ for the cold component of Al$_2$O$_3$ (Model A5 in Table \ref{Tdustmass}). We attempted a dust fitting with two cold dust components, MgSiO$_3$ and carbon, where we set their temperatures to $\geq$ 20\,K (because the infrared cirrus has a temperature below 20 K; \cite{reach95,reach00}). Doing this  results in significant dust masses from both types of dust (see Model A6 in Table \ref{Tdustmass}). The simple dust model by \citet{dwek07} predicts that 84\% of the dust is silicate,  10\% is carbon, and 6\% is other types for a 40\,M$_\odot$ progenitor star. The dust mass from Model A6 is  1.25$\pm$0.28\,M$_\odot$, which is between the dust masses assuming silicate (1.6$\pm$0.3\,M$_\odot$) and carbon dust (0.79$\pm$0.19\,M$_\odot$). 

The theoretical models of dust formation predict more than ten different grain species are formed in SN ejecta \citep{sluder18, marassi19, sarangi15}. Dust formation in SN Ib has been modeled by \citet{nozawa08}, who predicted FeS and Fe$_3$O$_4$ grains as well as the grains included in our fits. We did not experiment with using these two grain types since they produce strong dust features around 30 - 40 \mic, which are not observed in N132D. We also use Mg$_2$SiO$_4$ dust as a cold component, but the $\chi^2$ minimization strongly favors other dust composition such as MgSiO$_3$, carbon or Al$_2$O$_3$ as listed in Table \ref{Tdustmass}. We conclude the dust mass is 1.25$\pm$0.65 M$_\odot$ where 1.25 M$_\odot$ is from Model A6; the uncertainties are a combination of the lower limit of carbon dust (Model A3) and the upper limit of silicate (MgSiO$_3$) dust (Model A4) models.

%Table6
\begin{table*}
\footnotesize{
\caption[]{Dust Spectral fitting Results and Estimated Dust Mass.}\label{Tdustmass}
\begin{center}
\begin{tabular}{lccrrr}
\hline\hline
Model & $\Delta \chi^2$(=$\chi^2$/dof) &  18$\mu$m feature & other compositions (T, M$_d$) & cold (T, M$_d$) &total (M$_d$) \\
 &  &  dust (K, M$_\odot$) &  (K, M$_\odot$) & dust (K, M$_\odot$) & (M$_\odot$)  \\
&&&&\\
\hline
A1 & 3.34 & Carbon (107, 3.7E-4) & Mg$_2$SiO$_4$ (54, 5.1E-8), MgSiO$_3$ (38, 0.084)  &Al$_2$O$_3$ (25, 0.235) & 1.84\\
\hline
 A2 & 0.53  &{ MgSiO$_3$ } (700, 1.0E-7) & Mg$_2$SiO$_4$ (85, 3.6E-4), Al$_2$O$_3$ (40, 0.008) & Carbon (24, 0.77) & 0.78$\pm$0.15    \\  
    &       & ($>$300, $\pm$0.2E-7)  & ($\pm$3, $\pm$0.9E-4), ($\pm$2, $\pm$0.002)&($\pm$2, $\pm$0.15)&  \\
\hline
A3 & 0.40 & Mg$_2$SiO$_4$ (245, 1.78E-6) & MgSiO$_3$ (75, 6.80E-4), Al$_2$O$_3$ (44, 0.013 & Carbon (23, 0.779) & 0.79$\pm$0.19  \\  
    &       & ($\pm$14, 0.40E-6) & ($\pm$3, $\pm$2.3E-4), ($\pm$4, $\pm$0.007)  & ($\pm$3, $\pm$0.181)& \\
\hline
A4 & 0.45 & Mg$_2$SiO$_4$ (190, 4.5E-5) & Al$_2$O$_3$ (83, 3.7E-4), Carbon (48, 0.03) &MgSiO$_3$ (21, 1.565) &  1.60$\pm$0.30  \\  %Aug 23 
    &       & ($\pm$10, 1.3E-5) & ($\pm$7, $^{+5.8E-4}_{-0.3E-4}$), ($\pm$8, $\pm$0.03)  & ($\pm$2, $\pm$0.03)&\\

\hline
A5 & 0.40 &Mg$_2$SiO$_4$ (249, 1.6E-5) & MgSiO$_3$ (78, 4.4E-4),  Carbon (51, 0.020 )  & Al$_2$O$_3$ (26, 0.150) & 0.17$\pm$0.04 \\  %Aug 23
    &    &    ($\pm$5, $\pm$ 0.5E-6) & ($\pm$6, $\pm$3.3E-4),($\pm$6, 0.014)   & ($\pm$2, $\pm$0.03)& \\
    \hline
{\bf A6$^a$} & 0.52 & {\bf Mg$_2$SiO$_4$} (182, 6.8E-6) & {\bf Al$_2$O$_3$} (61, 1.4E-4) & {\bf MgSiO$_3$} (25$^b$, 0.863$\pm$0.150) & {\bf 1.25$\pm$0.28} \\
   &      & ($\pm$2, $\pm$0.4E-6)  & ($\pm$1, $\pm$1.4E-4) &  + {\bf C}(20$^b$, 0.386$\pm$0.234) & \\
\hline \hline
\end{tabular}
\end{center}
\footnotesize{$^a$The best values are marked in bold characters.}
\footnotesize{$^b$ The temperatures were set to be $\ge$20 K since the dust below 20 K belongs to cold ISM.}
}
\renewcommand{\baselinestretch}{0.8}
\end{table*}

%Figure16

%Figure17

We attempted to apply the three-component (unshocked, clumped, and diffuse) physical models of dust emission (called ``DINAMO") of \cite{priestley19, priestley22} to the N132D data. The model has been exclusively applied to the SNR Cas A \citep{priestley19, priestley22}. We calculated dust SEDs using the DINAMO model (Priestley et al. 2019) for a single grain size in each component, then fit the N132D data with the dust mass in each component as our three free parameters. Fitting was done using the Monte Carlo Markov chain code $`$emcee$'$ \citep{foreman-Mackey13}. However, we conclude that for N132D, the four far-IR data points from 100 to 250 \mic\ and the limited input parameters pertaining to its cold gas) were not adequate to disentangle the physical processes of heating and cooling that occur in the three phases. 

%Figure18
\begin{figure}
\includegraphics[scale=1,angle=0, width=8.5truecm]{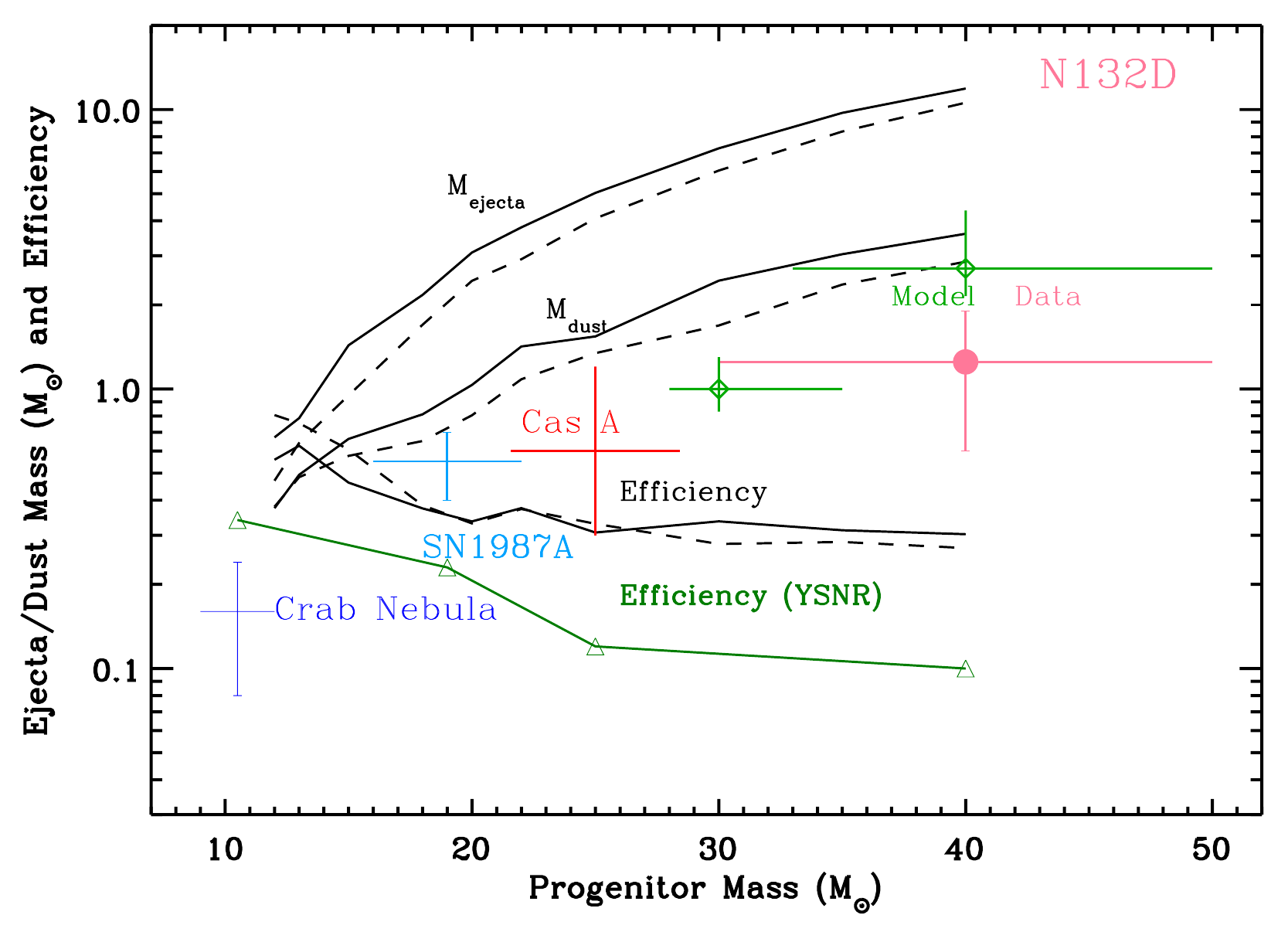}
\caption{Ejecta mass (M$_{ejecta}$), dust mass (M$_{dust}$), and dust formation efficiency (Efficiency = M$_{dust}$/M$_{ejecta}$) curves as a function of progenitor mass \citep{dwek07} with the dust mass of N132D. The dashed curves are for low metallicity. The observed dust mass of N132D (a filled circle in pink) is the results from the spectral fits in Table \ref{Tdustmass}. The upper and lower diamonds (green) are predicted dust masses from two different models (CE-0 and CE-2 with [Fe/H]=0 and [Fe/H]=-2, respectively) for 30-40 M$_\odot$ progenitor stars \citep{marassi19}. The errors were estimated by interpolation between the model grids. The dust masses of other YSNRs, namely Cas A, SN1987A, and the Crab Nebula, are also shown (Table \ref{tab:Tdustmassysnr}). The dust condensation efficiency is marked as $``$Efficiency (YSNR)" derived from the ratio of observed dust mass to total ejecta mass (M$_{ejecta}$). Both models and observations suggest that the dust mass in N132D is likely higher than the other three SNRs, but the uncertainty is large and depends on the dust composition.}
\label{dustmassprogenitor}
\end{figure}

%Figure19
\begin{figure*}
\includegraphics[scale=1,angle=0,width=8truecm]{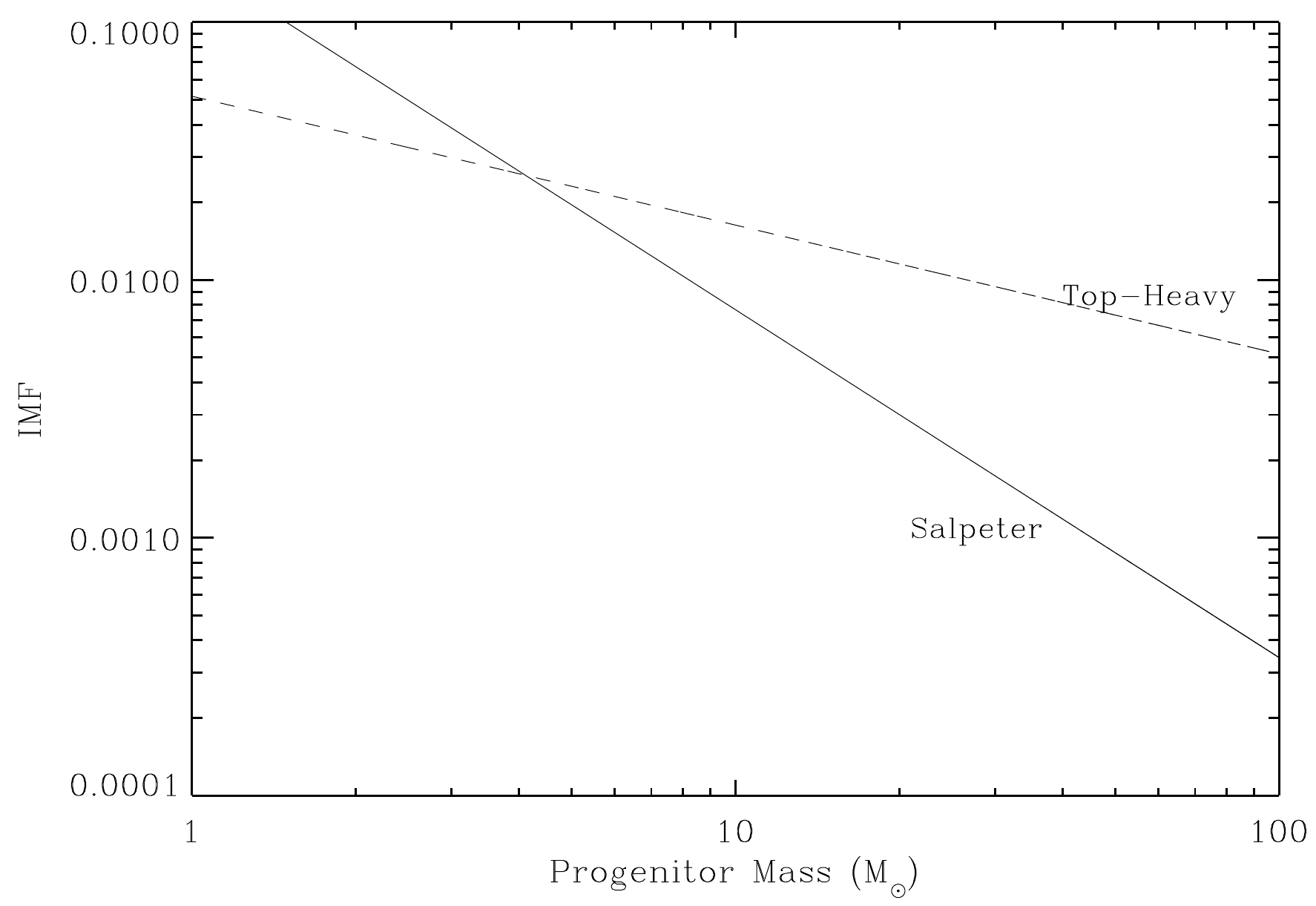}
\includegraphics[scale=1,angle=0,width=8truecm]{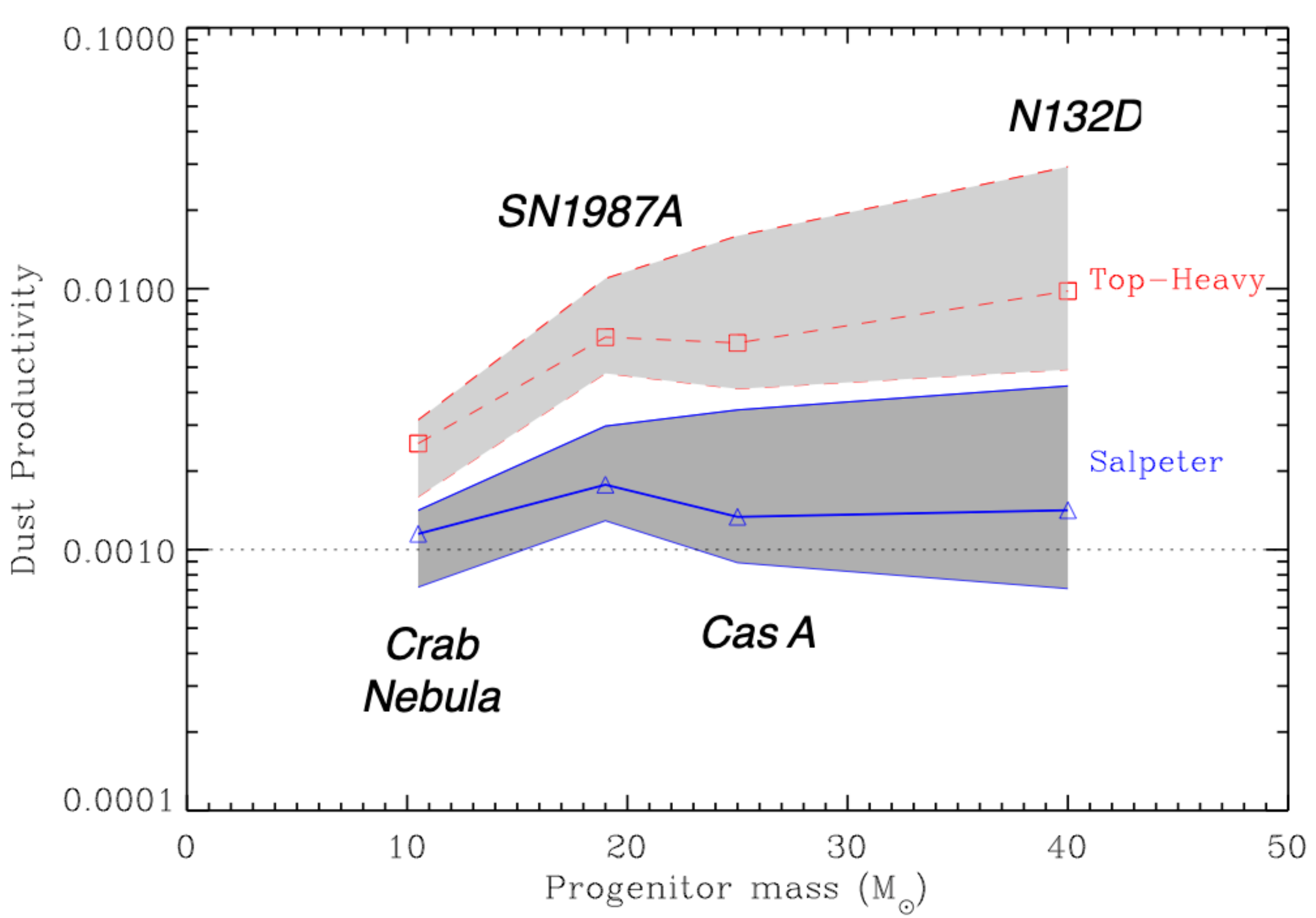}
\caption{The Salpeter and top-heavy IMFs {\it (left: a)} 
and dust productivity {\it (right: b)} are shown in solid and dashed lines,
based on the definitions in \cite{gall11}, and \cite{dwek11}. The dust
productivities are estimated based on the observed dust masses from 
the YSNRs, Crab Nebula, SN1987A, Cas A, and N132D (thick lines) with
squares symbols (Salpeter IMF) and triangles (top-heavy IMF) with
errors (grey)  The upper limits are from theoretical dust yields
\citep{dwek07}, which assume no dust destruction, and the lower limits are
from observations. The dotted line is the critical dust productivity
($\mu_D(\mathrm{crit})$, see the text for details).}
\label{imf}
\end{figure*}

\subsection{Progenitor mass of N132D}

Ground-based optical studies by \cite{sutherland95} and \cite{morse95} show that the O-rich ejecta filaments extend to a radius of 6 pc from the remnant center. HST WFPC2 images distinguish the ejecta from shocked circumstellar clouds \citep{morse96}. \cite{blair00}  suggested that the progenitor mass of N132D is between 30 and 35 M$_\odot$. The HST spectrographic observations found a silicon-to-oxygen abundance ratio ($\sim$0.01) which is consistent with the model of a progenitor of 50 M$_\odot$ \citep{france09}. Therefore, we use a progenitor mass of N132D of 40$\pm$10 M$_\odot$ (see Figure \ref{dustmassprogenitor}). Recent observations of SN 2019hgp show a rich set of carbon, oxygen, and neon emission lines, which suggest a massive progenitor star \citep{gal-yam21}. SN 2019hgp may be similar to the early stage of N132D.

Recently \citet{sharda20} carried out X-ray observations of N132D and detected Fe lines. They infer a low Fe abundance and a progenitor mass of 15 M$_\odot$. However, we detected many Fe lines from the CSM/ISM knots (such as Lasker's Bowl or West Complex in Table \ref{Tratioproperties}; see \cite{blair00}). These are not high-velocity ejecta knots. Thus we believe that contamination from CSM/ISM knots could have caused \citet{sharda20} to infer the lower progenitor mass. \citet{siegel21} suggest caution about using the X-ray Fe line as the sole diagnostic for typing a remnant. 

\subsection{Dust in N132D and in the early Universe}
\label{Sdustmass}

\subsubsection{Dust mass as a function of progenitor mass}

\begin{table*}
\caption[]{Summary of Dust Mass from \herschel\ - detected Young Supernova Remnants. } 
\begin{center}
\begin{tabular}{llllllllll}
\\
\hline \hline
SNR & dist & diameter&radius & age & dust mass & dominant & progenitor &ref. \\
    & (kpc) & ($'$) & (pc) & (yr)  & (M$_{\odot}$) &   dust type  & (M$_{\odot}$) \\
\hline
SN1987A & 51  & 0.02& 0.2 & 35 & 0.4-0.7 & carbon & 18-20: Type II   & 1 \\
Cas A (G111.7−2.1) & 3.4 &5 & 2& 350  & 0.3-0.5  &  silicate & 25-30: Type II/Ib & 2, 3 \\
\hline 
Crab Nebula (G184.6−5.8) & 2 & 7& 1.7& 968 & 0.1-0.54  & carbon & 10-12: Type IIP & 4, 5, 6 \\
G54.1+0.3 & 5 & 1.5& 1&2900 & {0.08-0.9} 
         & silicate & 15-35 (25$\pm$10): Type II$?$ &  6, 7 \\ %\cite{rho18}$^b$, \cite{temim17} \\
G21.5-0.9 & 4.6 &5 &3 & $<$1000 & 0.29$\pm$0.08 & ...& 10  & 8, 9\\
\hline
N132D &51 & 1.6 & 12.5& 2500 &1.25$\pm$0.65 &...&30-50: Ibc & 10, this paper\\
\hline \hline
\end{tabular}
\end{center}
\label{tab:Tdustmassysnr}
\footnotesize{$^a$
References are: (1) \cite{matsuura15}; (2) \cite{delooze17}; 
(3) \cite{rho08}; (4) \cite{gomez12}; 
(5) \cite{owen15}; (6) \cite{temim12}; (7) \cite{rho18}; (8) \cite{guest19}}; (9) \cite{chawner19}; (10) \cite{law20}
\end{table*}

We show predicted ejecta and dust masses as a function of progenitor mass and the observed dust masses of N132D in Figure \ref{dustmassprogenitor}. 
The amount of dust as a function of progenitor mass is given by \cite{dwek07}. Their calculation is based on the nucleosynthesis model and ejecta mass described in \cite{woosley95}. The dust masses of other YSNRs of the Crab Nebula, SN 1987A, and Cas A (see Table \ref{tab:Tdustmassysnr}) are also marked.

\citet{dwek07} modeled the mass of silicate dust, carbon dust, and other compositions (Ca-Ti-Al), and the total dust mass as a function of progenitor with the assumption of no dust destruction (see Figure \ref{dustmassprogenitor}). For a 35 M$_\odot$ progenitor, the total dust yield is 3.04 M$_\odot$, where silicate, carbon, and Ca-Ti-Al type dust are 2.56 (84\% of the total), 0.32 (10\%), 0.16 (6\%) M$_\odot$, respectively. In contrast, \citet{morgan03} predict the yield of carbon dust is similar to that of silicate dust. \citet{nozawa08} calculated a total dust mass of 1.5 M$_\odot$ for a type Ib SN model of which carbon dust is about half, 0.7 M$_\odot$. We obtain {\new a dust mass of} 0.6-1.9 M$_\odot$ for N132D. This is a factor of 1.3-2 times higher than those of the other four YSNRs (see Figure \ref{dustmassprogenitor}), whose progenitor masses are each lower than N132D. The estimated dust mass for N132D is comparable to some theoretical dust masses \citep{marassi19}. A dust mass of $\sim$1 M$_\odot$ per ccSN is needed to explain the dust observed in the early Universe \citep{dwek07}. N132D is likely an example with a higher than average dust mass for a young SN.

The progenitor and dust masses of Cas A, SN1987A, the Crab Nebula, G54.1-0.3, and G21.5-0.9, together with N132D are listed in Table \ref{tab:Tdustmassysnr}. 1E0102 (1E 0102-7219 or SNR B0102-72.3) \citep{rho09} is not included since its \herschel\ observation was not usable due to a bright \ion{H}{2} region near 1E0102 and {\it Herschel}'s limited spatial resolution. Thus the dust mass of 1E102 only using \spitzer\ data could not be compared with that of N132D. The progenitor and dust masses of G54.1+0.3 and G21.5-0.9 are similar to those of Cas A and the Crab Nebula,. The dust mass of N132D is up to a factor of two higher than those of Cas A, SN1987A, G54.1-0.3, and the Crab Nebula \citep{delooze17,matsuura11,indebetouw14,matsuura15,rho18,gomez12, owen15} as shown in Figure \ref{dustmassprogenitor}. The higher dust mass of N132D is consistent with the fact that its progenitor mass (30 - 50 M$_\odot$) is higher than those of the other YSNRs.

Dust formation models for higher mass progenitors are limited in number. \citet{marassi19} include models of SN Type Ib and show that dust masses of 1.4 - 3.4 M$_\odot$ (marked as diamonds in Figure \ref{dustmassprogenitor}) can be produced for 30 - 40 M$_\odot$ progenitor stars. Their models are for rotating progenitors; their model CE (CE-3) with $[\mathrm{Fe/H}] = -3$ for explosion properties calibrated to reproduce the $^{56}$Ni - M (progenitor mass) relation inferred from SN observations produces the highest dust mass. \citet{todini01} predicted that 30 - 35 M$_\odot$ progenitors produce a factor of 2 - 3 times higher dust masses than 25 M$_\odot$ progenitors and a factor of 4-5 times higher than 12 - 19 M$_\odot$ (SN IIP) progenitors; 30 M$_\odot$ progenitors predominantly produce carbon grains. \cite{marassi19} also find that carbon grains dominate with additional  Mg$_2$SiO$_4$ and MgSiO$_3$ present for a progenitor of 30 - 50 M$_\odot$ \citep[see Figure 11 of][]{marassi19}. N132D may be an example of this.

We examine the dust condensation efficiency for N132D. Figure \ref{dustmassprogenitor} shows the ejecta mass range is 9.72 - 11.88 M$_\odot$  for a 30 - 40 M$_\odot$ progenitor \citep{woosley95}. The maximum dust produced (assuming no dust destruction) is 3.03 -- 3.59 M$_\odot$ corresponding to a dust condensation efficiency range of 0.28 - 0.3 \citep{dwek07}. 

Detailed explosion and dust formation models accounting for rotation, different explosion energies, and abundances predict maximum dust masses of between 1.9 and 3.4 M$_\odot$ \citep{marassi19}, which corresponds to an efficiency of 0.20 - 0.28 for a 30 - 40   M$_\odot$ progenitor. We also derive dust condensation efficiencies (thick green line in Figure \ref{dustmassprogenitor}) from the observed dust mass and the ejecta mass from theoretical models depending on the progenitor mass. Using our derived dust mass for N132D (1.25$\pm$0.65 M$_\odot$) for a combination of silicate and carbon dust (Model A6 in Table \ref{Tdustmass}), and we estimate the dust condensation efficiency to be 0.10$\pm$0.05 using the dust mass divided by the gas mass of 11.88 M$_\odot$ for a 40 M$_\odot$ progenitor star (marked as a green curve in Fig.~\ref{dustmassprogenitor}). This is a factor of 2 - 3 smaller than efficiencies estimated from theoretical models when assuming no dust destruction \citep[marked as a light green curve in Fig.~\ref{dustmassprogenitor};][]{dwek07}.

\subsubsection{Dust production rate}

We derive a dust productivity ($\mu_D$) defined in \cite{gall11} and \cite{dwek07}  using the stellar initial functions (IMFs; $\phi_m$) and the dust mass ($M_d$ = $M_Z~\epsilon$ where $M_Z$ is the mass in metals and $\epsilon$ is a condensation efficiency). In other words, the dust productivity is the dust production rate per year and per ccSN. The dust productivity \citep[$\mu_D$; ][]{gall11} is given by:
$$\mu_D = \int_{m_l}^{m_u} \phi_m~ {{M_Z}\over{M_\odot}}~\epsilon(m)~dm$$

$$\mu_D~(obs) = \int_{m_l}^{m_u} \phi_m~{M_d~(obs)}~ dm$$

The observed dust mass, $M_d(\mathrm{obs})$, is from the four YSNRs, including our estimate for N132D, and the dust-mass yield (see $M_{dust}$ in Figure \ref{dustmassprogenitor}) is estimated from a theoretical model by \cite{dwek07}. We used the Salpeter and top-heavy IMFs \citep[the total sum of $\phi_m$ from $m_l = 0.1\,\mathrm{M}_\odot$ to $m_u = 100~\mathrm{M}_\odot$ is normalized to be 1;][]{salpeter55, kroupa01}, which are shown in Figure \ref{imf}a.

The dust productivities ($\mu_D(\mathrm{obs})$) for the progenitor masses of the four YSNRs are estimated from the observed dust masses and are shown in Figure~\ref{imf}b. The upper limits of $\mu_D$ are calculated using the theoretical dust yields \citep{dwek07} where no dust destruction is assumed, and the lower limits are calculated using the observed dust masses of the four YSNRs (0.39 M$_\odot$ using carbon dust for N132D). \cite{gall11} estimated the dust productivity using the dust observed in QSOs at $z \ge 6$ \citep{bertoldi03}. The dust production rate in the early Universe for 400 Myr after the Big Bang is $R_D = M_D/\delta t = 0.5~\mathrm{M}_\odot$ yr$^{-1} = \mu_D~\psi$ where $\psi$ (= 500 -1500 M$_\odot$ yr$^{-1}$) is the SFR \citep{bertoldi03, wang10}. This leads to $\mu_D$ = 10$^{-3}$ (= 0.5 / 500), which is the minimum dust productivity, which we call a critical dust productivity ($= \mu_D(\mathrm{crit})$; marked as a dotted line in Fig.~\ref{imf}b), required to explain the dust in the early Universe \citep[see][for details]{gall11, dwek07}. 

The dust productivities using the top-heavy IMF are higher than
$\mu_D(\mathrm{crit})$. The dust productivities using the Salpeter IMF are
higher, but their lower limits are below $\mu_D(\mathrm{crit})$.
\cite{dwek11} showed that AGB stars are only minor contributors to the dust
abundance in the early Universe. Figure \ref{imf}b shows that for a 
top-heavy IMF, the dust productivity ($\mu_D$) is higher than the critical
dust productivity ($\mu_D (crit)$). In particular, the implied $\mu_D$ from
the dust mass of N132D for 40 M$_\odot$ progenitors is a factor of 5 - 20
higher than the dust productivity required to explain the dust mass in the
early Universe. Using a Salpeter IMF, $\mu_D$ is higher than $\mu_D
(crit)$; however, after accounting for the errors, the lower limit of
$\mu_D$ is below $\mu_D (crit)$. In summary, the estimated dust
productivity implies that the dust from ccSNe can explain the dust observed
in the early Universe when we assume a top-heavy IMF. 

Our conclusion, based in part on the observations reported here, that dust produced by ccSNe can account for a large amount of dust observed in the early Universe, considers the uncertainties associated with dust survival. The reason is because the dust mass observed in the young SNR N132D is after the dust destruction occurs in reverse shocks that traverse the ejecta. The dust destruction may continue, but molecule reformation in the post-shock gas can reduce the dust destruction \citep{biscaro14, matsuura19}. Early studies suggested that the lifetimes of the major dust materials against destruction are short compared to the time scales for the dust formation \citep{jones96} and imply 45-100\% destruction in \citep{nozawa07, bianchi07}. However, more recent work \citep{silvia12, nath08, jones11} indicate lower destruction percentages (1-50\%). In addition, hydrodynamical calculations show that up to 20 - 50\%  of the dust can survive \citep{slavin20}. Adding to the uncertainties, dust destruction rates vary depending on dust composition, grain size, and reverse-shock properties. For example, carbon dust is more resilient against the supernova shock than silicate dust \citep{silvia12, jones11, kirchschlager19, priestley21, slavin20}.

Clearly, N132D alone does not resolve the issue of dust production in the early Universe. Further observational and theoretical work is required. The JWST has recently detected galaxies in the early Universe at $z$ $>$ 12 \citep{pontoppidan22,finkelstein22,treu22}. The nature of the dust in them and how that dust was produced will be subjects of great interest. The observations reported here of N132D, with its high progenitor mass and high dust mass, raise the possibility that ccSNe associated with high mass progenitors could make a significant contribution to the dust observed in early galaxies, despite their being only a small percentage of the IMF compared to SNe with smaller progenitor masses (e.g., SNe IIP).

{\vskip 0.4truecm}
JWST can resolve the ejecta and dust continuum structures in N132D on a spatial resolution similar to the HST and may find a direct correlation between them, as \spitzer\ revealed the one-to-one correlation between the Ar ejecta and 21\mic-dust in Cas A  \citep{rho08}. JWST is also important in unambiguously identifying dust features and understanding the dust composition. Most importantly, spectral coverage in far-IR with a higher spatial resolution, such as would be possible with the Origins Space Telescope \citep{leisawitz21}, together with JWST, is vital in distinguishing between complex dust models, including multi-composition dust populations, where the shape of the continuum can constrain the composition and yield accurate dust masses. Detailed dust formation models for massive stars ($> 25~ \mathrm{M}_\odot$) including SN Type Ib or Ic, and including all dust compositions, are also needed.

\subsubsection*{Conclusions}  
1. We present {\it Spitzer}, {\it WISE}, and \herschel\ observations of the young supernova remnant (SNR) N132D in the Large Magellanic Cloud, including 3-40 \mic\ {\it Spitzer} IRS mapping, {\bf 12 \mic\ {\it WISE}} and 70, 100, 160, 250, 350, and 500 \mic\ \herschel\ photometry. The [\ion{Ne}{3}] at 15.5 \mic\ and [\ion{O}{4}] 26 \mic, and [\ion{Si}{2}] 34.8 \mic\ maps reveal a spatial distribution of infrared-emitting ejecta coinciding with the optical and X-ray ejecta at the center and the two CSM/ISM knots of Lasker's Bowl and the West Complex.
{\it WISE} 12 \mic\ image is essentially a \neiif\ map that shows both dominant ejecta emission at the center and shocked ISM emission at the shell.

2. The line width of [\ion{Ne}{3}] is larger than the instrumental spectral resolution. The implied velocity is 3167$\pm$60 and 3620$\pm$47 km s$^{-1}$ for the NE ejecta and central ejecta, respectively.

3. Ejecta spectra of N132D are remarkably similar to those of another YSNR, 1E0102, in the Ne and O lines and continuum shapes. N132D shows a higher ratio of [\ion{Ne}{2}]/[\ion{Ne}{3}] than 1E0102. Shock modeling of Ne ejecta emission suggests a gas temperature of $\sim500$ K and densities in the range $1000-3000$ cm$^{-3}$.

4. The Lasker's Bowl and West Complex regions contain dense CSM knots from pre-SN mass loss which emit strong [\ion{Fe}{2}]/[\ion{S}{3}] emission at 18 \mic\ as well as weaker neon and oxygen emission. They show bright \feiif,  \ariif, \siiif, and \niciif\ lines. Diagnostics of the \feiif\ lines using line ratios 17.9/5.35 \mic\ and 17.9/26 \mic\ imply densities of 2000 - 12000 cm$^{-3}$ of the CSM knots and temperatures of $(6 - 13)\times10^4$ K.

5. The PACS and SPIRE images of N132D show infrared emission at the center and in the shell. The cold dust at the center coincides with the optical/X-ray/infrared ejecta, implying that there is freshly formed dust there. The continuum from the ejecta shows an 18 \mic\ dust feature.

6. We performed spectral fitting to the IRS dust continuum and \herschel\ far-IR photometry. The spectra are fitted with the compositions of MgSiO$_3$ and Mg$_2$SiO$_4$, carbon, and Al$_2$O$_3$. The  18 \mic-dust feature requires forsterite grains. The derived dust mass for the ejecta in the central region is 1.25$\pm$0.65 M$_\odot$ using a combination of cold silicate carbon dust, which implies that the range of possible dust mass is 0.6 - 1.9 M$_\odot$. The dust mass of N132D may be higher than those of other YSNRs and exceed the typical expectations of dust mass per ccSN. This may be due to N132D's high progenitor mass of  30 - 50 M$_\odot$.

7. N132D is suggested to be the result of the  explosion of a massive star with a progenitor mass of  30 - 50 M$_\odot$. We estimate the dust condensation efficiency in N132D to be $\sim$11 \%, lower than for lower mass progenitor stars. The dust productivities of N132D and other YSNRs using a top-heavy IMF are higher than the critical dust productivity $\mu_D(\mathrm{crit})$,  suggesting that dust from  ccSNe could explain the amount of dust present in the early Universe.  However, it is necessary to expand the sample of galaxies in the early Universe since the dust productivity depends on the SFR, dust mass measurements, and the evolution time scale. It is also vital to have more samples of nearby ccSNe and YSNRs with different progenitor masses and accurate measurements of their dust masses for more convincing tests of whether ccSNe are the main dust producers in the early Universe.

\acknowledgements

We thank Felix Priestley for running the DINAMO models and guiding us to run the code, Charles Law and Dan Milisavljevic for providing the velocity map of fast O-rich knots, and
Achim Tappe for updating temperature-dependent atomic data of Fe.
J.\,R. would like to thank T. R. Geballe for carefully reading the manuscript and helping to clarify the text.
We thank the anonymous referee for insightful and detailed comments.  
{\it Herschel} is an ESA space observatory with science instruments provided by European-led Principal Investigator consortia and with important participation from NASA.
J.\,R. and A.\,P.\,R. acknowledge support from NASA ADAP grants NNX12AG97G and 80NSSC20K0449.

\bibliography{msrefsall}

\begin{thebibliography}{}
\expandafter\ifx\csname natexlab\endcsname\relax\def\natexlab#1{#1}\fi

\bibitem[{{Bautista}(2004)}]{bautista04}
{Bautista}, M.~A. 2004, \aap, 420, 763

\bibitem[{{Begemann} {et~al.}(1997){Begemann}, {Dorschner}, {Henning},
  {Mutschke}, {G{\"u}rtler}, {K{\"o}mpe}, \& {Nass}}]{begemann97}
{Begemann}, B., {Dorschner}, J., {Henning}, T., {et~al.} 1997, \apj, 476, 199

\bibitem[{{Behar} {et~al.}(2001){Behar}, {Rasmussen}, {Griffiths}, {Dennerl},
  {Audard}, {Aschenbach}, \& {Brinkman}}]{behar01}
{Behar}, E., {Rasmussen}, A.~P., {Griffiths}, R.~G., {et~al.} 2001, \aap, 365,
  L242

\bibitem[{{Bertoldi} {et~al.}(2003){Bertoldi}, {Carilli}, {Cox}, {Fan},
  {Strauss}, {Beelen}, {Omont}, \& {Zylka}}]{bertoldi03}
{Bertoldi}, F., {Carilli}, C.~L., {Cox}, P., {et~al.} 2003, \aap, 406, L55

\bibitem[{{Bianchi} \& {Schneider}(2007)}]{bianchi07}
{Bianchi}, S., \& {Schneider}, R. 2007, \mnras, 378, 973

\bibitem[{{Biscaro} \& {Cherchneff}(2014)}]{biscaro14}
{Biscaro}, C., \& {Cherchneff}, I. 2014, \aap, 564, A25

\bibitem[{{Blair} {et~al.}(2000){Blair}, {Morse}, {Raymond}, {Kirshner},
  {Hughes}, {Dopita}, {Sutherland}, {Long}, \& {Winkler}}]{blair00}
{Blair}, W.~P., {Morse}, J.~A., {Raymond}, J.~C., {et~al.} 2000, \apj, 537, 667

\bibitem[{{Bohren} \& {Huffman}(1985)}]{bohren83}
{Bohren}, C.~F., \& {Huffman}, D.~R. 1985, Absorption and Scattering of Light
  by Small Particles, 571 (New York: Wiley), 571

\bibitem[{{Borkowski} {et~al.}(2007){Borkowski}, {Hendrick}, \&
  {Reynolds}}]{borkowski07}
{Borkowski}, K.~J., {Hendrick}, S.~P., \& {Reynolds}, S.~P. 2007, \apjl, 671,
  L45

\bibitem[{{Boulanger} {et~al.}(1996){Boulanger}, {Abergel}, {Bernard},
  {Burton}, {Desert}, {Hartmann}, {Lagache}, \& {Puget}}]{boulanger96}
{Boulanger}, F., {Abergel}, A., {Bernard}, J.~P., {et~al.} 1996, \aap, 312, 256

\bibitem[{{Chawner} {et~al.}(2019){Chawner}, {Marsh}, {Matsuura}, {Gomez},
  {Cigan}, {De Looze}, {Barlow}, {Dunne}, {Noriega-Crespo}, \&
  {Rho}}]{chawner19}
{Chawner}, H., {Marsh}, K., {Matsuura}, M., {et~al.} 2019, \mnras, 483, 70

\bibitem[{{Dayal} {et~al.}(2022){Dayal}, {Ferrara}, {Sommovigo}, {Bouwens},
  {Oesch}, {Smit}, {Gonzalez}, {Schouws}, {Stefanon}, {Kobayashi}, {Bremer},
  {Algera}, {Aravena}, {Bowler}, {da Cunha}, {Fudamoto}, {Graziani}, {Hodge},
  {Inami}, {De Looze}, {Pallottini}, {Riechers}, {Schneider}, {Stark}, \&
  {Endsley}}]{dayal22}
{Dayal}, P., {Ferrara}, A., {Sommovigo}, L., {et~al.} 2022, \mnras, 512, 989

\bibitem[{{De Looze} {et~al.}(2017){De Looze}, {Barlow}, {Swinyard}, {Rho},
  {Gomez}, {Matsuura}, \& {Wesson}}]{delooze17}
{De Looze}, I., {Barlow}, M.~J., {Swinyard}, B.~M., {et~al.} 2017, \mnras, 465,
  3309

\bibitem[{{Dickel} \& {Milne}(1995)}]{dickel95}
{Dickel}, J.~R., \& {Milne}, D.~K. 1995, \aj, 109, 200

\bibitem[{{Dopita} {et~al.}(2018){Dopita}, {Vogt}, {Sutherland}, {Seitenzahl},
  {Ruiter}, \& {Ghavamian}}]{dopita18}
{Dopita}, M.~A., {Vogt}, F. P.~A., {Sutherland}, R.~S., {et~al.} 2018, \apjs,
  237, 10

\bibitem[{{Dwek} \& {Cherchneff}(2011)}]{dwek11}
{Dwek}, E., \& {Cherchneff}, I. 2011, \apj, 727, 63

\bibitem[{{Dwek} {et~al.}(2007){Dwek}, {Galliano}, \& {Jones}}]{dwek07}
{Dwek}, E., {Galliano}, F., \& {Jones}, A.~P. 2007, \apj, 662, 927

\bibitem[{{Ferland} {et~al.}(2017){Ferland}, {Chatzikos}, {Guzm{\'a}n},
  {Lykins}, {van Hoof}, {Williams}, {Abel}, {Badnell}, {Keenan}, {Porter}, \&
  {Stancil}}]{ferland17}
{Ferland}, G.~J., {Chatzikos}, M., {Guzm{\'a}n}, F., {et~al.} 2017, RMxAA, 53,
  385

\bibitem[{{Finkelstein} {et~al.}(2022){Finkelstein}, {Bagley}, {Haro},
  {Dickinson}, {Ferguson}, {Kartaltepe}, {Papovich}, {Burgarella}, {Kocevski},
  {Huertas-Company}, {Iyer}, {Koekemoer}, {Larson}, {P{\'e}rez-Gonz{\'a}lez},
  {Rose}, {Tacchella}, {Wilkins}, {Chworowsky}, {Medrano}, {Morales},
  {Somerville}, {Aaron Yung}, {Fontana}, {Giavalisco}, {Grazian}, {Grogin},
  {Kewley}, {Kirkpatrick}, {Kurczynski}, {Lotz}, {Pentericci}, {Pirzkal},
  {Ravindranath}, {Ryan}, {Trump}, {Yang}, {Almaini}, {Amor{\'\i}n},
  {Annunziatella}, {Backhaus}, {Barro}, {Behroozi}, {Bell}, {Bhatawdekar},
  {Bisigello}, {Bromm}, {Buat}, {Buitrago}, {Calabr{\`o}}, {Casey},
  {Castellano}, {Ch{\'a}vez Ortiz}, {Ciesla}, {Cleri}, {Cohen}, {Cole},
  {Cooke}, {Cooper}, {Cooray}, {Costantin}, {Cox}, {Croton}, {Daddi},
  {Dav{\'e}}, {de La Vega}, {Dekel}, {Elbaz}, {Estrada-Carpenter}, {Faber},
  {Fern{\'a}ndez}, {Finkelstein}, {Freundlich}, {Fujimoto},
  {Garc{\'\i}a-Argum{\'a}nez}, {Gardner}, {Gawiser}, {G{\'o}mez-Guijarro},
  {Guo}, {Hamblin}, {Hamilton}, {Hathi}, {Holwerda}, {Hirschmann}, {Hutchison},
  {Jaskot}, {Jha}, {Jogee}, {Juneau}, {Jung}, {Kassin}, {Bail}, {Leung},
  {Lucas}, {Magnelli}, {Mantha}, {Matharu}, {McGrath}, {McIntosh}, {Merlin},
  {Mobasher}, {Newman}, {Nicholls}, {Pandya}, {Rafelski}, {Ronayne}, {Santini},
  {Seill{\'e}}, {Shah}, {Shen}, {Simons}, {Snyder}, {Stanway}, {Straughn},
  {Teplitz}, {Vanderhoof}, {Vega-Ferrero}, {Wang}, {Weiner}, {Willmer},
  {Wuyts}, {Zavala}, \& {The Ceers Team:}}]{finkelstein22}
{Finkelstein}, S.~L., {Bagley}, M.~B., {Haro}, P.~A., {et~al.} 2022, \apjl,
  940, L55

\bibitem[{{Foreman-Mackey} {et~al.}(2013){Foreman-Mackey}, {Hogg}, {Lang}, \&
  {Goodman}}]{foreman-Mackey13}
{Foreman-Mackey}, D., {Hogg}, D.~W., {Lang}, D., \& {Goodman}, J. 2013, \pasp,
  125, 306

\bibitem[{{France} {et~al.}(2009){France}, {Beasley}, {Keeney}, {Danforth},
  {Froning}, {Green}, \& {Shull}}]{france09}
{France}, K., {Beasley}, M., {Keeney}, B.~A., {et~al.} 2009, \apjl, 707, L27

\bibitem[{{Gal-Yam} {et~al.}(2021){Gal-Yam}, {Bruch}, {Schulze}, {Yang},
  {Perley}, {Irani}, {Sollerman}, {Kool}, {Soumagnac}, {Yaron}, {Strotjohann},
  {Zimmerman}, {Barbarino}, {Kulkarni}, {Kasliwal}, {De}, {Yao}, {Fremling},
  {Yan}, {Ofek}, {Fransson}, {Filippenko}, {Zheng}, {Brink}, {Copperwheat},
  {Foley}, {Brown}, {Siebert}, {Leloudas}, {Cabrera-Lavers}, {Garcia-Alvarez},
  {Marante-Barreto}, {Frederick}, {Hung}, {Wheeler}, {Vinko}, {Thomas},
  {Graham}, {Duev}, {Drake}, {Dekany}, {Bellm}, {Rusholme}, {Shupe},
  {Andreoni}, {Sharma}, {Riddle}, {van Roestel}, \& {Knezevic}}]{gal-yam21}
{Gal-Yam}, A., {Bruch}, R., {Schulze}, S., {et~al.} 2021, arXiv e-prints,
  arXiv:2111.12435

\bibitem[{{Galavis} {et~al.}(1995){Galavis}, {Mendoza}, \&
  {Zeippen}}]{galavis95}
{Galavis}, M.~E., {Mendoza}, C., \& {Zeippen}, C.~J. 1995, \aaps, 111, 347

\bibitem[{{Galav{\'\i}s} {et~al.}(1998){Galav{\'\i}s}, {Mendoza}, \&
  {Zeippen}}]{galavis98}
{Galav{\'\i}s}, M.~E., {Mendoza}, C., \& {Zeippen}, C.~J. 1998, \aaps, 133, 245

\bibitem[{{Gall} {et~al.}(2011){Gall}, {Hjorth}, \& {Andersen}}]{gall11}
{Gall}, C., {Hjorth}, J., \& {Andersen}, A.~C. 2011, \aapr, 19, 43

\bibitem[{{Gomez} {et~al.}(2012){Gomez}, {Krause}, {Barlow}, {Swinyard},
  {Owen}, {Clark}, {Matsuura}, {Gomez}, {Rho}, {Besel}, {Bouwman}, {Gear},
  {Henning}, {Ivison}, {Polehampton}, \& {Sibthorpe}}]{gomez12}
{Gomez}, H.~L., {Krause}, O., {Barlow}, M.~J., {et~al.} 2012, \apj, 760, 96

\bibitem[{{Guest} {et~al.}(2019){Guest}, {Safi-Harb}, \& {Tang}}]{guest19}
{Guest}, B.~T., {Safi-Harb}, S., \& {Tang}, X. 2019, \mnras, 482, 1031

\bibitem[{{Hartigan} {et~al.}(1987){Hartigan}, {Raymond}, \&
  {Hartmann}}]{hartigan87}
{Hartigan}, P., {Raymond}, J., \& {Hartmann}, L. 1987, \apj, 316, 323

\bibitem[{{Hewitt} {et~al.}(2009){Hewitt}, {Rho}, {Andersen}, \&
  {Reach}}]{hewitt09}
{Hewitt}, J.~W., {Rho}, J., {Andersen}, M., \& {Reach}, W.~T. 2009, \apj, 694,
  1266

\bibitem[{{Houck} {et~al.}(2004){Houck}, {Roellig}, {van Cleve}, {Forrest},
  {Herter}, {Lawrence}, {Matthews}, {Reitsema}, {Soifer}, {Watson}, {Weedman},
  {Huisjen}, {Troeltzsch}, {Barry}, {Bernard-Salas}, {Blacken}, {Brandl},
  {Charmandaris}, {Devost}, {Gull}, {Hall}, {Henderson}, {Higdon}, {Pirger},
  {Schoenwald}, {Sloan}, {Uchida}, {Appleton}, {Armus}, {Burgdorf},
  {Fajardo-Acosta}, {Grillmair}, {Ingalls}, {Morris}, \& {Teplitz}}]{houck04}
{Houck}, J.~R., {Roellig}, T.~L., {van Cleve}, J., {et~al.} 2004, \apjs, 154,
  18

\bibitem[{{Hwang} {et~al.}(1993){Hwang}, {Hughes}, {Canizares}, \&
  {Markert}}]{hwang93}
{Hwang}, U., {Hughes}, J.~P., {Canizares}, C.~R., \& {Markert}, T.~H. 1993,
  \apj, 414, 219

\bibitem[{{Indebetouw} {et~al.}(2014){Indebetouw}, {Matsuura}, {Dwek},
  {Zanardo}, {Barlow}, {Baes}, {Bouchet}, {Burrows}, {Chevalier}, {Clayton},
  {Fransson}, {Gaensler}, {Kirshner}, {Laki{\'c}evi{\'c}}, {Long}, {Lundqvist},
  {Mart{\'{\i}}-Vidal}, {Marcaide}, {McCray}, {Meixner}, {Ng}, {Park},
  {Sonneborn}, {Staveley-Smith}, {Vlahakis}, \& {van Loon}}]{indebetouw14}
{Indebetouw}, R., {Matsuura}, M., {Dwek}, E., {et~al.} 2014, \apjl, 782, L2

\bibitem[{{Isaak} {et~al.}(2002){Isaak}, {Priddey}, {McMahon}, {Omont},
  {Peroux}, {Sharp}, \& {Withington}}]{isaak02}
{Isaak}, K.~G., {Priddey}, R.~S., {McMahon}, R.~G., {et~al.} 2002, \mnras, 329,
  149

\bibitem[{{Jones} \& {Nuth}(2011)}]{jones11}
{Jones}, A.~P., \& {Nuth}, J.~A. 2011, \aap, 530, A44

\bibitem[{{Jones} {et~al.}(1996){Jones}, {Tielens}, \& {Hollenbach}}]{jones96}
{Jones}, A.~P., {Tielens}, A.~G.~G.~M., \& {Hollenbach}, D.~J. 1996, \apj, 469,
  740

\bibitem[{{Kirchschlager} {et~al.}(2019){Kirchschlager}, {Schmidt}, {Barlow},
  {Fogerty}, {Bevan}, \& {Priestley}}]{kirchschlager19}
{Kirchschlager}, F., {Schmidt}, F.~D., {Barlow}, M.~J., {et~al.} 2019, \mnras,
  489, 4465

\bibitem[{{Kotak} {et~al.}(2006){Kotak}, {Meikle}, {Pozzo}, {van Dyk},
  {Farrah}, {Fesen}, {Filippenko}, {Foley}, {Fransson}, {Gerardy}, {Hoeflich},
  {Lundqvist}, {Mattila}, {Sollerman}, \& {Wheeler}}]{kotak06}
{Kotak}, R., {Meikle}, P., {Pozzo}, M., {et~al.} 2006, \apjl, 651, L117

\bibitem[{{Kroupa}(2001)}]{kroupa01}
{Kroupa}, P. 2001, \mnras, 322, 231

\bibitem[{{Lagache} {et~al.}(1998){Lagache}, {Abergel}, {Boulanger}, \&
  {Puget}}]{lagache98}
{Lagache}, G., {Abergel}, A., {Boulanger}, F., \& {Puget}, J.~L. 1998, \aap,
  333, 709

\bibitem[{{Lai} {et~al.}(2020){Lai}, {Smith}, {Baba}, {Spoon}, \&
  {Imanishi}}]{lai20}
{Lai}, T. S.~Y., {Smith}, J.~D.~T., {Baba}, S., {Spoon}, H. W.~W., \&
  {Imanishi}, M. 2020, \apj, 905, 55

\bibitem[{{Laporte} {et~al.}(2017){Laporte}, {Ellis}, {Boone}, {Bauer},
  {Qu{\'e}nard}, {Roberts-Borsani}, {Pell{\'o}}, {P{\'e}rez-Fournon}, \&
  {Streblyanska}}]{laporte17}
{Laporte}, N., {Ellis}, R.~S., {Boone}, F., {et~al.} 2017, \apjl, 837, L21

\bibitem[{{Lasker}(1980)}]{lasker80}
{Lasker}, B.~M. 1980, \apj, 237, 765

\bibitem[{{Law} {et~al.}(2020){Law}, {Milisavljevic}, {Patnaude}, {Plucinsky},
  {Gladders}, {Schmidt}, {Sravan}, {Banovetz}, {Sano}, {McGraw}, {Takahashi},
  \& {Orlando}}]{law20}
{Law}, C.~J., {Milisavljevic}, D., {Patnaude}, D.~J., {et~al.} 2020, \apj, 894,
  73

\bibitem[{{Leisawitz} {et~al.}(2021){Leisawitz}, {Amatucci}, {Allen},
  {Arenberg}, {Armus}, {Battersby}, {Bauer}, {Bell}, {Benford}, {Bergin},
  {Booth}, {Bradford}, {Bradley}, {Carey}, {Carter}, {Cooray}, {Corsetti},
  {Dewell}, {DiPirro}, {Drake}, {East}, {Ennico}, {Feller}, {Flores},
  {Fortney}, {Granger}, {Greene}, {Howard}, {Kataria}, {Knight}, {Lawrence},
  {Lightsey}, {Mather}, {Meixner}, {Melnick}, {McMurtry}, {Milam}, {Moseley},
  {Narayanan}, {Nordt}, {Padgett}, {Pontoppidan}, {Pope}, {Rafanelli},
  {Redding}, {Rieke}, {Roellig}, {Sakon}, {Sandin}, {Sandstrom}, {Sengupta},
  {Sheth}, {Sokolsky}, {Staguhn}, {Steeves}, {Stevenson}, {Su}, {Vieira},
  {Webster}, {Wiedner}, {Wright}, {Wu}, {Yanatsis}, {Zmuidzinas}, \& {Origins
  Space Telescope Mission Concept and Study Team}}]{leisawitz21}
{Leisawitz}, D., {Amatucci}, E., {Allen}, L., {et~al.} 2021, Journal of
  Astronomical Telescopes, Instruments, and Systems, 7, 011014

\bibitem[{{Marassi} {et~al.}(2019){Marassi}, {Schneider}, {Limongi}, {Chieffi},
  {Graziani}, \& {Bianchi}}]{marassi19}
{Marassi}, S., {Schneider}, R., {Limongi}, M., {et~al.} 2019, \mnras, 484, 2587

\bibitem[{{Markwardt}(2009)}]{markwardt09}
{Markwardt}, C.~B. 2009, in Astronomical Society of the Pacific Conference
  Series, Vol. 411, Astronomical Data Analysis Software and Systems XVIII, ed.
  D.~A. {Bohlender}, D.~{Durand}, \& P.~{Dowler}, 251

\bibitem[{{Matsuura} {et~al.}(2011){Matsuura}, {Dwek}, {Meixner}, {Otsuka},
  {Babler}, {Barlow}, {Roman-Duval}, {Engelbracht}, {Sandstrom},
  {Laki{\'c}evi{\'c}}, {van Loon}, {Sonneborn}, {Clayton}, {Long}, {Lundqvist},
  {Nozawa}, {Gordon}, {Hony}, {Panuzzo}, {Okumura}, {Misselt}, {Montiel}, \&
  {Sauvage}}]{matsuura11}
{Matsuura}, M., {Dwek}, E., {Meixner}, M., {et~al.} 2011, Science, 333, 1258

\bibitem[{{Matsuura} {et~al.}(2015){Matsuura}, {Dwek}, {Barlow}, {Babler},
  {Baes}, {Meixner}, {Cernicharo}, {Clayton}, {Dunne}, {Fransson}, {Fritz},
  {Gear}, {Gomez}, {Groenewegen}, {Indebetouw}, {Ivison}, {Jerkstrand},
  {Lebouteiller}, {Lim}, {Lundqvist}, {Pearson}, {Roman-Duval}, {Royer},
  {Staveley-Smith}, {Swinyard}, {van Hoof}, {van Loon}, {Verstappen}, {Wesson},
  {Zanardo}, {Blommaert}, {Decin}, {Reach}, {Sonneborn}, {Van de Steene}, \&
  {Yates}}]{matsuura15}
{Matsuura}, M., {Dwek}, E., {Barlow}, M.~J., {et~al.} 2015, \apj, 800, 50

\bibitem[{{Matsuura} {et~al.}(2019){Matsuura}, {De Buizer}, {Arendt}, {Dwek},
  {Barlow}, {Bevan}, {Cigan}, {Gomez}, {Rho}, {Wesson}, {Bouchet}, {Danziger},
  \& {Meixner}}]{matsuura19}
{Matsuura}, M., {De Buizer}, J.~M., {Arendt}, R.~G., {et~al.} 2019, \mnras,
  482, 1715

\bibitem[{{Micha{\l}owski}(2015)}]{michalowski15}
{Micha{\l}owski}, M.~J. 2015, \aap, 577, A80

\bibitem[{{Millard} {et~al.}(2021){Millard}, {Ravi}, {Rho}, \&
  {Park}}]{millard21}
{Millard}, M.~J., {Ravi}, A.~P., {Rho}, J., \& {Park}, S. 2021, \apjs, 257, 36

\bibitem[{{Morgan} \& {Edmunds}(2003)}]{morgan03}
{Morgan}, H.~L., \& {Edmunds}, M.~G. 2003, \mnras, 343, 427

\bibitem[{{Morse} {et~al.}(1995){Morse}, {Winkler}, \& {Kirshner}}]{morse95}
{Morse}, J.~A., {Winkler}, P.~F., \& {Kirshner}, R.~P. 1995, \aj, 109, 2104

\bibitem[{{Morse} {et~al.}(1996){Morse}, {Blair}, {Dopita}, {Hughes},
  {Kirshner}, {Long}, {Raymond}, {Sutherland}, \& {Winkler}}]{morse96}
{Morse}, J.~A., {Blair}, W.~P., {Dopita}, M.~A., {et~al.} 1996, \aj, 112, 509

\bibitem[{{Nath} {et~al.}(2008){Nath}, {Laskar}, \& {Shull}}]{nath08}
{Nath}, B.~B., {Laskar}, T., \& {Shull}, J.~M. 2008, \apj, 682, 1055

\bibitem[{{Nozawa} {et~al.}(2007){Nozawa}, {Kozasa}, {Habe}, {Dwek}, {Umeda},
  {Tominaga}, {Maeda}, \& {Nomoto}}]{nozawa07}
{Nozawa}, T., {Kozasa}, T., {Habe}, A., {et~al.} 2007, \apj, 666, 955

\bibitem[{{Nozawa} {et~al.}(2003){Nozawa}, {Kozasa}, {Umeda}, {Maeda}, \&
  {Nomoto}}]{nozawa03}
{Nozawa}, T., {Kozasa}, T., {Umeda}, H., {Maeda}, K., \& {Nomoto}, K. 2003,
  \apj, 598, 785

\bibitem[{{Nozawa} {et~al.}(2008){Nozawa}, {Kozasa}, {Tominaga}, {Sakon},
  {Tanaka}, {Suzuki}, {Nomoto}, {Maeda}, {Umeda}, {Limongi}, \&
  {Onaka}}]{nozawa08}
{Nozawa}, T., {Kozasa}, T., {Tominaga}, N., {et~al.} 2008, \apj, 684, 1343

\bibitem[{{Owen} \& {Barlow}(2015)}]{owen15}
{Owen}, P.~J., \& {Barlow}, M.~J. 2015, \apj, 801, 141

\bibitem[{{Poglitsch} {et~al.}(2010){Poglitsch}, {Waelkens}, {Geis},
  {Feuchtgruber}, {Vandenbussche}, {Rodriguez}, {Krause}, {Renotte}, {van
  Hoof}, {Saraceno}, {Cepa}, {Kerschbaum}, {Agn{\`e}se}, {Ali}, {Altieri},
  {Andreani}, {Augueres}, {Balog}, {Barl}, {Bauer}, {Belbachir}, {Benedettini},
  {Billot}, {Boulade}, {Bischof}, {Blommaert}, {Callut}, {Cara}, {Cerulli},
  {Cesarsky}, {Contursi}, {Creten}, {De Meester}, {Doublier}, {Doumayrou},
  {Duband}, {Exter}, {Genzel}, {Gillis}, {Gr{\"o}zinger}, {Henning},
  {Herreros}, {Huygen}, {Inguscio}, {Jakob}, {Jamar}, {Jean}, {de Jong},
  {Katterloher}, {Kiss}, {Klaas}, {Lemke}, {Lutz}, {Madden}, {Marquet},
  {Martignac}, {Mazy}, {Merken}, {Montfort}, {Morbidelli}, {M{\"u}ller},
  {Nielbock}, {Okumura}, {Orfei}, {Ottensamer}, {Pezzuto}, {Popesso},
  {Putzeys}, {Regibo}, {Reveret}, {Royer}, {Sauvage}, {Schreiber}, {Stegmaier},
  {Schmitt}, {Schubert}, {Sturm}, {Thiel}, {Tofani}, {Vavrek}, {Wetzstein},
  {Wieprecht}, \& {Wiezorrek}}]{pogslitsch10}
{Poglitsch}, A., {Waelkens}, C., {Geis}, N., {et~al.} 2010, \aap, 518, L2

\bibitem[{{Pontoppidan} {et~al.}(2022){Pontoppidan}, {Barrientes}, {Blome},
  {Braun}, {Brown}, {Carruthers}, {Coe}, {DePasquale}, {Espinoza}, {Marin},
  {Gordon}, {Henry}, {Hustak}, {James}, {Jenkins}, {Koekemoer}, {LaMassa},
  {Law}, {Lockwood}, {Moro-Martin}, {Mullally}, {Pagan}, {Player}, {Proffitt},
  {Pulliam}, {Ramsay}, {Ravindranath}, {Reid}, {Robberto}, {Sabbi}, {Ubeda},
  {Balogh}, {Flanagan}, {Gardner}, {Hasan}, {Meinke}, \&
  {Nota}}]{pontoppidan22}
{Pontoppidan}, K.~M., {Barrientes}, J., {Blome}, C., {et~al.} 2022, \apjl, 936,
  L14

\bibitem[{{Priestley} {et~al.}(2022){Priestley}, {Arias}, {Barlow}, \& {De
  Looze}}]{priestley22}
{Priestley}, F.~D., {Arias}, M., {Barlow}, M.~J., \& {De Looze}, I. 2022,
  \mnras, 509, 3163

\bibitem[{{Priestley} {et~al.}(2019){Priestley}, {Barlow}, \& {De
  Looze}}]{priestley19}
{Priestley}, F.~D., {Barlow}, M.~J., \& {De Looze}, I. 2019, \mnras, 485, 440

\bibitem[{{Priestley} {et~al.}(2021){Priestley}, {Chawner}, {Matsuura}, {De
  Looze}, {Barlow}, \& {Gomez}}]{priestley21}
{Priestley}, F.~D., {Chawner}, H., {Matsuura}, M., {et~al.} 2021, \mnras, 500,
  2543

\bibitem[{{Ramsbottom} {et~al.}(2007){Ramsbottom}, {Hudson}, {Norrington}, \&
  {Scott}}]{ramsbottom07}
{Ramsbottom}, C.~A., {Hudson}, C.~E., {Norrington}, P.~H., \& {Scott}, M.~P.
  2007, \aap, 475, 765

\bibitem[{{Reach} \& {Rho}(2000)}]{reach00}
{Reach}, W.~T., \& {Rho}, J. 2000, \apj, 544, 843

\bibitem[{{Reach} {et~al.}(1995){Reach}, {Dwek}, {Fixsen}, {Hewagama},
  {Mather}, {Shafer}, {Banday}, {Bennett}, {Cheng}, {Eplee}, {Leisawitz},
  {Lubin}, {Read}, {Rosen}, {Shuman}, {Smoot}, {Sodroski}, \&
  {Wright}}]{reach95}
{Reach}, W.~T., {Dwek}, E., {Fixsen}, D.~J., {et~al.} 1995, \apj, 451, 188

\bibitem[{{Rho} {et~al.}(2001){Rho}, {Jarrett}, {Cutri}, \& {Reach}}]{rho01}
{Rho}, J., {Jarrett}, T.~H., {Cutri}, R.~M., \& {Reach}, W.~T. 2001, \apj, 547,
  885

\bibitem[{{Rho} {et~al.}(2009){Rho}, {Reach}, {Tappe}, {Hwang}, {Slavin},
  {Kozasa}, \& {Dunne}}]{rho09}
{Rho}, J., {Reach}, W.~T., {Tappe}, A., {et~al.} 2009, \apj, 700, 579

\bibitem[{{Rho} {et~al.}(2008){Rho}, {Kozasa}, {Reach}, {Smith}, {Rudnick},
  {DeLaney}, {Ennis}, {Gomez}, \& {Tappe}}]{rho08}
{Rho}, J., {Kozasa}, T., {Reach}, W.~T., {et~al.} 2008, \apj, 673, 271

\bibitem[{{Rho} {et~al.}(2018){Rho}, {Gomez}, {Boogert}, {Smith}, {Lagage},
  {Dowell}, {Clark}, {Peeters}, \& {Cami}}]{rho18}
{Rho}, J., {Gomez}, H.~L., {Boogert}, A., {et~al.} 2018, \mnras, 479, 5101

\bibitem[{{Salpeter}(1955)}]{salpeter55}
{Salpeter}, E.~E. 1955, \apj, 121, 161

\bibitem[{{Sano} {et~al.}(2020){Sano}, {Plucinsky}, {Bamba}, {Sharda},
  {Filipovi{\'c}}, {Law}, {Alsaberi}, {Yamane}, {Tokuda}, {Acero}, {Sasaki},
  {Vink}, {Inoue}, {Inutsuka}, {Shimoda}, {Tsuge}, {Fujii}, {Voisin}, {Maxted},
  {Rowell}, {Onishi}, {Kawamura}, {Mizuno}, {Yamamoto}, {Tachihara}, \&
  {Fukui}}]{sano20}
{Sano}, H., {Plucinsky}, P.~P., {Bamba}, A., {et~al.} 2020, \apj, 902, 53

\bibitem[{{Sarangi} \& {Cherchneff}(2015)}]{sarangi15}
{Sarangi}, A., \& {Cherchneff}, I. 2015, \aap, 575, A95

\bibitem[{{Sharda} {et~al.}(2020){Sharda}, {Gaetz}, {Kashyap}, \&
  {Plucinsky}}]{sharda20}
{Sharda}, P., {Gaetz}, T.~J., {Kashyap}, V.~L., \& {Plucinsky}, P.~P. 2020,
  \apj, 894, 145

\bibitem[{{Shull} \& {McKee}(1979)}]{shull79}
{Shull}, J.~M., \& {McKee}, C.~F. 1979, \apj, 227, 131

\bibitem[{{Siegel} {et~al.}(2021){Siegel}, {Dwarkadas}, {Frank}, \&
  {Burrows}}]{siegel21}
{Siegel}, J., {Dwarkadas}, V.~V., {Frank}, K.~A., \& {Burrows}, D.~N. 2021,
  arXiv e-prints, arXiv:2109.01157

\bibitem[{{Silvia} {et~al.}(2012){Silvia}, {Smith}, \& {Shull}}]{silvia12}
{Silvia}, D.~W., {Smith}, B.~D., \& {Shull}, J.~M. 2012, \apj, 748, 12

\bibitem[{{Slavin} {et~al.}(2020){Slavin}, {Dwek}, {Mac Low}, \&
  {Hill}}]{slavin20}
{Slavin}, J.~D., {Dwek}, E., {Mac Low}, M.-M., \& {Hill}, A.~S. 2020, \apj,
  902, 135

\bibitem[{{Sluder} {et~al.}(2018){Sluder}, {Milosavljevi{\'c}}, \&
  {Montgomery}}]{sluder18}
{Sluder}, A., {Milosavljevi{\'c}}, M., \& {Montgomery}, M.~H. 2018, \mnras,
  480, 5580

\bibitem[{{Smith} {et~al.}(2007){Smith}, {Draine}, {Dale}, {Moustakas},
  {Kennicutt}, {Helou}, {Armus}, {Roussel}, {Sheth}, {Bendo}, {Buckalew},
  {Calzetti}, {Engelbracht}, {Gordon}, {Hollenbach}, {Li}, {Malhotra},
  {Murphy}, \& {Walter}}]{smith07}
{Smith}, J.~D.~T., {Draine}, B.~T., {Dale}, D.~A., {et~al.} 2007, \apj, 656,
  770

\bibitem[{{Sommovigo} {et~al.}(2022){Sommovigo}, {Ferrara}, {Pallottini},
  {Dayal}, {Bouwens}, {Smit}, {da Cunha}, {De Looze}, {Bowler}, {Hodge},
  {Inami}, {Oesch}, {Endsley}, {Gonzalez}, {Schouws}, {Stark}, {Stefanon},
  {Aravena}, {Graziani}, {Riechers}, {Schneider}, {van der Werf}, {Algera},
  {Barrufet}, {Fudamoto}, {Hygate}, {Labb{\'e}}, {Li}, {Nanayakkara}, \&
  {Topping}}]{sommovigo22}
{Sommovigo}, L., {Ferrara}, A., {Pallottini}, A., {et~al.} 2022, \mnras,
  arXiv:2202.01227

\bibitem[{{Spilker} {et~al.}(2018){Spilker}, {Aravena}, {B{\'e}thermin},
  {Chapman}, {Chen}, {Cunningham}, {De Breuck}, {Dong}, {Gonzalez}, {Hayward},
  {Hezaveh}, {Litke}, {Ma}, {Malkan}, {Marrone}, {Miller}, {Morningstar},
  {Narayanan}, {Phadke}, {Sreevani}, {Stark}, {Vieira}, \&
  {Wei{\ss}}}]{spilker18}
{Spilker}, J.~S., {Aravena}, M., {B{\'e}thermin}, M., {et~al.} 2018, Science,
  361, 1016

\bibitem[{{Sutherland} {et~al.}(1993){Sutherland}, {Bicknell}, \&
  {Dopita}}]{sutherland93}
{Sutherland}, R.~S., {Bicknell}, G.~V., \& {Dopita}, M.~A. 1993, \apj, 414, 510

\bibitem[{{Sutherland} \& {Dopita}(1995)}]{sutherland95}
{Sutherland}, R.~S., \& {Dopita}, M.~A. 1995, \apj, 439, 381

\bibitem[{{Swinyard} {et~al.}(2010){Swinyard}, {Ade}, {Baluteau}, {Aussel},
  {Barlow}, {Bendo}, {Benielli}, {Bock}, {Brisbin}, {Conley}, {Conversi},
  {Dowell}, {Dowell}, {Ferlet}, {Fulton}, {Glenn}, {Glauser}, {Griffin},
  {Griffin}, {Guest}, {Imhof}, {Isaak}, {Jones}, {King}, {Leeks}, {Levenson},
  {Lim}, {Lu}, {Makiwa}, {Naylor}, {Nguyen}, {Oliver}, {Panuzzo},
  {Papageorgiou}, {Pearson}, {Pohlen}, {Polehampton}, {Pouliquen},
  {Rigopoulou}, {Ronayette}, {Roussel}, {Rykala}, {Savini}, {Schulz},
  {Schwartz}, {Shupe}, {Sibthorpe}, {Sidher}, {Smith}, {Spencer}, {Trichas},
  {Triou}, {Valtchanov}, {Wesson}, {Woodcraft}, {Xu}, {Zemcov}, \&
  {Zhang}}]{swinyard10}
{Swinyard}, B.~M., {Ade}, P., {Baluteau}, J.~P., {et~al.} 2010, \aap, 518, L4

\bibitem[{{Tacchella} {et~al.}(2022){Tacchella}, {Finkelstein}, {Bagley},
  {Dickinson}, {Ferguson}, {Giavalisco}, {Graziani}, {Grogin}, {Hathi},
  {Hutchison}, {Jung}, {Koekemoer}, {Larson}, {Papovich}, {Pirzkal},
  {Rojas-Ruiz}, {Song}, {Schneider}, {Somerville}, {Wilkins}, \&
  {Yung}}]{sandro22}
{Tacchella}, S., {Finkelstein}, S.~L., {Bagley}, M., {et~al.} 2022, \apj, 927,
  170

\bibitem[{{Tappe} {et~al.}(2012){Tappe}, {Rho}, {Boersma}, \&
  {Micelotta}}]{tappe12}
{Tappe}, A., {Rho}, J., {Boersma}, C., \& {Micelotta}, E.~R. 2012, \apj, 754,
  132

\bibitem[{{Tappe} {et~al.}(2006){Tappe}, {Rho}, \& {Reach}}]{tappe06}
{Tappe}, A., {Rho}, J., \& {Reach}, W.~T. 2006, \apj, 653, 267

\bibitem[{{Temim} {et~al.}(2017){Temim}, {Dwek}, {Arendt}, {Borkowski},
  {Reynolds}, {Slane}, {Gelfand}, \& {Raymond}}]{temim17}
{Temim}, T., {Dwek}, E., {Arendt}, R.~G., {et~al.} 2017, \apj, 836, 129

\bibitem[{{Temim} {et~al.}(2012){Temim}, {Sonneborn}, {Dwek}, {Arendt},
  {Gehrz}, {Slane}, \& {Roellig}}]{temim12}
{Temim}, T., {Sonneborn}, G., {Dwek}, E., {et~al.} 2012, \apj, 753, 72

\bibitem[{{Todini} \& {Ferrara}(2001)}]{todini01}
{Todini}, P., \& {Ferrara}, A. 2001, \mnras, 325, 726

\bibitem[{{Treu} {et~al.}(2022){Treu}, {Roberts-Borsani}, {Bradac}, {Brammer},
  {Fontana}, {Henry}, {Mason}, {Morishita}, {Pentericci}, {Wang}, {Acebron},
  {Bagley}, {Bergamini}, {Belfiori}, {Bonchi}, {Boyett}, {Boutsia},
  {Calabr{\'o}}, {Caminha}, {Castellano}, {Dressler}, {Glazebrook}, {Grillo},
  {Jacobs}, {Jones}, {Kelly}, {Leethochawalit}, {Malkan}, {Marchesini},
  {Mascia}, {Mercurio}, {Merlin}, {Nanayakkara}, {Nonino}, {Paris},
  {Poggianti}, {Rosati}, {Santini}, {Scarlata}, {Shipley}, {Strait}, {Trenti},
  {Tubthong}, {Vanzella}, {Vulcani}, \& {Yang}}]{treu22}
{Treu}, T., {Roberts-Borsani}, G., {Bradac}, M., {et~al.} 2022, \apj, 935, 110

\bibitem[{{Vogt} \& {Dopita}(2011)}]{vogt11}
{Vogt}, F., \& {Dopita}, M.~A. 2011, \apss, 331, 521

\bibitem[{{Wang} {et~al.}(2010){Wang}, {Carilli}, {Neri}, {Riechers}, {Wagg},
  {Walter}, {Bertoldi}, {Menten}, {Omont}, {Cox}, \& {Fan}}]{wang10}
{Wang}, R., {Carilli}, C.~L., {Neri}, R., {et~al.} 2010, \apj, 714, 699

\bibitem[{{Woosley} \& {Weaver}(1995)}]{woosley95}
{Woosley}, S.~E., \& {Weaver}, T.~A. 1995, \apjs, 101, 181

\bibitem[{{Wright} {et~al.}(2010){Wright}, {Eisenhardt}, {Mainzer}, {Ressler},
  {Cutri}, {Jarrett}, {Kirkpatrick}, {Padgett}, {McMillan}, {Skrutskie},
  {Stanford}, {Cohen}, {Walker}, {Mather}, {Leisawitz}, {Gautier}, {McLean},
  {Benford}, {Lonsdale}, {Blain}, {Mendez}, {Irace}, {Duval}, {Liu}, {Royer},
  {Heinrichsen}, {Howard}, {Shannon}, {Kendall}, {Walsh}, {Larsen}, {Cardon},
  {Schick}, {Schwalm}, {Abid}, {Fabinsky}, {Naes}, \& {Tsai}}]{wright10}
{Wright}, E.~L., {Eisenhardt}, P. R.~M., {Mainzer}, A.~K., {et~al.} 2010, \aj,
  140, 1868

\end{thebibliography}

\end{document}